\documentclass[
    aps,
    reprint,
    nofootinbib,
    longbibliography,
    superscriptaddress
]{revtex4-2}

\usepackage[english]{babel}
\usepackage[utf8x]{inputenc}
\usepackage[T1]{fontenc}

\usepackage{amsmath}
\usepackage{xcolor}
\usepackage[colorlinks,allcolors=violet]{hyperref}
\usepackage{cleveref}
\usepackage{graphicx}
\usepackage{booktabs}
\usepackage{glossaries}
\usepackage{xspace}
\usepackage{stix}
\usepackage{physics}
\usepackage[retain-unity-mantissa=false]{siunitx}
\usepackage[frozencache,cachedir=./]{minted}

\renewcommand{\qc}{\,\text{,}}
\newcommand{\qs}{\,\text{.}}

\newcommand{\psd}[1]{\opbraces{S_{#1}}}

\newcommand{\filter}{{\mathbf{F}}}
\newcommand{\delay}[1]{{\mathbf{D}_{#1}}}
\newcommand{\ddelay}[1]{{\dot{\mathbf{D}}_{#1}}}

\newcommand{\cmplx}[1]{\opbraces{\text{#1}}}
\newcommand{\ft}[1]{\opbraces{\mathcal{F}\left[{#1}\right]}}
\newcommand{\timestamp}[1]{\opbraces{\mathbf{T}_{#1}}}
\newcommand{\dtimestamp}[1]{\opbraces{\dot{\mathbf{T}}_{#1}}}

\DeclareSIUnit\year{yr}
\DeclareSIUnit\cycle{cycle}

\newcommand{\pytdi}{\textsc{PyTDI}\xspace}
\newcommand{\lisanode}{\textsc{LISANode}\xspace}
\newcommand{\lisainstrument}{\textsc{LISA Instrument}\xspace}
\newcommand{\lisaorbits}{\textsc{LISA Orbits}\xspace}
\newcommand{\lisagwresponse}{\textsc{LISA GW Response}\xspace}
\newcommand{\lisaglitch}{\textsc{LISA Glitch}\xspace}
\newcommand{\cpp}{C\nolinebreak\hspace{-.05em}\raisebox{.4ex}{\tiny\bf +}\nolinebreak\hspace{-.10em}\raisebox{.4ex}{\tiny\bf +}\xspace}

\newacronym{aei}{AEI}{Albert Einstein Institute}
\newacronym{gfz}{GFZ}{Deutsches GeoForschungsZentrum}

\newacronym{em}{EM}{electromagnetic}
\newacronym{gw}{GW}{gravitational wave}
\newacronym{asd}{ASD}{amplitude spectral density}
\newacronym{psd}{PSD}{power spectral density}
\newacronym{csd}{CSD}{cross spectral density}

\newacronym{grace-fo}{GRACE-FO}{Gravity Recovery and Climate Experiment-Follow-On}
\newacronym{lisa}{LISA}{the Laser Interferometer Space Antenna}
\newacronym{esa}{ESA}{the European Space Agency}
\newacronym{cdf}{CDF}{Concurrent Design Facility}
\newacronym{gaia}{GAIA}{Global Astrometric Interferometer for Astrophysics}
\newacronym{ldpg}{LDPG}{LISA data-processing group}
\newacronym{ldc}{LDC}{LISA data challenge}
\newacronym{inrep}{INREP}{initial noise-reduction pipeline}
\newacronym{wg}{WG}{working group}
\newacronym{aligo}{aLIGO}{advanced Laser Interferometer Gravitational-Wave Observatory}
\newacronym{emri}{EMRI}{Extreme mass-ratio inspiral}
\newacronym{pta}{PTA}{pulsar timing array}
\newacronym{lfn}{LFN}{laser frequency noise}
\newacronym{ob}{OB}{optical bench}
\newacronym{isi}{ISI}{interspacecraft interferometer}
\newacronym{tmi}{TMI}{test-mass interferometer}
\newacronym{rfi}{RFI}{reference interferometer}

\newacronym{tdi}{TDI}{time-delay interferometry}
\newacronym{tdir}{TDI-R}{time-delay interferometric ranging}

\newacronym{ppr}{PPR}{proper pseudo-range}
\newacronym{mpr}{MPR}{measured pseudo-range}

\newacronym[longplural={movable optical subassemblies}]{mosa}{MOSA}{movable optical subassembly}
\newacronym{uso}{USO}{ultrastable oscillator}
\newacronym{scet}{SCET}{spacecraft elapsed time}
\newacronym{eom}{EOM}{electro-optical modulator}
\newacronym{adc}{ADC}{analog-to-digital converter}
\newacronym{dac}{DAC}{digital-to-analog converter}
\newacronym{fpga}{FPGA}{field programmable gate array}
\newacronym{dsp}{DSP}{digital signal processing}
\newacronym{prn}{PRN}{pseudo-random noise}
\newacronym{dfacs}{DFACS}{drag-free attitude control system}
\newacronym{grs}{GRS}{gravitational reference sensor}
\newacronym{dws}{DWS}{differential wavefront sensing}
\newacronym{oms}{OMS}{optical metrology system}
\newacronym{pll}{PLL}{phase-locked loop}
\newacronym{dpll}{DPLL}{digital phase-locked loop}
\newacronym{dll}{DLL}{delay-locked loop}
\newacronym{nco}{NCO}{numerically controlled oscillator}
\newacronym{pir}{PIR}{phase increment register}
\newacronym{pa}{PA}{phase accumulator}
\newacronym{lut}{LUT}{look-up table}
\newacronym{qpd}{QPD}{quadrant photodiode}
\newacronym{fds}{FDS}{frequency distribution system}
\newacronym{ttl}{TTL}{tilt-to-length}
\newacronym{hatt}{HATT}{high accuracy time transfer}
\newacronym{dsn}{DSN}{deep-space network}
\newacronym{estrack}{ESTRACK}{European Space Tracking}
\newacronym{icrs}{ICRS}{international celestial reference system}
\newacronym{icrf}{ICRF}{international celestial reference frame}
\newacronym{bcrs}{BCRS}{barycentric celestial reference system}
\newacronym{gcrs}{GCRS}{geocentric celestial reference system}
\newacronym{srs}{SRS}{spacecraft reference system}
\newacronym{lpsd}{LPSD}{log-scale power spectral density}

\newacronym{rin}{RIN}{relative intensity noise}

\newacronym{tcb}{TCB}{barycentric coordinate time}
\newacronym{tcg}{TCG}{geocentric coordinate time}
\newacronym{tps}{TPS}{spacecraft proper time}
\newacronym{the}{THE}{on-board clock time}

\newacronym{fir}{FIR}{finite impulse response}
\newacronym{snr}{SNR}{signal-to-noise ratio}

\newacronym{xo}{XO}{crystal oscillator}
\newacronym{vcxo}{VCXO}{voltage controlled crystal oscillator}
\newacronym{tcxo}{TCXO}{temperature compensated crystal oscillator}
\newacronym{ocxo}{OCXO}{oven controlled crystal oscillator}
\newacronym{afs}{AFS}{atomic frequency standard}
\newacronym{dut}{DUT}{device under test}
\newacronym{lo}{LO}{local oscillator}

\newacronym{gps}{GPS}{Global Positioning System}
\newacronym{gnss}{GNSS}{Global Navigation(al) Satellite System}
\newacronym{glonass}{GLONASS}{Globalnaja nawigazionnaja sputnikowaja sistema}
\newacronym{aces}{ACES}{Atomic Clock Ensemble in Space}
\newacronym{iss}{ISS}{International Space Station}

\newacronym{cbe}{CBE}{current best estimate}

\begin{document}

%% Preamble

\title{Unified model for the LISA measurements and instrument simulations}

\author{Jean-Baptiste Bayle}
\email{j2b.bayle@gmail.com}
\affiliation{University of Glasgow, Glasgow G12 8QQ, United Kingdom}

\author{Olaf Hartwig}
\affiliation{SYRTE, Observatoire  de  Paris,  Universit\'e  PSL, CNRS,  Sorbonne  Universit\'e,  LNE,  61 avenue de l’Observatoire 75014  Paris,  France}
\affiliation{Max-Planck-Institut für Gravitationsphysik (Albert-Einstein-Institut),
Callinstraße 38, 30167 Hannover, Germany}

\date{\today}



%% Abstract
\begin{abstract}
\glsentryshort{lisa} is a space-based \si{\milli\hertz} gravitational-wave observatory, with a planned launch in 2034. It is expected to be the first detector of its kind, and will present unique challenges in instrumentation and data analysis. An accurate preflight simulation of \glsentryshort{lisa} data is a vital part of the development of both the instrument and the analysis methods. The simulation must include a detailed model of the full measurement and analysis chain, capturing the main features that affect the instrument performance and processing algorithms. Here, we propose a new model that includes, for the first time, proper relativistic treatment of reference frames with realistic orbits; a model for onboard clocks and clock synchronization measurements; proper modeling of total laser frequencies, including laser locking, frequency planning and Doppler shifts; better treatment of onboard processing and updated noise models. We then introduce two implementations of this model, \lisanode and \lisainstrument. We demonstrate that \glsentryshort{tdi} processing successfully recovers gravitational-wave signals from the significantly more realistic and complex simulated data. \lisanode and \lisainstrument are already widely used by the \glsentryshort{lisa} community and, for example, currently provide the mock data for the LISA Data Challenges.
\end{abstract}
\maketitle

\section{Introduction}
\label{introduction}

Following the opening of the gravitational Universe by the many observations of ground-based gravitational-wave detectors~\cite{LIGOScientific:2016aoc,LIGOScientific:2016sjg,LIGOScientific:2017bnn,LIGOScientific:2017vox,LIGOScientific:2017ycc,LIGOScientific:2017vwq,LIGOScientific:2018mvr,LIGOScientific:2020aai,LIGOScientific:2020stg,LIGOScientific:2020zkf,LIGOScientific:2020iuh,LIGOScientific:2020ibl,LIGOScientific:2021usb,LIGOScientific:2021qlt,LIGOScientific:2021djp}, \gls{esa} has selected \gls{lisa} as the L3 mission. \Gls{lisa} is a space-borne gravitational-wave observatory sensitive to gravitational signals between \SI{0.1}{\milli\hertz} and \SI{1}{\hertz}, where we expect a large diversity of sources, ranging from supermassive black-hole binaries, quasi-monochromatic Galactic binaries, extreme mass-ratio inspirals, and stellar-mass binaries~\cite{LISA:2017pwj}. In addition to these expected sources, a number of potential signals might be detected, including stochastic gravitational-wave signals from the early Universe, cusps and kinks of cosmic strings and other unmodeled burst sources. Precise measurements of the source parameters will help answer many astrophysical and cosmological questions, as well as constrain models beyond the general theory of relativity.

Achieving these outstanding science objectives will present challenges in both instrumentation and data analysis. Contrary to ground-based gravitational-wave observatories, \gls{lisa} is expected to be signal dominated, with tens of thousands of sources of different kind present in the \gls{lisa} band at all times. Telling all of these sources apart and estimating their parameters requires novel approaches to data analysis (explored in the context of the \gls{lisa} Data Challenges), the development and testing of which necessitates realistic simulated data. In addition, \gls{lisa} will make use of sophisticated noise reduction algorithms to reject the most dominant instrumental noise sources. The core of these algorithms is known as \gls{tdi}, in which multiple data streams are combined with appropriate time shifts to generate virtual equal-arm interferometers in postprocessing. Understanding how different noise sources couple into the data is crucial to guide the development of such noise-reduction pipelines. Finally, with a planned launch in 2034, the \gls{lisa} mission is currently preparing for adoption. The development of a simulation model is needed to support these activities, validate the instrument design and ensure that the science objectives can be achieved.

To fulfill these objectives, one needs to capture in the simulation model the main features that affect the instrument performance and processing algorithms. The simulated data should be representative of the time series we will receive from the real instrument. Therefore, we focus in this paper on a time-domain instrument model. In addition, we must be able to simulate several years of data in a reasonable time to evaluate different instrument configurations for full mission duration, currently planned as \SI{4}{\year}~\cite{LISA:2017pwj}; this makes a detailed engineering-level simulation unfeasible.

This instrument model builds on a legacy of previous constellation-level \gls{lisa} simulators. \textsc{LISA Simulator} was developed to quickly generate measurement data~\cite{Cornish:2003tz,Rubbo:2003ap}. The simulator worked exclusively in the frequency domain and was based on transfer functions for a simple instrumental model. \textsc{Synthetic LISA} was a Python-based simulator that worked in the time domain and used an idealized (and now out-of-date) instrumental configuration to study the performance of noise reduction algorithms for a constellation with time-varying arm lengths~\cite{Vallisneri:2004bn}. \textsc{TDISim} was a prototype \gls{tdi} simulation tool programmed in Matlab. The simulation fully operated in the time domain and performed both data generation and \gls{tdi}, including for the first time the updated split-interferometry optical bench design and a simplified state-space model for the motion of the test mass and the spacecraft~\cite{Otto:2015erp}.

We based our simulation efforts on \textsc{LISACode}, which was initiated with the similar ambition to include most of the ingredients that were thought to influence \gls{lisa}'s performance at the time~\cite{Petiteau:2008zz}. Since then, developments in the instrument and mission design revealed new important effects that must be included in the simulations. The model that we propose in this paper is an attempt to extend \textsc{LISACode}'s model to capture those effects.

\Cref{sec:framework-and-conventions} introduces the conventions we use, and, for the first time, a description of the time frames relevant for \gls{lisa} instrument simulations. In \cref{sec:optical-simulation}, we describe the optical simulation model, which includes the up-to-date split-interferometry optical bench design. Contrary to previous simulators, we properly model the total laser frequencies, as well as realistic orbits and any Doppler effects arising from differential spacecraft motion. We also account for the sideband modulations used to correct for clock errors. Then, we describe in \cref{sec:phase-readout} the readout of the interferometric beatnotes and how it is affected by imperfections of onboard clocks. Our treatment of the onboard processing is presented in \cref{sec:onboard-processing}; here, we also give the equations for the final phasemeter readouts. In \cref{sec:locking}, we describe how we model laser locking, and its impact on the measurements. Lastly, in \cref{sec:pseudoranging}, we give a high-level model of the pseudoranging measurements that are used to estimate the arm lengths. We then introduce \lisanode and \lisainstrument, two implementations of this simulation model, and discuss their performances in \cref{sec:implementation}. Finally, in \cref{sec:results}, we show simulation results and highlight the main features that differ from previously simulated data. We demonstrate that despite the added complexity, we can recover gravitational signals using the latest noise-reduction algorithms. We conclude in \cref{sec:conclusion}.

\section{Framework and conventions}
\label{sec:framework-and-conventions}

\subsection{Constellation overview}

\Gls{lisa} is an almost equilateral triangle, composed of 3 identical spacecraft, which we label 1, 2, 3 clockwise when looking down at their solar panels. These spacecraft exchange laser beams, which are combined on optical benches inside \glspl{mosa}.

\begin{figure}
	\centering
	\includegraphics[width=\columnwidth]{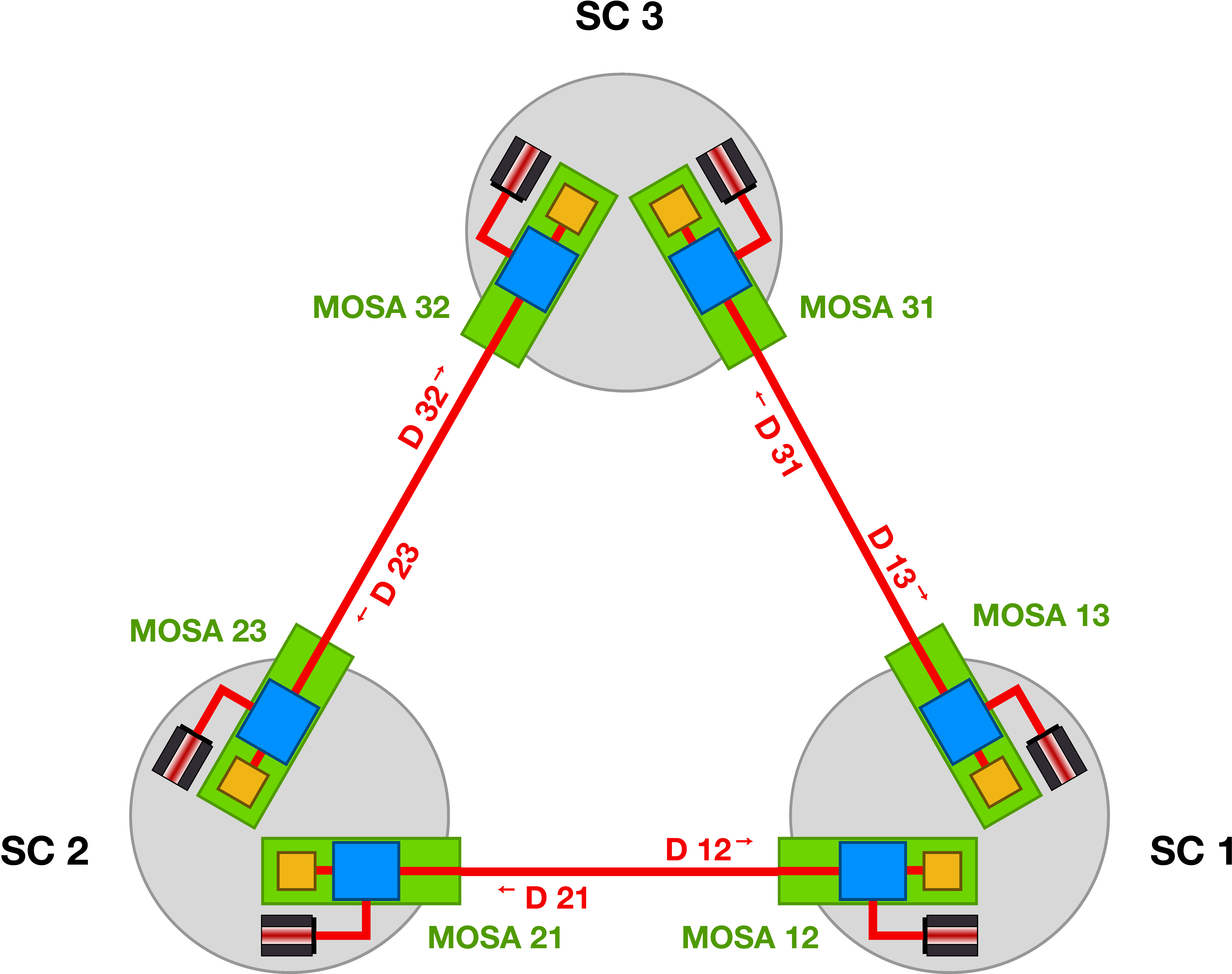}
	\caption{Labeling conventions used for spacecraft, light travel times, lasers, \glspl{mosa}, and interferometric measurements (from~\cite{Bayle:2021mue}).}
	\label{fig:labeling}
\end{figure}

To uniquely label these \glspl{mosa}, we use two indices. The first one is that of the spacecraft the \gls{mosa} is mounted, while the second index is that of the spacecraft the \gls{mosa} is pointing to. Most components of interest (such as the optical benches, test masses, etc.) can be uniquely associated to one of the \glspl{mosa}, in which case we use the same two indices. Elements that exist only once onboard a spacecraft, such as the \glspl{uso}, are indexed by that spacecraft index. These labeling conventions, which are largely based on the proposed unified conventions of the \gls{lisa} consortium~\cite{Hartwig:2020tdu}, are summarized in \cref{fig:labeling}.

Quantities that describe a process that involves the propagation between two spacecraft will be interpreted as being associated with the spacecraft in which the quantity is measured. For example, the gravitational-wave signal observed in the interferometer on \gls{mosa}~12 will be indexed with the same indices~12. The same convention applies to the propagation delay of a beam arriving on spacecraft~1 from spacecraft~2, which will be labeled by the indices~12.

In this paper, we often derive equations for a specific spacecraft or \gls{mosa}. The expressions for the other 2 spacecraft or the other 5 \glspl{mosa} can then be deduced by combining cyclic permutations $\{1\rightarrow2$, $2\rightarrow3$, $3\rightarrow1\}$, and reflections $\{1\rightarrow1, 2\rightarrow3, 3\rightarrow2\}$.

\subsection{Time coordinate frames}

The instrumental simulation mostly concerns itself with the physics inside a spacecraft (e.g., the evolution of laser beam phases and their interferometric beatnotes), which is best modeled in the three \glsfirst{tps}. These time frames are defined as the times shown by perfect clocks comoving with the spacecraft centers of mass. We denote them with $\tau_1$, $\tau_2$, and $\tau_3$.

The \glspl{tps} are idealized timescales, which cannot be realized in practice. All measurements instead refer to an imperfect on-board timer, which represents an approximation of the associated \gls{tps}. We denote these three onboard clock time frames as $\hat{\tau}_1$, $\hat{\tau}_2$, and $\hat{\tau}_3$.

Finally, processes on the Solar-system scale are modeled according to a global time frame, such as the \glsfirst{tcb}, denoted $t$. This is the case for the spacecraft orbits or the gravitational waveforms. Our instrumental simulation does not make a direct use of the \gls{tcb}. Instead, we rely on external tools, such as \textsc{LISA Orbits}~\cite{lisaorbits}, to directly compute quantities expressed in the \glspl{tps}.

In general, signals are expressed in their \textit{natural} time coordinate. E.g., laser beam phases and beatnotes are expressed in the \gls{tps} of the spacecraft housing the laser. It is sometimes useful to express a signal in a different time coordinate. To prevent confusion, we will use the same symbol but add a superscript denoting the time coordinate. For example, a phase $\phi$ could be expressed as a function of the \gls{tps}~1, writing $\phi^{\tau_1}(x)$, or as a function of the clock time of that spacecraft, writing $\phi^{\hat\tau_1}(x)$. Note that the symbol used for the function argument is arbitrary, and does not specify the reference frame. We will often use $\tau$ without subscripts as a generic function argument.

Conversions between time coordinates can easily be expressed with these conventions. For example, $\tau_1^{\hat\tau_1}(\tau)$ is the \gls{tps} as a function of the clock time onboard spacecraft~1. Trivially,
\begin{equation}
    t^t(\tau) = \tau_1^{\tau_1}(\tau)
    = \hat\tau_1^{\hat\tau_1}(\tau) = \tau
    \qs
    \label{eq:time-coordinate-conversion-identity}
\end{equation}

It is often useful to model the deviation of the onboard clock time with respect to the associated \gls{tps}. We adopt the notation
\begin{subequations}
    \begin{align}
    \tau_1^{\hat\tau_1}(\tau) &= \tau + \delta \tau_1^{\hat\tau_1}(\tau)
    \qc
    \\
    \hat\tau_1^{\tau_1}(\tau) &= \tau + \delta \hat\tau_1^{\tau_1}(\tau)
    \qs
    \end{align}
    \label{eq:time-deviation-notation}
\end{subequations}

One important class of signals we study are phases $\phi$ of electromagnetic waves. As scalar quantities, these are invariant under coordinate transformations, such that they transform from one time frame to another using a simple time shift,
\begin{equation}
    \phi^{\tau_1}(\tau)
    = \phi^{\hat\tau_1}(\hat\tau_1^{\tau_1}(\tau))
    \qs
	\label{eq:phase-time-coordinate-conversion}
\end{equation}

\section{Optical simulation}
\label{sec:optical-simulation}

In this section, we derive the model for the generation and propagation of the laser beams, as well as their interference at the photodiodes.

\subsection{Optical bench design}

As illustrated in \cref{fig:labeling}, each spacecraft hosts two optical benches. We usually refer to one optical bench as the \textit{local} optical bench; the other optical bench hosted by the same spacecraft as the \textit{adjacent} optical bench; we call the \textit{distant} optical bench the one situated on the spacecraft exchanging light with the local optical bench. Each optical bench is associated with a laser source, a \gls{grs} containing a free-falling test mass, and telescope to send and collect light to and from distant spacecraft.

\begin{figure}
	\centering
	\includegraphics[width=\columnwidth]{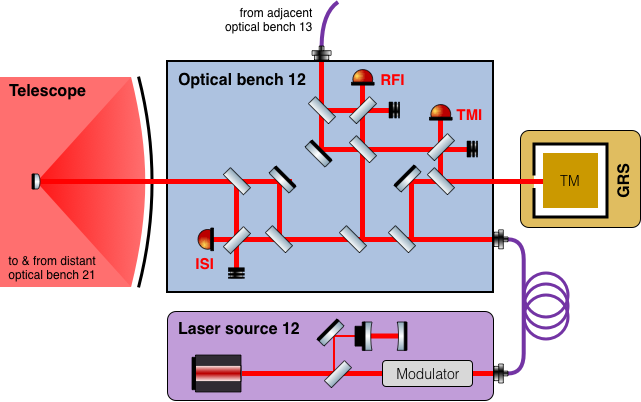}
	\caption{Schematics of the optical design implemented in the simulation, along with notations for the laser beam and beatnote total phases, here for \gls{mosa}~12.}
	\label{fig:optical-design}
\end{figure}

Laser beams are combined in 3 different heterodyne interferometers. The \gls{isi} mixes the local beam with the distant beam (coming from the distant optical bench); the \gls{tmi} mixes the local and adjacent beams, after it has bounced on the local test mass; and the \gls{rfi} mixes the local and adjacent beams without interaction with the test mass. \Cref{fig:optical-design} gives an overview of the optical bench~12.

In reality, each single interferometer output is implemented using redundant balanced detection with four \glspl{qpd}. We do not simulate balanced detection, and only consider a single data stream for each interferometer. Additional readouts related to the laser beams alignment, such as the \gls{dws}, are not included in the model presented here. We are currently working to implement them in the simulation by propagating additional independent variables representing the different beam tilts. We plan to describe this model in more detail in a follow-up paper.

\subsection{Laser beam model}
\label{sec:laser-beam-model}

\subsubsection{Simple laser beam}

We use a number of assumptions to model the information carried by the electromagnetic field of a laser beam (in all generality, these are two 3-vector fields).

We work in the plane-wave approximation, and assume that any effects due to wavefront imperfections can be modeled as equivalent longitudinal path length variations. In addition, we neglect effects related to the fields polarization, and assume that the waves propagate in a perfect vacuum, such that we only model the scalar electric field amplitude (we do not model the magnetic field amplitude, as it is completely determined by the electrical field amplitude~\cite{Wolski:2011fy}).

At any fixed point inside a spacecraft, the complex amplitude of the electromagnetic field associated with a laser beam can be written as
\begin{equation}
	E(\tau) = E_0(\tau) e^{i 2\pi \Phi(\tau)}
	\qc
	\label{eq:electromagnetic-signal}
\end{equation}
where $\Phi(\tau)$ is the total phase in units of cycles.

\Gls{lisa} ultimately measures phase differences, such that we do not simulate the field amplitude term $E_0$, but only the phase $\Phi(\tau)$. We expect couplings between the field amplitude and the measured phase difference (e.g., the relative intensity noise~\cite{Wissel:2022bpi}). We currently do not model these effects, but assume that they can be modeled as equivalent phase noise in the final readout.

\subsubsection{Phase or frequency?}

The optical frequency of the lasers is around $\nu_0=\SI{281.6}{\tera\hertz}$, such that the total phase increases quickly with time. This makes using it challenging for numerical simulations, as any variable representing the total phase will either numerically overflow when using fixed-point arithmetic, or eventually suffer an unacceptable loss of precision when using floating-point arithmetic\footnote{For a precision better than a micro-cycle, a 64-bit integer representing the total phase will overflow every \SI{0.07}{\second}.}.

To avoid these issues, we simulate frequencies instead of phase (given by $\nu = \dot{\Phi}$, since we express the phase in units of cycles), which are controlled to remain at the same order of magnitude during the whole mission duration. However, modeling the propagation of laser beams is often easier in phase. Therefore, we will derive most of the equations of this paper both in units of phase and frequency.

\subsubsection{Two-variable decomposition}
\label{sec:two-variable-decomposition}

In \gls{lisa}, effects on the laser beams come into play at completely different timescales and dynamic ranges. On the one hand, some effects modulate the frequency of our beams on a time scale of the orbital revolution around the Sun, which lies well outside our measurement frequency band (below \SI{E-4}{\hertz}). These effects tend to have large dynamic ranges; for instance, the Doppler shifts caused by the relative spacecraft motion can fluctuate by several megahertz over the mission duration.

On the other hand, we want to track small phase or frequency fluctuations within our measurement band (from \SI{E-4}{\hertz} and up to \SI{1}{\hertz}), caused by gravitational-wave signals and instrumental noises. These fluctuations have a much smaller amplitude. The laser noise being the dominant effect, causing the heterodyne beatnote frequency to shift by about a few tens of Hertz, while gravitational waves typically cause frequency shifts of a few hundreds of nanohertz.

To address this problem, we model these different effects independently. We decompose the laser beam frequency into one constant and two variables,
\begin{equation}
    \nu(\tau) = \nu_0 + \nu^o(\tau) + \nu^\epsilon(\tau)
    \qs
    \label{eq:two-variable-beam-frequency}
\end{equation}
The large frequency offsets $\nu^o(\tau)$ are used to represent frequency-plan offsets and Doppler shifts, both on the order of megahertz, as well as the gigahertz sideband frequency offsets. The small frequency fluctuations $\nu^\epsilon(\tau)$, on the other hand, are used to describe gravitational signals and noises. A simple laser beam would therefore be entirely represented by the couple $\qty{\nu^o(\tau), \nu^\epsilon(\tau)}$.

Alternatively, we can express \cref{eq:two-variable-beam-frequency} in phase units by writing the total phase as
\begin{equation}
    \Phi(\tau) = \nu_0 \tau + \phi^o(\tau) + \phi^\epsilon(\tau) + \phi^0
    \qc
    \label{eq:two-variable-beam-phase}
\end{equation}
where the definitions of large phase drifts $\phi^o(\tau)$ and small phase fluctuations $\phi^\epsilon(\tau)$ follow from \cref{eq:two-variable-beam-frequency},
\begin{equation}
    \nu^o(\tau) = \dot{\phi}^o\qty(\tau)
    \qand
    \nu^\epsilon(\tau) = \dot{\phi}^\epsilon \qty(\tau)
    \qs
    \label{eq:two-variable-beam-phase-frequency-relationship}
\end{equation}
As we simulate frequencies, we do not track the initial phase of the laser beam $\phi^0 \in [0, 2\pi]$ in the following.

Let us stress that this decomposition is entirely artificial. In reality, we will only have access to the total phase or frequency. Therefore, to produce data representative of the real instrument telemetry, we always compute the total phase or frequency as the final simulation output.

\subsubsection{Modulated beams}

In \gls{lisa}, laser beams are phase-modulated using a gigahertz signal derived from the local clock. The electric field reads
\begin{equation}
	E(\tau) = E_0 e^{i 2\pi \Phi_\text{c}(\tau)} e^{i m \cos(2 \pi \Phi_m(\tau))}
	\qc
\end{equation}
where $m$ is the modulation depth; $\Phi_\text{c}(\tau)$ is the total phase of the carrier, and $\Phi_m(\tau)$ is the total phase of the modulating signal, both expressed in cycles.

The Jacobi-Anger expansion lets us write the previous expression using Bessel functions. Because the modulation depth $m \approx 0.15$ is small~\cite{Otto:2015erp}, we can further expand the result to first order in $m$ and write the complex field amplitude as the sum
\begin{equation}
    E(\tau) \approx E_0 \qty(e^{i 2\pi \Phi_\text{c}(\tau)}
    + i \frac{m}{2} \qty[e^{i 2\pi \Phi_{\text{sb}^+}(\tau)} + e^{i 2\pi \Phi_{\text{sb}^-}(\tau)}])
	\qc
\end{equation}
where we have defined the upper and lower sideband phases,
\begin{subequations}
    \begin{align}
    	\Phi_{\text{sb}^+}(\tau) &= \Phi_\text{c}(\tau) + \Phi_m(\tau)
    	\qc
    	\\
    	\Phi_{\text{sb}^-}(\tau) &= \Phi_\text{c}(\tau) - \Phi_m(\tau)
    	\qs
    \end{align}
    \label{eq:sideband-phase-decomposition}
\end{subequations}
The modulated laser beam can then be written as the superposition of carrier, upper sideband, and lower sideband,
\begin{equation}
	E(\tau) \approx E_\text{c}(\tau) + E_{\text{sb}^+}(\tau) + E_{\text{sb}^-}(\tau)
	\qs
\end{equation}
For the purpose of our simulation, the information content of the upper and lower sidebands are almost identical (one difference is that they lie at a different frequencies, and are thus affected differently by Doppler shifts). We make the assumption that they can be combined in such a way that we can treat them as one signal. Therefore, we only simulate the upper sideband. For clarity, we drop the sign in all sideband indices and simply use \textit{sb} when we refer to the upper sideband.

We apply the same two-variable decomposition to the sideband total frequency. Ultimately, each modulated laser beam is then implemented using 4 quantities,
\begin{equation}
	\nu(\tau) \equiv \qty{\nu_\text{c}^o(\tau), \nu_\text{c}^\epsilon(\tau), \nu_\text{sb}^o(\tau), \nu_\text{sb}^\epsilon(\tau)}
	\qc
\end{equation}
where $\nu_\text{c}^o$ and $\nu_\text{sb}^o$ are the carrier and sideband frequency offsets, respectively, and $\nu_\text{c}^\epsilon$ and $\nu_\text{sb}^\epsilon$ are the carrier and sideband frequency fluctuations.

\subsection{Local beams}
\label{sec:local-beams}

\subsubsection{Local beam at laser source}

As illustrated in \cref{fig:optical-design}, optical bench~12 has an associated laser source. We call \textit{local beam} the modulated beam produced by this laser source. We denote the total phase and frequency of the carrier as $\Phi_{12,\text{c}}(\tau)$ and $\nu_{12,\text{c}}(\tau)$, respectively. Similarly, the sideband total phase and frequency are denotes as $\Phi_{12,\text{sb}}(\tau)$ and $\nu_{12,\text{sb}}(\tau)$. All these signals are functions of the \glsfirst{tps} $\tau_1$.

The total phase $\Phi_{12,\text{c}}(\tau) = \nu_0 \tau + \phi_{12,\text{c}}^o(\tau) + \phi_{12,\text{c}}^\epsilon(\tau)$ of the carrier is decomposed in terms of drifts and fluctuations, with
\begin{subequations}
\begin{align}
    \phi_{12,\text{c}}^o(\tau) &= \int_{\tau_{1,0}}^{\tau}{O_{12}(\tau') \dd{\tau'}}
    \qc
    \\
    \phi_{12,\text{c}}^\epsilon(\tau) &= p_{12}(\tau)
    \qc
\end{align}
\end{subequations}
where $O_{12}(\tau)$ is the carrier frequency offset for this laser source with respect to the central frequency $\nu_0$, and $p_{12}(\tau)$ is the laser source phase fluctuations expressed in cycles.

As explained in \cref{sec:locking}, $p_{ij}(\tau)$ can either describe the noise $N^p_{ij}(\tau)$ of a cavity-stabilized laser (c.f.~\cref{sec:noise-models}) or the fluctuations resulting from an offset frequency lock. Likewise, $O_{12}(\tau)$ is either set as an offset from the nominal frequency\footnote{In the current mission baseline, there is no way to measure or set the absolute laser frequency with high precision. Therefore, the values set in the simulation cannot be accessed in reality.}, or computed based on the locking conditions.

In terms of frequency, we simply have
\begin{subequations}
	\begin{align}
	\nu_{12,\text{c}}^o(\tau) &= O_{12}(\tau)
	\qc \\
	\nu_{12,\text{c}}^\epsilon(\tau) &= \dot p_{12}(\tau)
	\qs
	\end{align}
\end{subequations}

Let us now look at the sideband, which is derived from the local clock. As described in detail in \cref{sec:phase-readout}, the modulating signal inherits any \gls{uso} timing errors $q_1$, such that we have
\begin{equation}
    \Phi_{12,m}(\tau) = \nu_{12}^m \vdot (\tau + q_1^o(\tau) + q_1^\epsilon(\tau) + M_{12}(\tau))
    \label{eq:total-phase-modulating-signal}
\end{equation}
for the total phase of the modulating signal. Here, $\nu_{12}^m = \SI{2.4}{\giga\hertz}$ is the constant \textit{nominal} frequency of the modulating signal on optical bench~12. We use the same modulation frequency for all optical benches indexed cyclically (12, 23, 31), while the remaining ones (13, 32, 21) are instead at \SI{2.401}{\giga\hertz}. The modulation noise term $M_{12}(\tau)$ accounts for any additional imperfections (either in the electrical frequency conversion to \SI{2.4}{\giga\hertz} or the optical modulation).

The total phase of the modulating signal can then be decomposed into
\begin{subequations}
	\begin{align}
	\phi_{12}^o(\tau) &= \nu_{12}^m \vdot (\tau + q_1^o(\tau))
	\qc \\
	\phi_{12,m}^\epsilon(\tau) &= \nu_{12}^m \vdot (q_1^\epsilon(\tau) + M_{12}(\tau))
	\qs
	\end{align}
\end{subequations}

Inserting these terms in \cref{eq:sideband-phase-decomposition}, we get the phase and frequency offsets and fluctuations for the local sideband,
\begin{subequations}
	\begin{align}
	\phi_{12,\text{sb}}^o(\tau) &= \int_{\tau_{1,0}}^{\tau}{O_{12}(\tau')} \dd{\tau'} + \nu_{12}^m (\tau + q_1^o(\tau))
	\qc
	\\
	\phi_{12,\text{sb}}^\epsilon(\tau) &= p_{12}(\tau) + \nu_{12}^m (q_1^\epsilon(\tau) + M_{12}(\tau))
	\qc
	\end{align}
\end{subequations}
and
\begin{subequations}
	\begin{align}
	\nu_{12,\text{sb}}^o(\tau) &= O_{12}(\tau) + \nu_{12}^m (1 + \dot q_1^o(\tau))
	\qc
	\\
	\nu_{12,\text{sb}}^\epsilon(\tau) &= \dot p_{12}(\tau) + \nu_{12}^m ( \dot q_1(\tau) + \dot M_{12}(\tau) )
	\qs
	\end{align}
\end{subequations}
Note that there is only one clock per spacecraft, such that we use the same $q_1$ for sideband beams on both optical benches on spacecraft~1.

\subsubsection{Local beams at the interspacecraft and reference interferometer photodiodes}

As shown in \cref{fig:optical-design}, local beams propagate in the local optical bench~12 and interfere at the \gls{isi}, \gls{tmi}, and \gls{rfi} photodiodes. In our simulations, we neglect any phase term due to the propagation time. However, all beams pick up a generic optical path length noise term $N^\text{ob}(\tau)$ (different for each interferometer), which models all optical path length variations due to, e.g., jitters of optical components in the path of the laser beams. By convention, we choose that a positive value of the optical path length noise term corresponds to a decrease in the actual optical path length, which in turn corresponds to a positive shift in phase or frequency.

Therefore, we write the phase drifts and fluctuations of the local beams at the \gls{isi} and \gls{rfi} photodiodes (valid for both carriers and sidebands) as
\begin{subequations}
\begin{align}
    \phi^o_{\text{isi/rfi}_{12} \leftarrow 12}(\tau) &= \phi^o_{12}(\tau)
    \qc
    \\
    \phi^\epsilon_{\text{isi/rfi}_{12} \leftarrow 12}(\tau) &= \phi^\epsilon_{12}(\tau) + \frac{\nu_0}{c} N^\text{ob}_{\text{isi/rfi}_{12} \leftarrow 12}(\tau)
    \qs
\end{align}
\end{subequations}
Equivalently, the frequency offsets and fluctuations of the same beams read
\begin{subequations}
\begin{align}
    \nu^o_{\text{isi/rfi}_{12} \leftarrow 12}(\tau) &= \nu^o_{12}(\tau)
    \qc
    \\
    \nu^\epsilon_{\text{isi/rfi}_{12} \leftarrow 12}(\tau) &= \nu^\epsilon_{12}(\tau) + \frac{\nu_0}{c} \dot{N}^\text{ob}_{\text{isi/rfi}_{12} \leftarrow 12}(\tau)
    \qs
\end{align}
\end{subequations}

\subsubsection{Local beam at the test-mass interferometer photodiode}

The local beam reflects off the test mass before impinging on the \gls{tmi} photodiode. As a consequence, it couples to the test-mass motion.

In reality, the motion of the test mass and spacecraft will be coupled by the \gls{dfacs}. The spacecraft motion is expected to be suppressed in on-ground processing \cite{Otto:2015erp}. For our purposes, we simply assume that the spacecraft (and the associated optical benches) perfectly follows a geodesic.

\begin{figure}[t]
	\centering
	\includegraphics[width=0.5\columnwidth]{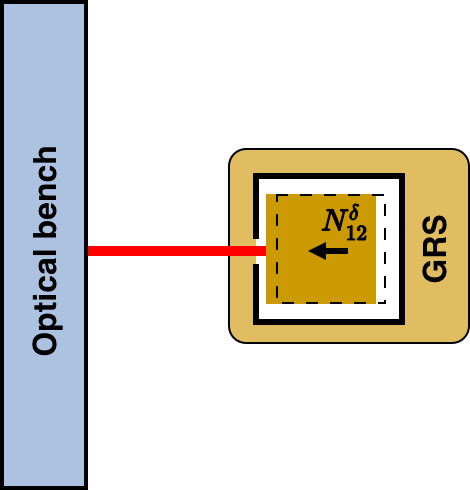}
	\caption{Definition of line-of-sight displacement of the test mass. Positive values of $N_{12}^\delta$ indicate a motion \textit{towards} the measuring optical bench.}
	\label{fig:test-mass-motion}
\end{figure}

The laser beam then picks up an additional noise term $N^\delta_{23}(\tau)$ due to any deviation in the motion of the test-mass from geodesic, caused by spurious forces (c.f., \cref{sec:noise-models}). This noise represents the movement of the test mass \textit{towards} the measuring optical bench, such that a positive value corresponds to a decrease in path length (see \cref{fig:test-mass-motion}), and thus a positive phase shift. The noise term enter with a factor 2, since the beam travels to the test mass and back.

Therefore, at the \gls{tmi} photodiode, the phase components of the local beam (carrier and sideband) read
\begin{subequations}
\begin{align}
    \phi^o_{\text{tmi}_{12} \leftarrow 12}(\tau) &= \phi^o_{12}(\tau)
    \qc
    \\
    \phi^\epsilon_{\text{tmi}_{12} \leftarrow 12}(\tau) &= \phi^\epsilon_{12}(\tau) + \frac{\nu_0}{c} \qty(N^\text{ob}_{\text{tmi}_{12} \leftarrow 12}(\tau) + 2 N^\delta_{12}(\tau))
    \qc
\end{align}
\end{subequations}
while the frequency offsets and fluctuations read
\begin{subequations}
\begin{align}
    \nu^o_{\text{tmi}_{12} \leftarrow 12}(\tau) &= \nu^o_{12}(\tau)
    \qc
    \\
    \nu^\epsilon_{\text{tmi}_{12} \leftarrow 12}(\tau) &= \nu^\epsilon_{12}(\tau)  + \frac{\nu_0}{c} \qty(\dot{N}^\text{ob}_{\text{tmi}_{12} \leftarrow 12}(\tau) + 2 \dot{N}^\delta_{12}(\tau))
    \qs
\end{align}
\end{subequations}

\subsection{Adjacent beams}
\label{sec:adjacent-beams}

In this section, we study the propagation of a modulated laser beam generated by laser source~13 (attached to the adjacent optical bench), which travels through the optical fiber to the local optical bench~12, to finally interfere on the \gls{tmi} and \gls{rfi} photodiodes (see \cref{fig:optical-design}). We call it \textit{adjacent beam}. We express all phase and frequency quantities as functions of the \gls{tps} $\tau_1$.

Similarly to local beams, we neglect the propagation time for the adjacent beams, and model fluctuations in the optical path length by a noise term $N^\text{ob}(\tau)$. We model any nonreciprocal noise terms related to the propagation through the optical fibres by the backlink noise term $N^\text{bl}_{12 \leftarrow 13}(\tau)$, expressed as an equivalent path length change.

Therefore, the phase drifts and fluctuations of adjacent beams (carrier and sideband) at the \gls{tmi} and \gls{rfi} photodiodes read
\begin{subequations}
\begin{align}
    \phi^o_{\text{tmi/rfi}_{12} \leftarrow 13}(\tau) ={}& \phi^o_{13}(\tau)
    \qc
    \\
    \begin{split}
        \phi^\epsilon_{\text{tmi/rfi}_{12} \leftarrow 13}(\tau) ={}& \phi^\epsilon_{13}(\tau) + \frac{\nu_0}{c} N^\text{bl}_{12 \leftarrow 13}(\tau)
        \\
        &+ \frac{\nu_0}{c} N^\text{ob}_{\text{tmi/rfi}_{12} \leftarrow 13}
        \qc
    \end{split}
\end{align}
\end{subequations}
where $\phi^o_{13}(\tau)$ and $\phi^\epsilon_{13}(\tau)$ are the phase drifts and fluctuations of the laser beam produced by laser source~13, respectively. The equivalent frequency quantities are
\begin{subequations}
\begin{align}
    \nu^o_{\text{tmi/rfi}_{12} \leftarrow 13}(\tau) ={}& \nu^o_{13}(\tau)
    \qc
    \\
    \begin{split}
        \nu^\epsilon_{\text{tmi/rfi}_{12} \leftarrow 13}(\tau) ={}& \nu^\epsilon_{13}(\tau) + \frac{\nu_0}{c} \dot{N}^\text{bl}_{12 \leftarrow 13}(\tau)
        \\
        & + \frac{\nu_0}{c} \dot{N}^\text{ob}_{\text{tmi/rfi}_{12} \leftarrow 13} \qs
    \end{split}
\end{align}
\end{subequations}

\subsection{Distant beams}
\label{sec:distant-beams}

Finally, we study the propagation of a modulated laser beam generated by laser source~21 (attached to the distant optical bench), which travels roughly \num{2.5}~million kilometers in free space before it reaches the local optical bench~12. This \textit{distant beam} eventually interferes on the \gls{isi} photodiode, see \cref{fig:optical-design}.

\subsubsection{Interspacecraft propagation}

As described in \cref{sec:laser-beam-model}, modulated beams are represented as the superposition of simple beams, each treated independently. Consequently, the same propagation equations apply to both carrier and sideband beams.

We shall derive the expression of a simple laser beam's phase $\Phi_{12\leftarrow 21}(\tau)$ and frequency $\nu_{12\leftarrow 21}(\tau)$ measured on receiver optical bench~12 (expressed in comoving time coordinate $\tau_1$) as a function of the same beam's phase $\Phi_{21}(\tau)$ and frequency $\nu_{21}(\tau)$ measured on emitter optical bench~21 (expressed in comoving time coordinate $\tau_2$). We write
\begin{equation}
	\Phi_{12 \leftarrow 21}(\tau) = \Phi_{21}(\tau - d_{12}(\tau))
	\qc
	\label{eq:propagated-phase-definition}
\end{equation}
where $d_{12}(\tau)$ is the \gls{ppr}, which includes not only the light time of flight, but also conversions between reference frames associated to $\tau_1$ and $\tau_2$.

Since we model small in-band and large out-of-band effects independently, we need to decompose the \gls{ppr} in a similar manner. We define $d_{12}^o(\tau)$ as slowly varying \gls{ppr} offsets (e.g., due to constant path lengths and variations in orbital motion, relativistic effects, and coordinate transformations), and $d_{12}^\epsilon(\tau)$ as small in-band \gls{ppr} fluctuations.

In our simulation, we only consider the effect of gravitational waves and neglect any other small in-band fluctuations of the \glspl{ppr} (such as spacecraft jitter motion or variations of the interplanetary medium optical index). Therefore, if $H_{12}(\tau)$ denotes the integrated fluctuations of the \gls{ppr} due to gravitational waves measured on \gls{mosa}~12, we have $d_{12}^\epsilon(\tau) = H_{12}(\tau)$. The total \gls{ppr} now reads
\begin{equation}
	d_{12}(\tau) = d_{12}^o(\tau) + H_{12}(\tau)
	\qs
	\label{eq:ppr-decomposition}
\end{equation}

Applying this decomposition to \cref{eq:two-variable-beam-phase,eq:propagated-phase-definition}, we have
\begin{equation}
	\begin{split}
		\Phi_{12 \leftarrow 21}(\tau) ={}&
		\nu_0 \vdot \qty(\tau - d_{12}^o(\tau) - H_{12}(\tau))
		\\
		& + \phi_{21}^o(\tau - d_{12}^o(\tau) - H_{12}(\tau))
		\\
		& + \phi_{21}^\epsilon(\tau - d_{12}^o(\tau) - H_{12}(\tau))
		\qs
	\end{split}
\end{equation}
We expand the previous equation to first order in both the small fluctuations $H_{12}(\tau)$, and $\phi_{21}^\epsilon(\tau)$ and neglect any second-order cross terms,
\begin{equation}
	\begin{split}
		\Phi_{12 \leftarrow 21}(\tau) ={}&
		\nu_0 \vdot \qty(\tau - d_{12}^o(\tau) - H_{12}(\tau))
		\\
		& + \phi_{21}^o(\tau - d_{12}^o(\tau))
		\\
		& - \nu_{21}^o(\tau - d_{12}^o(\tau)) H_{12}(\tau)
		\\
		& + \phi_{21}^\epsilon(\tau - d_{12}^o(\tau))
		\qs
	\end{split}
	\label{eq:general-propagated-phase}
\end{equation}

We can again write the previous quantity as the sum of large phase drifts and small phase fluctuations, $\Phi_{12 \leftarrow 21}(\tau) = \nu_0 \tau + \phi_{12 \leftarrow 21}^o(\tau) + \phi_{12 \leftarrow 21}^\epsilon(\tau)$, with
\begin{subequations}
\begin{align}
	\phi_{12 \leftarrow 21}^o(\tau) ={}& \phi_{21}^o(\tau - d_{12}^o(\tau)) - \nu_0 d_{12}^o(\tau)
	\qc
	\label{eq:propagated-phase-drifts}
	\\
	\begin{split}
	    \phi_{12 \leftarrow 21}^\epsilon(\tau) ={}&
	    \phi_{21}^\epsilon(\tau - d_{12}^o(\tau))
	    \\
	    & - \qty[\nu_0 + \nu_{21}^o(\tau - d_{12}^o(\tau))] H_{12}(\tau)
	    \qs
	\end{split}
	\label{eq:propagated-phase-fluctuations}
\end{align}
\end{subequations}
We write the equivalent instantaneous frequency $\nu_{12 \leftarrow 21}(\tau) = \nu_0 + \nu_{12 \leftarrow 21}^o(\tau) + \nu_{12 \leftarrow 21}^\epsilon(\tau)$ as the sum of a large frequency offsets and small frequency fluctuations,
\begin{subequations}
\begin{align}
    \begin{split}
        \nu_{12 \leftarrow 21}^o(\tau) ={}&
        \nu_{21}^o(\tau - d_{12}^o(\tau))\qty(1 - \dot{d}_{12}^o(\tau))
        \\
        & - \nu_0\dot{d}_{12}^o (\tau)
        \qc
    \end{split}
    \label{eq:propagated-frequency-offsets}
    \\
    \begin{split}
        \nu_{12 \leftarrow 21}^\epsilon(\tau) ={}&
        \nu_{21}^\epsilon(\tau - d_{12}^o(\tau)) \qty(1 - \dot{d}_{12}^o\qty(\tau))
        \\
        & - [\nu_0 + \nu_{21}^o(\tau - d_{12}^o(\tau))] \dot H_{12}(\tau)
        \qs
    \end{split}
    \label{eq:propagated-frequency-fluctuations}
\end{align}
\end{subequations}
Here, we have neglected first-order terms in $\dot \nu_A^o  H_{12}(\tau)$, so these equations are only valid if the laser frequency is evolving slowly. This is discussed in more detail in \cref{sec:dotnu-discussion}.

\subsubsection{Distant beams at the interspacecraft interferometer photodiode}

The received distant beam propagates inside the optical bench to interfere with the local beam at the \gls{isi} photodiode. As for the other beams, we only add a generic optical path length noise term $N^\text{ob}_{\text{isi}_{12}\leftarrow 21}(\tau)$.

We write the phase drifts and fluctuations of the distant beam at the \gls{isi} photodiode (valid for both carrier and sideband) as
\begin{subequations}
\begin{align}
    \phi^o_{\text{isi}_{12} \leftarrow 21}(\tau) ={}&
    \phi_{21}^o(\tau - d_{12}^o(\tau)) - \nu_0 d_{12}^o(\tau)
    \qc
    \\
    \begin{split}
        \phi^\epsilon_{\text{isi}_{12} \leftarrow 21}(\tau) ={}&
        \phi_{21}^\epsilon(\tau - d_{12}^o(\tau))
        \\
        & - [\nu_0 + \nu_{21}^o(\tau - d_{12}^o(\tau))] H_{12}(\tau)
        \\
        & + \frac{\nu_0}{c} N^\text{ob}_{\text{isi}_{12} \leftarrow 21}(\tau)
        \qs
    \end{split}
\end{align}
\end{subequations}

Equivalently, the frequency offsets and fluctuations read
\begin{subequations}
\begin{align}
    \begin{split}
        \nu^o_{\text{isi}_{12} \leftarrow 21}(\tau) ={}&
        \nu_{21}^o(\tau - d_{12}^o(\tau)) \qty(1 - \dot{d}_{12}^o\qty(\tau))
        \\
        & - \nu_0 \dot{d}_{12}^o\qty(\tau)
        \qc
    \end{split}
    \\
    \begin{split}
    	\nu^\epsilon_{\text{isi}_{12} \leftarrow 21}(\tau) ={}&
    	\nu_{21}^\epsilon(\tau - d_{12}^o(\tau)) \qty(1 - \dot{d}_{12}^o\qty(\tau))
    	\\
    	& - [\nu_0 + \nu_{21}^o(\tau - d_{12}^o(\tau))] \dot H_{12}(\tau)
    	\\
    	& + \frac{\nu_0}{c} \dot{N}^\text{ob}_{\text{isi}_{12} \leftarrow 21}(\tau)
        \qs
	\end{split}
\end{align}
\end{subequations}

\subsection{Interferometers}
\label{sec:interferometers}

\subsubsection{Beatnote for simple beams}

Using definitions given in \cref{eq:electromagnetic-signal}, let us write the complex amplitude for two simple beams 1 and 2 interfering at a photodiode,
\begin{subequations}
\begin{align}
    \cmplx{E}_1(\tau) &= E_{1,0}(\tau) e^{i 2\pi \Phi_1(\tau)}
    \qc
    \\
    \cmplx{E}_2(\tau) &= E_{2,0}(\tau) e^{i 2\pi \Phi_2(\tau)}
    \qs
\end{align}
\end{subequations}
We ignore any effects due to spatial dimensions of the beam or the photodiode, and assume that such effects will be modeled as either an equivalent phase error in the readout signal, or as an independent quantity\footnote{For example, \gls{dws} could be modeled as a direct measurement of beam tilt angles, with all beam angles represented by independent variables.}.

The power of the total electromagnetic field measured near the photodiode is
\begin{equation}
    P(\tau) \propto \abs{\cmplx{E}_1(\tau) + \cmplx{E}_2(\tau)}^2
    \qs
\end{equation}
Substituting the expressions of the two beams yields
\begin{equation}
    \begin{split}
        P(\tau) \propto{}&
        \abs{E_{1,0}(\tau)}^2 + \abs{E_{2,0}(\tau)}^2
        \\
        & + 2 E_{1,0}(\tau) E_{2,0}(\tau) \cos(2\pi (\Phi_1(\tau) - \Phi_2(\tau)))
        \qs
    \end{split}
    \label{eq:photodiode-power}
\end{equation}

The power near the photodiode has an oscillating component with a total phase of $\Phi_\text{PD}(\tau) = \Phi_1(\tau) - \Phi_2(\tau)$. We call this signal the \textit{beatnote}.

Let us use the two-variable representation described in \cref{eq:two-variable-beam-phase},
\begin{subequations}
	\begin{align}
    	\Phi_1(\tau) &= \nu_0 \tau + \phi^o_1(\tau) + \phi^\epsilon_1(\tau)
    	\qc
    	\\
    	\Phi_2(\tau) &= \nu_0 \tau + \phi^o_2(\tau) + \phi^\epsilon_2(\tau)
    	\qc
	\end{align}
\end{subequations}
to express the total phase of the beatnote as the sum of large phase drifts and small phase fluctuations,
\begin{equation}
    \Phi_\text{PD}(\tau) = \phi^o_\text{PD}(\tau) + \phi^\epsilon_\text{PD}(\tau)
    \qc
    \label{eq:two-variable-photodiode-phase}
\end{equation}
with
\begin{subequations}
	\begin{align}
    	\phi^o_\text{PD}(\tau) &= \phi^o_1(\tau) - \phi^o_2(\tau)
    	\qc
    	\label{eq:two-variable-photodiode-phase-drifts}
    	\\
    	\phi^\epsilon_\text{PD}(\tau) &= \phi^\epsilon_1(\tau) - \phi^\epsilon_2(\tau)
    	\qs
    	\label{eq:two-variable-photodiode-phase-fluctuations}
	\end{align}
\end{subequations}

We simulate the equivalent instantaneous frequency defined as $\nu_\text{PD}(\tau) =  \dot{\Phi}_\text{PD}(\tau)$. It can be written as
\begin{equation}
    \nu_\text{PD}(\tau) = \nu^o_\text{PD}(\tau) + \nu^\epsilon_\text{PD}(\tau)
    \qc
\end{equation}
where the beatnote frequency offsets $\nu^o_\text{PD}(\tau)$ and the beatnote frequency fluctuations $\nu^\epsilon_\text{PD}(\tau)$ are defined by
\begin{subequations}
	\begin{align}
	\nu^o_\text{PD}(\tau) &= \nu^o_1(\tau) - \nu^o_2(\tau)
	\qc
	\\
	\nu^\epsilon_\text{PD}(\tau) &= \nu^\epsilon_1(\tau) - \nu^\epsilon_2(\tau)
	\qs
	\end{align}
\label{eq:two-variable-photodiode-frequency}
\end{subequations}

\subsubsection{Beatnote polarity}

A closer look at \cref{eq:photodiode-power} shows that we do not have direct access to the total phase of the beatnote $\Phi_\text{PD}(\tau)$, but only measure its cosine value. Therefore, the total phase can only be known up to a sign and a multiple of $2 \pi$.

Physically, this sign ambiguity corresponds to the fact that the electrical signal does not contain any information about which of the two interfering laser beams is of higher frequency. In practice, however, the beatnote polarity can be determined at all times by applying a known frequency offset on the local laser beam and observing the resulting change in the beatnote frequency. In addition, once all lasers are locked, the beatnote polarities can simply be read from the frequency plan, as described in \cref{sec:locking}.

Therefore, we do not include the beatnote polarity ambiguity in our optical models, and we will instead assume that it is solved directly by the phasemeter, or in a first processing step on ground.

\subsubsection{Beatnotes for modulated beams}

We now study the electromagnetic field of two interfering modulated beams, labeled $k=1,2$. As derived in \cref{sec:laser-beam-model}, we write both modulated beams as the sum of three independent simple beams, namely the carriers and the upper and lower sidebands,
\begin{equation}
    \cmplx{E}_{k}(\tau) = \cmplx{E}_{k,\text{c}}(\tau) + \cmplx{E}_{k,\text{sb}^+}(\tau) + \cmplx{E}_{k,\text{sb}^-}(\tau)
    \qc
\end{equation}
with total phases
\begin{subequations}
	\begin{align}
	\Phi_{k,\text{c}}(\tau) &= \nu_0 \tau + \Phi_{k,\text{c}}^o(\tau) + \Phi_{k,\text{c}}^\epsilon(\tau)
	\qc
	\\
	\Phi_{k,\text{sb}^+}(\tau) &= \nu_0 \tau + \Phi_{k,\text{sb}^+}^o(\tau) + \Phi_{k,\text{sb}^+}^\epsilon(\tau)
	\qc
	\\
	\Phi_{k,\text{sb}^-}(\tau) &= \nu_0 \tau + \Phi_{k,\text{sb}^-}^o(\tau) + \Phi_{k,\text{sb}^-}^\epsilon(\tau)
	\qs
	\end{align}
\end{subequations}
or the equivalent instantaneous frequencies
\begin{subequations}
	\begin{align}
	\nu_{k,\text{c}}(\tau) &= \nu_0 + \nu_{k,\text{c}}^o(\tau) + \nu_{k,\text{c}}^\epsilon(\tau)
	\qc
	\\
	\nu_{k,\text{sb}^+}(\tau) &= \nu_0 + \nu_{k,\text{sb}^+}^o(\tau) + \nu_{k,\text{sb}^+}^\epsilon(\tau)
	\qc
	\\
	\nu_{k,\text{sb}^-}(\tau) &= \nu_0 + \nu_{k,\text{sb}^-}^o(\tau) + \nu_{k,\text{sb}^-}^\epsilon(\tau)
	\qs
	\end{align}
\end{subequations}

The total power at the photodiode reads
\begin{equation}
    \begin{split}
    \abs{\cmplx{E}_1(\tau) + \cmplx{E}_2(\tau)}^2 ={}&
    \big| \cmplx{E}_{1,\text{c}}(\tau) + \cmplx{E}_{1,\text{sb}^+}(\tau) + \cmplx{E}_{1,\text{sb}^-}(\tau)
    \\
    &+ \cmplx{E}_{2,\text{c}}(\tau) + \cmplx{E}_{2,\text{sb}^+}(\tau) + \cmplx{E}_{2,\text{sb}^-}(\tau) \big|^2
    \qs
    \end{split}
\end{equation}
Expanding this expression yields cross terms between all 6 terms, which correspond to beatnotes at their difference frequencies.

Because the sidebands are modulated at a frequency of about \SI{2.4}{\giga\hertz}, most of these beatnote frequencies lie far outside of the phasemeters measurement bandwidth (approximately \SI{5}{\mega\hertz} to \SI{25}{\mega\hertz}).

Only three beatnotes lie inside this region,
\begin{itemize}
	\item The carrier-carrier beatnote,
	\begin{subequations}
		\begin{align}
		\Phi_{\text{PD},\text{c}}(\tau) &= \Phi_{1,\text{c}}(\tau) - \Phi_{2,\text{c}}(\tau)
		\qc
		\\
		\nu_{\text{PD},\text{c}}(\tau) &= \nu_{1,\text{c}}(\tau) - \nu_{2,\text{c}}(\tau)
		\qc
		\end{align}
	\end{subequations}
	\item The upper sideband-upper sideband beatnote,
	\begin{subequations}
		\begin{align}
		\Phi_{\text{PD},\text{sb}^+}(\tau) &= \Phi_{1,\text{sb}^+}(\tau) - \Phi_{2,\text{sb}^+}(\tau)
		\qc
		\\
		\nu_{\text{PD},\text{sb}^+}(\tau) &= \nu_{1,\text{sb}^+}(\tau) - \nu_{2,\text{sb}^+}(\tau)
		\qc
		\end{align}
	\end{subequations}
	\item The lower sideband-lower sideband beatnote,
	\begin{subequations}
		\begin{align}
		\Phi_{\text{PD},\text{sb}^-}(\tau) &= \Phi_{1,\text{sb}^-}(\tau) - \Phi_{2,\text{sb}^-}(\tau)
		\qc
		\\
		\nu_{\text{PD},\text{sb}^-}(\tau) &= \nu_{1,\text{sb}^-}(\tau) - \nu_{2,\text{sb}^-}(\tau)
		\qc
		\end{align}
	\end{subequations}
\end{itemize}
Because the sidebands of the lasers on \glspl{mosa} 12, 23, and 31 (respectively 13, 32, and 21) are offset by \SI{2.4}{\giga\hertz} (respectively, \SI{2.401}{\giga\hertz}), and because we always interfere beams from both types of \gls{mosa}, these three beatnotes will always be offset by \SI{1}{\mega\hertz}. Therefore, they can be tracked individually by the phasemeter.

Each of these beatnote frequencies can be decomposed again as a sum of large frequency offsets and small fluctuations, and we recover equations similar to \cref{eq:two-variable-photodiode-frequency}. Therefore, the carrier and sideband parts of a modulated laser beam can be implemented as three distinct beams in the simulation, from which we form three beatnotes.

As described in the previous sections, we only include the carrier and upper-sideband laser beams in our model; as a consequence, we only compute the carrier-carrier and the upper sideband-upper sideband beatnotes.

\subsubsection{Interspacecraft, test-mass, and reference interferometer beatnotes}

To obtain the beatnote phases (or frequencies) measured by the \gls{isi}, \gls{tmi}, and \gls{rfi}, we can substitute in the previous equations the phases (or frequencies) of the interfering beams.

As discussed above, the beatnote polarities are arbitrary. As a convention, we will always write the beatnote phase (and frequencies) as the difference of the distant or adjacent beam phase (or frequency) and the local beam phase (or frequency),
\begin{subequations}
\begin{align}
    \phi(\tau) &= \phi_\text{distant/adjacent}(\tau) - \phi_\text{local}(\tau)
    \qc
    \\
    \nu(\tau) &= \nu_\text{distant/adjacent}(\tau) - \nu_\text{local}(\tau)
    \qs
\end{align}
\end{subequations}
Following the optical-bench design of \cref{fig:optical-design}, we have the following beatnote phase offsets and fluctuations, for both carriers and sidebands,
\begin{subequations}
\begin{align}
    \phi_{\text{isi}_{12}}(\tau) &= \phi_{\text{isi}_{12} \leftarrow 21}(\tau) - \phi_{\text{isi}_{12} \leftarrow 12}(\tau)
    \qc
    \\
    \phi_{\text{tmi}_{12}}(\tau) &= \phi_{\text{tmi}_{12} \leftarrow 13}(\tau) - \phi_{\text{tmi}_{12} \leftarrow 12}(\tau)
    \qc
    \\
    \phi_{\text{rfi}_{12}}(\tau) &= \phi_{\text{rfi}_{12} \leftarrow 13}(\tau) - \phi_{\text{rfi}_{12} \leftarrow 12}(\tau)
    \qc
\label{eq:beatnote-phases}
\end{align}
\end{subequations}
and similarly for beatnote frequencies.

\section{Phase readout, frequency distribution, and clock error modelling}
\label{sec:phase-readout}

\begin{figure}
    \centering
    \includegraphics[width=\columnwidth]{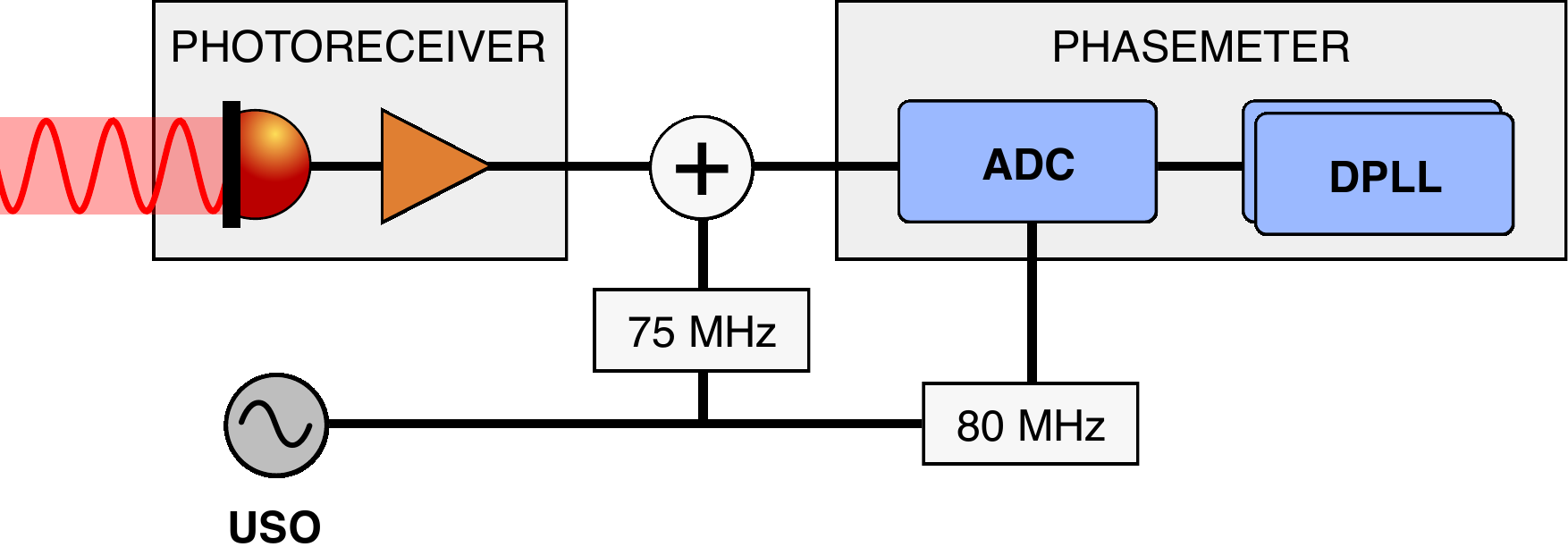}
    \caption{Overview of the phase readout chain of \gls{lisa}, adapted from~\cite{Barke:2015srr}. The photoreceiver converts the optical signal to electric signals, which are then digitized by the \gls{adc}. The intrinsic timing jitter of the \gls{adc} is corrected by a \SI{75}{\mega\hertz} pilot tone superimposed to the photoreceiver signal. The phasemeter \glspl{dpll} then tracks the beatnote phases.}
    \label{fig:readout-chain}
\end{figure}

We show in \cref{fig:readout-chain} an overview of the \gls{lisa} phase readout chain, adapted from  \cite{Barke:2015srr} (where technical details on the phase readout and frequency distribution system can be found). The optical beatnotes are converted to electrical signals by photoreceivers, which are then digitized by an \gls{adc}. The phase of these digital signals are then tracked by \glspl{dpll}.

The phasemeter is driven by an \SI{80}{\mega\hertz} clock signal. Inside the phasemeter, the \gls{adc} samples the electrical beatnotes at the same rate, with an additional timing jitter intrinsic to the \gls{adc}. This \gls{adc} jitter results in a phase error in the measured beatnotes, which is expected above the requirements. To correct for the \gls{adc} jitter, an additional periodic pilot tone signal at \SI{75}{\mega\hertz} is derived from the on-board clock and superimposed on each electrical signal fed to the phasemeter. The pilot tone phase is tracked alongside the main beatnotes in dedicated \gls{dpll} channels. By comparing the measured pilot tone phase against its nominal \SI{75}{\mega\hertz} value, the \gls{adc} jitter can be corrected in the main beatnotes, such that the pilot tone becomes the effective reference clock signal for the phase measurements.

\subsection{Readout noise}

We directly simulate the optical beatnote frequencies. Our simulated electrical signals are therefore the same quantities, with the addition of a \textit{readout noise} term $N^\text{ro}(\tau)$. This readout noise accounts for both shot noise and any errors due to front-end electronic in the photoreceivers; refer to \cref{sec:noise-models} for more details.

\subsection{Phasemeter and pilot tone}

Because the photoreceiver signals are already simulated as discrete beatnote frequency samples, we do not directly simulate the digitization process of the \gls{adc} nor the phase tracking by the \glspl{dpll}.

Furthermore, we do not simulate the pilot tone correction, but assume that it perfectly removes the \gls{adc} jitter. We do account for timing errors in the pilot tone itself, which are also expected above the requirements. These pilot tone errors will be corrected using the sidebands introduced in \cref{sec:laser-beam-model}. Refer to the next \cref{sec:phase-readout} for how we model the pilot tone and sideband signals.

% In this section we describe how we model the onboard clocks, and how their timing signals are distributed through the constellation.  Note that the specific numerical values of the different frequencies used in the LISA phase readout and frequency distribution chains are not finalized and should be seen as placeholders.

\subsection{Frequency distribution and clock signals}

\subsubsection{Frequency distribution scheme}

\begin{figure}
	\centering
	\includegraphics[width=\columnwidth]{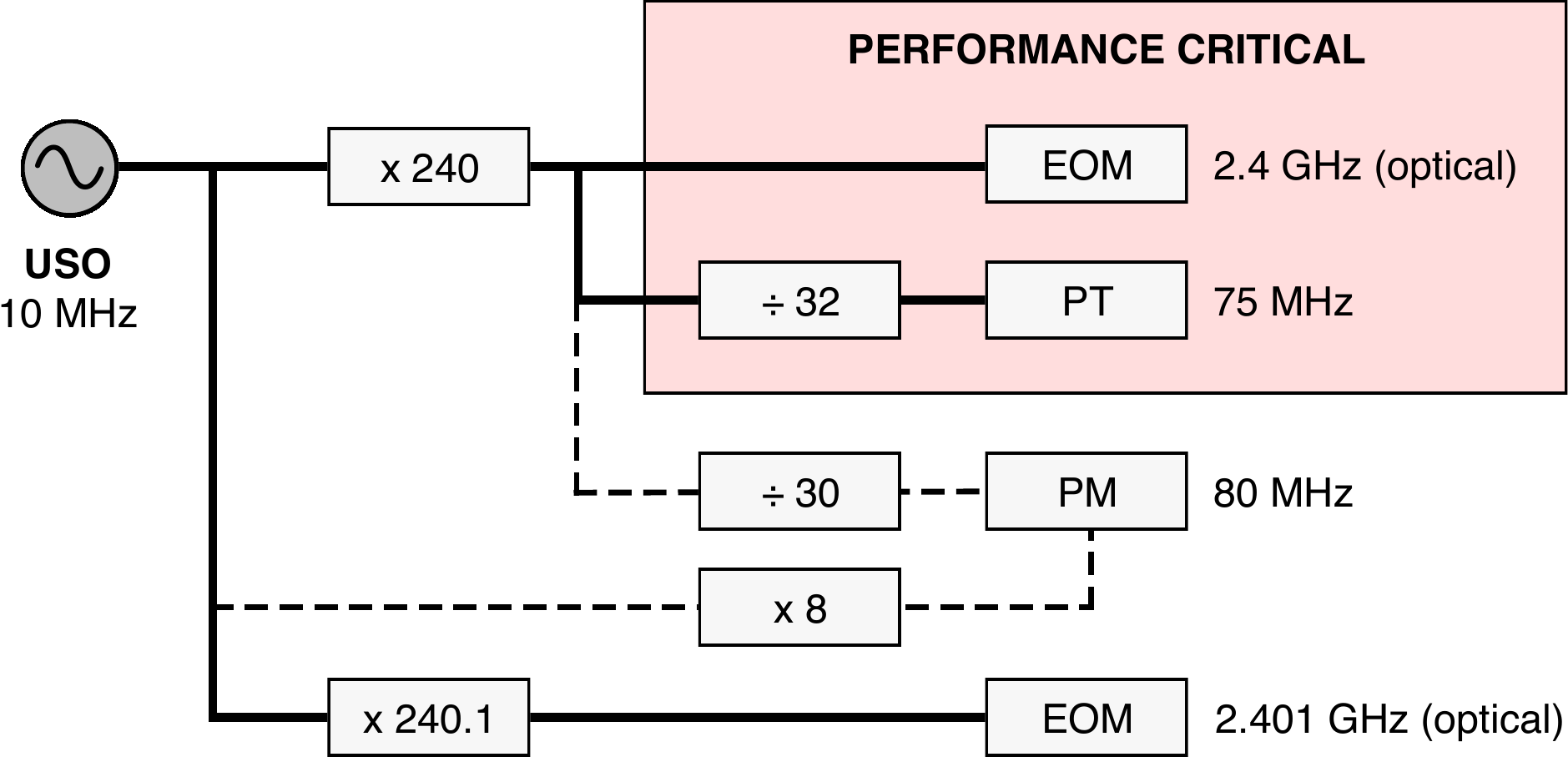}
	\caption{Overview over the USO frequency distribution on one optical bench, based on~\cite{Barke:2015srr}. Not depicted is a possible electrical comparison between the \SI{2.4}{\giga\hertz} and \SI{2.401}{\giga\hertz} signals. Note that the phasemeter clock is not performance critical, and could be synthesized from either the \SI{2.4}{\giga\hertz} signal or directly from the \gls{uso}, indicated by two possible connections in dotted lines.}
	\label{fig:fds-overview}
\end{figure}

Most subsystems on board \gls{lisa} are driven by timing signals derived from the \gls{uso}. In our simulation model, we focus on processes for which timing is performance critical, which are summarized in \cref{fig:fds-overview}.

Following the current mission design, each \gls{lisa} spacecraft uses one dedicated clock (realized by an \gls{uso}), from which all timing signals are derived. As described above, the timing reference for all phasemeter measurements is the pilot tone, which is derived from the \gls{uso} by first up-converting its nominal frequency\footnote{No final decision has been made on the precise \gls{uso} frequency that will be used for \gls{lisa}.} to \SI{2.4}{\giga\hertz}, and then converting that signal to the desired $\nu_\text{PT} = \SI{75}{\mega\hertz}$ using frequency dividers. This conversion chain allows for a very stable phase relationship between the electrical pilot tone and the \SI{2.4}{\giga\hertz} optical sideband \cite{Barke:2015srr}, which are used in postprocessing to reduce the timing errors of the pilot tone itself~\cite{Hellings:1996he,Hellings:2000cc,Tinto:2002de,Tinto:2018kij,Hartwig:2020tdu,Hartwig:2022yqw}.

The \SI{2.401}{\giga\hertz} sidebands used on right \glspl{mosa}, on the other hand, are less stable with respect to the pilot tone. This is acceptable, as additional clock noise in this signal can be corrected for using the sideband beatnotes in the \gls{rfi}~\cite{Hartwig:2020tdu,Hartwig:2022yqw}.

Lastly, any errors in the \SI{80}{\mega\hertz} phasemeter clock are also corrected by the pilot tone correction, such that it is not performance critical and could either be directly synthesized from the \gls{uso} or from the \SI{2.4}{\giga\hertz} signal. This choice is currently  irrelevant for our simulation since we directly simulate the pilot tone as reference clock for all measurements.

\subsubsection{Clock-signal model}

We model the pilot tone signal as a periodic signal of the form
\begin{equation}
	V_\text{PT}(\tau) = \cos(2\pi \nu_\text{PT}[\tau + q_1(\tau)])
	\qs
\end{equation}
Here, $q_1(\tau)$ describes the timing deviations of the pilot tone generated on spacecraft~1 with respect to the \gls{tps}~$\tau_1$, expressed in the latter. Note that the dominant noise source in the pilot tone generation is the \gls{uso} itself~\cite{Barke:2015srr}, such that we assume the statistical properties of the pilot tone noise to be identical to those of the \gls{uso} noise.

We further decompose $q_1(\tau)$ using two time series,
\begin{equation}
    q_1(\tau) = q_1^o(\tau) + q_1^\epsilon(\tau)
    \qc
    \label{eq:clock-two-variable-definition}
\end{equation}
to model large deterministic effects (such as clock frequency offsets and drifts) and small in-band stochastic fluctuations. As before, we do not simulate the \SI{75}{\mega\hertz} signal itself, but only $q_1^o(\tau)$ and $q_1^\epsilon(\tau)$ (or rather $\dot q_1^o(\tau)$ and $\dot q_1^\epsilon(\tau)$ as the pilot tones fractional frequency fluctuations).

The clock signal is used to create the sidebands, as described in \cref{sec:local-beams}. The total phase of the sideband modulation signals is modeled as
\begin{equation}
	\nu^m_{12} \vdot (\tau + q_1(\tau) + M_{12}(\tau))
	\qs
\end{equation}
Here, $\nu_{12}^m$ is the constant \textit{nominal} frequency\footnote{By definition, these frequencies are at their nominal values. The real modulation signals will have a frequency offset due to the terms $q_1$ and $M_{12}$ in \cref{eq:total-phase-modulating-signal}.} of the modulating signal on optical bench~12. Imperfections in the frequency conversion between the pilot tone and the sidebands are modeled by an additional modulation noise term $M_{12}(\tau)$.

\subsection{Timer model}
\label{sec:timer-model}

In order to model timestamping and pseudoranging (c.f.,~\cref{sec:pseudoranging}), we not only need the frequency fluctuations of the local clock, but also the time shown by each spacecraft timer. These times must be tracked down to at least \si{\nano\second}-precision while reaching values of around \SI{E8}{s} at the end of the 10 years of extended mission. The use of double-precision floating-point numbers is not compatible with such a dynamic range. Therefore, we simulate offsets of that timer relative to the associated \gls{tps} $\delta\hat\tau_1^{\tau_1}(\tau) \equiv \delta\hat\tau_1(\tau)$, called timer deviations, which evolve slowly with time. The total clock time\footnote{This timescale will be realized in practice by the so-called \gls{scet}, which is the only timescale directly available onboard the satellites.} $\hat\tau_1^{\tau_1}(\tau)$ as a function of the \gls{tps} can then be computed by
\begin{equation}
	\hat\tau_1^{\tau_1}(\tau) = \tau + \delta\hat\tau_1(\tau)
	\qs
\label{eq:spacecraft-clock-time}
\end{equation}
Timer deviations are closely related to the clock timing jitter,
\begin{equation}
	\delta\hat\tau_1(\tau) = q_1(\tau) + \delta\hat\tau_{1,0}
	\qs
	\label{eq:time-deviation-definition}
\end{equation}
In this equation, $\delta\hat\tau_{1,0}$ accounts for the fact that we don't know the true time $\tau_{1,0}$ at which we turn on the timer, i.e., we can't directly relate the initial phase of the clock signal $q_1(\tau_{1,0})$ to any external time frame.
% Using \cref{eq:clock-fractional-frequency-fluctuations}, we obtain
% \begin{equation}
% 	\delta\hat\tau_i(\tau) = \delta\hat\tau_{1,0} + \int_{\tau_{1,0}}^\tau{N^q_i(\tau') \dd{\tau'}} + y_{0,i} \tau + \frac{1}{2} y_{1,i} \tau^2 + \frac{1}{3} y_{2,i} \tau^3
% 	\label{eq:total-timer-deviation}
% \end{equation}
% for the timer deviations, which corresponds to
% \begin{itemize}
% 	\item a constant time offset $\delta\hat\tau_{1,0}$,
% 	\item a random walk following the frequency fluctuations $N^q_i(\tau)$,
% 	\item a linear drift due to the frequency offset $y_{0,i}$,
% 	\item a quadratic component due to the linear frequency drift $y_{1,i}$, and
% 	\item a cubic component due to the quadratic frequency drift $y_{2,i}$.
% \end{itemize}
% Note that the deterministic frequency drift and offset are the most relevant parts for the long term evolution; however, the stochastic part is typically not a white noise, and can reach values larger than our desired \si{\nano\second}-timing accuracy. As such, we consider the total $\delta\hat\tau_i(\tau)$ for computing timing errors.

\subsection{Signal sampling}
\label{sec:signal-sampling}

\subsubsection{Signal sampling in terms of phase}

%In the simulation, we simulate interspacecraft signal propagation, gravitational waves, and orbits according to the \gls{tcb}, whereas all on-board physics is modeled according to the \gls{tps} $\tau_i$ of each spacecraft $i$.

The photoreceiver signals recorded, say, on spacecraft~1, are generated according to the \gls{tps}~$\tau_1$. The measurements that are eventually telemetered, however, are recorded and timestamped with clock time~$\hat{\tau}_1$. As a consequence, we need to resample the photoreceiver signals from the \gls{tps} to the clock time frame.

If a photoreceiver signal $\Phi_\text{PD}$ is expressed in terms of phase, this can be achieved following \cref{eq:phase-time-coordinate-conversion},
\begin{equation}
	\Phi_\text{PD}^{\hat\tau_1}(\tau) = \Phi_\text{PD}^{\tau_1}(\tau_1^{\hat\tau_1}(\tau))
	\qs
	\label{eq:time-sampling-error-general}
\end{equation}
Therefore, we need to compute the \gls{tps} $\tau_1^{\hat\tau_1}(\tau)$ given a given clock time $\tau$. This quantity can be computed by writing \cref{eq:spacecraft-clock-time} evaluated at $\tau_1^{\hat\tau_1}(\tau)$,
\begin{equation}
	\hat\tau_1^{\tau_1}(\tau_1^{\hat\tau_1}(\tau)) = \tau_1^{\hat\tau_1}(\tau) + \delta\hat\tau_1(\tau_1^{\hat\tau_1}(\tau))
	\qs
\end{equation}
We use \cref{eq:phase-time-coordinate-conversion} to re-write the left-hand side, which gives, after rearranging,
\begin{equation}
	\tau_1^{\hat\tau_1}(\tau) = \tau - \delta\hat\tau_1(\tau_1^{{\hat\tau_1}}(\tau))
	\qs
	\label{eq:timer-deviations-implicit-equation}
\end{equation}

We can solve this implicit equation for $\tau_1^{\hat\tau_1}(\tau)$ iteratively, by computing
\begin{subequations}
\begin{align}
	\delta\hat\tau_1^{(0)}(\tau) &= \delta\hat\tau_1(\tau)
	\qc
	\\
	\delta\hat\tau_1^{(n+1)}(\tau) &= \delta\hat\tau_1(\tau - \delta\hat\tau_1^{(n)}(\tau))
	\qc
\end{align}
\end{subequations}
such that
\begin{equation}
	\lim_{n \rightarrow \infty} \delta\hat\tau_1^{(n)}(\tau) = \delta\hat\tau_1(\tau_1^{{\hat\tau_1}}(\tau))
	\qs
\end{equation}
Since the timer deviations are evolving slowly, the iteration converges quickly. In our simulations, we stop after two iterations, such that
\begin{equation}
	\tau_1^{\hat\tau_1}(\tau) \approx \tau - \delta\hat\tau_1^{(2)}(\tau)
	\qs
	\label{eq:timer-deviations-first-order}
\end{equation}

We can then plug the previous equation in \cref{eq:time-sampling-error-general} to write all frame-independent measurements as a functions of the correct recording times $\Phi_\text{PD}^{\hat\tau_1}(\tau)$, given the same quantities expressed in the \gls{tps}. We find
\begin{equation}
	\Phi_\text{PD}^{\hat\tau_1}(\tau) \approx \Phi_\text{PD}^{\tau_1}(\tau - \delta\hat\tau_1^{(2)}(\tau))
	\qs
	\label{eq:resampling-first-order}
\end{equation}
This operation can be implemented with time-varying fractional delay filter (interpolation).

We introduce the timestamping operator $\timestamp{1}$, which shifts a signal $s(\tau)$ from the \gls{tps} to the clock time of spacecraft~1. Formally, its action is given by
\begin{equation}
	\timestamp{1} s(\tau) = s(\tau - \delta\hat\tau_1^{(2)}(\tau))
	\qs
	\label{eq:timestamping-operator}
\end{equation}
Using this shorthand notation, \cref{eq:resampling-first-order} now reads
\begin{equation}
	\Phi_\text{PD}^{\hat\tau_1}(\tau) = \Phi_\text{PD}^{\tau_1}(\tau_1^{\hat\tau_1}(\tau)) \approx \timestamp{1} \Phi_\text{PD}^{\tau_1}(\tau)
	\qs
	\label{eq:resampling-first-order-shorthand}
\end{equation}

Note that this is only valid for measurements expressed in phase, as frequencies are not frame-independent quantities.

\subsubsection{Sampling errors in terms of frequency}

The effect of sampling can also be expressed in terms of total frequency, where it manifests itself as a Doppler-like frequency shift.

In the following paragraph, we compute frequencies by taking the derivative of phase with respect to the clock time, since this is the time reference that the phasemeter will use to measure the signal frequency. From \cref{eq:time-sampling-error-general}, and denoting function composition as $(\Phi^{\tau_1}_\text{PD} \circ \tau_1^{\hat\tau_1})(\tau) = \Phi^{\tau_1}_\text{PD}(\tau_1^{\hat\tau_1}(\tau))$, we have
\begin{equation}
	\nu_\text{PD}^{\hat\tau_1}(\tau) = \dv{\Phi_\text{PD}^{\hat\tau_1}}{\tau} \qty(\tau)
	= \dv{(\Phi^{\tau_1}_\text{PD} \circ \tau_1^{\hat\tau_1})}{\tau} \qty(\tau)
	\qs
\end{equation}
Using the chain rule,
\begin{align}
	\nu_\text{PD}^{\hat\tau_1}(\tau) = \nu_\text{PD}^{\tau_1}(\tau_1^{\hat\tau_1}(\tau)) \times \dv{\tau_1^{\hat\tau_1}}{\tau} \qty(\tau)
	\qs
\end{align}

To compute the derivative of $\tau_1^{\hat\tau_1}(\tau)$, let us differentiate the defining implicit \cref{eq:timer-deviations-implicit-equation},
\begin{align}
\begin{split}
	\dv{\tau_1^{\hat\tau_1}}{\tau}(\tau) \qty(\tau) &= 1 - \dv{(\delta\hat\tau_1 \circ \tau_1^{\hat\tau_1})}{\tau} \qty(\tau)
	\\
	&= 1 - \dv{\delta\hat\tau_1}{\tau}\qty(\tau_1^{\hat\tau_1}(\tau)) \times \dv{\tau_1^{\hat\tau_1}}{\tau}\qty(\tau)
	\qs
\end{split}
\end{align}
Using \cref{eq:time-deviation-definition}, we find $\dv*{\delta\hat\tau_1}{\tau}(\tau) = \dot q_1(\tau)$. Inserting this identity, we can rearrange the previous equation to get
\begin{equation}
	\dv{\tau_1^{\hat\tau_1}}{\tau}(\tau) \qty(\tau) = \frac{1}{1 + \dot q_1(\tau_1^{\hat\tau_1}(\tau))}
	\qc
	\label{eq:dtau-freq-inverse}
\end{equation}
which finally yields for the total frequency,
\begin{equation}
	\nu_\text{PD}^{\hat\tau_1}(\tau) = \frac{\nu_\text{PD}^{\tau_1}(\tau_1^{\hat\tau_1}(\tau))}{1 + \dot q_1(\tau_1^{\hat\tau_1}(\tau))}
	\approx \timestamp{1} \qty[ \frac{ \nu_\text{PD}^{\tau_1}(\tau)}{1 + \dot{q}_1(\tau)}]
	\qs
	\label{eq:sampling-error-total-frequency}
\end{equation}

\subsubsection{Sampling in two-variable decomposition}

We now want to describe the effect of timing errors in the framework of two-variable decomposition. This will allow us to split the sampling errors derived previously into large deterministic offsets in the measurement timestamps, and small stochastic fluctuations that enter as an additional noise term. The latter represent what is often referred to as \textit{clock noise}~\cite[e.g.,][]{Otto:2015erp}.

However, we want to make it clear once more that this decomposition is entirely artificial. Both slow drifts and in-band clock noise describe the same physical process, namely the instability of the \gls{uso}, on different time scales.

The sampling process applies to the total phase of each photoreceiver signal, given by \cref{eq:two-variable-photodiode-phase} as
\begin{equation}
\begin{split}
	\Phi_\text{PD}^{\hat\tau_1}(\tau) &= \Phi_\text{PD}^{\tau_1}(\tau_1^{\hat{\tau}_1}(\tau))
	\\
	&= \phi^{o}_\text{PD}(\tau_1^{\hat{\tau}_1}(\tau)) + \phi^\epsilon_\text{PD}(\tau_1^{\hat{\tau}_1}(\tau))
	\qs
\end{split}
\label{eq:photodiode-signal-in-the}
\end{equation}
Since $\phi^o_\text{PD}(\tau)$ is very quickly evolving, small (first-order) timing fluctuations in $\tau_1^{\hat{\tau}_1}(\tau)$ must appear in the measurement described by $\phi^\epsilon_\text{PD}(\tau)$. Thus, we must account for the cross coupling between $\phi^o_\text{PD}(\tau)$ and $\phi^\epsilon_\text{PD}(\tau)$, and we cannot simply time shift both components individually.

We can insert \cref{eq:time-deviation-definition,eq:clock-two-variable-definition} into \cref{eq:timer-deviations-implicit-equation} to get
\begin{equation}
    \tau_1^{\hat{\tau}_1}(\tau) = \tau - \delta\hat\tau_{1,0} - q_1^o(\tau_1^{\hat{\tau}_1}(\tau) ) - q_1^\epsilon(\tau_1^{\hat{\tau}_1}(\tau))
    \qs
\end{equation}
We model clock noise fluctuations $\dot q_1^\epsilon$ as band-limited noise, such that they remain small and we can expand the $\phi^{o}$ term in \cref{eq:photodiode-signal-in-the} to first order in $\dot q_1^\epsilon$,
\begin{equation}
\begin{split}
		\Phi_\text{PD}^{\hat\tau_1}(\tau) ={}& \phi^{o}_\text{PD}(\tau - \delta\hat\tau_{1,0} - q_1^o(\tau_1^{\hat{\tau}_1}(\tau))) + \phi^\epsilon_\text{PD}(\tau_1^{\hat{\tau}_1}(\tau))
		\\
		&- \nu^{o}_\text{PD}(\tau - \delta\hat\tau_{1,0} - q_1^o(\tau_1^{\hat{\tau}_1}(\tau))) q_1^\epsilon(\tau_1^{\hat{\tau}_1}(\tau))
		\qs
\end{split}
\end{equation}
Finally, we obtain the two variable-decomposition for the resampled photoreceiver phase.
\begin{subequations}
\begin{align}
	\phi^{\hat\tau_1,o}_\text{PD}(\tau) &\approx \phi^{o}_\text{PD}(\tau - \delta\hat\tau_{1,0} - \timestamp{1} q_1^o(\tau))
	\qc
	\label{eq:timestamping-errors-in-phase-drifts}
	\\
	\begin{split}
    	\phi^{\hat\tau_1,\epsilon}_\text{PD}(\tau) &\approx \timestamp{1} \phi^\epsilon_\text{PD}(\tau) - \nu^{o}_\text{PD}(\tau - \delta\hat\tau_{1,0} - \timestamp{1} q_1^o(\tau))
    	\\
    	&\qquad \times \timestamp{1} q_1^\epsilon(\tau)
    	\qs
	\end{split}
	\label{eq:timestamping-errors-in-phase-fluctuations}
\end{align}
\end{subequations}

For frequency data, we start with \cref{eq:sampling-error-total-frequency}, and decompose again clock noise $\dot q_1$ into two variables, as explained in \cref{sec:phase-readout}. We then expand it to first order in $\dot q_1^\epsilon$ to get
\begin{equation}
\begin{split}
    &\frac{1}{1 + \dot q_1^o(\tau_1^{\hat\tau_1}(\tau)) + \dot q_1^\epsilon(\tau_1^{\hat\tau_1}(\tau))}
	\approx
	\\
	&\qquad\qquad\qquad \frac{1}{1 + \dot q_1^o(\tau_1^{\hat\tau_1}(\tau))} - \frac{\dot q_1^\epsilon(\tau_1^{\hat\tau_1}(\tau))}{[1 + \dot q_1^o(\tau_1^{\hat\tau_1}(\tau))]^2}
	\qs
\end{split}
\end{equation}

So in total, we have
\begin{equation}
\begin{split}
	\nu_\text{PD}^{\hat\tau_1}(\tau) \approx{}& \nu_\text{PD}^{\tau_1}(\tau_1^{\hat{\tau}_1}(\tau))
	\\
	& \times \qty[\frac{1}{1 + \dot q_1^o(\tau_1^{\hat\tau_1}(\tau))} - \frac{\dot q_1^\epsilon(\tau_1^{\hat\tau_1}(\tau))}{[1 + \dot q_1^o(\tau_1^{\hat\tau_1}(\tau))]^2}]
	\qs
\end{split}
\end{equation}
We now expand $\nu_\text{PD}^{\tau_1}(\tau) = \nu_\text{PD}^{\tau_1,o}(\tau) + \nu_\text{PD}^{\tau_1,\epsilon}(\tau)$, and neglect the small coupling of $q_1^\epsilon(\tau)$ to the already small fluctuations $\nu_\text{PD}^{\tau_1,\epsilon}(\tau)$. We collect the terms to express the photodiode signal offsets $\nu_\text{PD}^{\hat\tau_1,o}(\tau)$ and fluctuations $\nu_\text{PD}^{\hat\tau_1,\epsilon}(\tau)$ after shifting to the clock time, using \cref{eq:resampling-first-order},
\begin{subequations}
\begin{align}
	\nu^{\hat\tau_1,o}_\text{PD}(\tau) &\approx \frac{\timestamp{1} \nu_\text{PD}^{\tau_1,o}(\tau))}{1 + \timestamp{1} \dot q_1^o(\tau)}
	\qc
	\label{eq:timestamping-errors-in-frequency-offsets}
	\\
	\nu^{\hat\tau_1,\epsilon}_\text{PD}(\tau) &\approx
	\frac{\timestamp{1} \nu_\text{PD}^{\tau_1,\epsilon}(\tau)}{1 + \timestamp{1} \dot q_1^o(\tau)} - \frac{\timestamp{1}\nu_\text{PD}^{\tau_1,o}(\tau) \timestamp{1} \dot q_1^\epsilon(\tau)}{[1 + \timestamp{1} \dot q_1^o(\tau)]^2}
	\qs
	\label{eq:timestamping-errors-in-frequency-fluctuations}
\end{align}
\label{eq:timestamping-errors-in-frequency}
\end{subequations}

To simplify further our equations, we define the frequency timestamping operator, which includes the rescaling by $1 + \dot{q}_1^o$. It is formally defined by its action on a signal $s(\tau)$,
\begin{equation}
    \dtimestamp{1} s(\tau) = \timestamp{1} \qty[\frac{s(\tau)}{1 + \dot q_1^o(\tau)}]
    = \frac{\timestamp{1} s(\tau)}{1 + \timestamp{1} \dot q_1^o(\tau)}
    \qs
\label{eq:timestamping-operator-frequency}
\end{equation}
Now, photoreceiver frequency signals in the clock time frame of spacecraft~1 read
\begin{subequations}
\begin{align}
	\nu^{\hat\tau_1,o}_\text{PD}(\tau) &\approx \dtimestamp{1} \nu_\text{PD}^{\tau_1,o}(\tau)
	\qc
	\\
	\nu^{\hat\tau_1,\epsilon}_\text{PD}(\tau) &\approx \dtimestamp{1}[
	\nu_\text{PD}^{\tau_1,\epsilon}(\tau) - \frac{\nu_\text{PD}^{\tau_1,o}(\tau) \dot q_1^\epsilon(\tau)}{1 + \dot q_1^o(\tau)}]
	\qs
\end{align}
\label{eq:the-photodiode-signal}
\end{subequations}

\section{Onboard processing}
\label{sec:onboard-processing}

\begin{figure*}
    \centering
    \includegraphics[width=\textwidth]{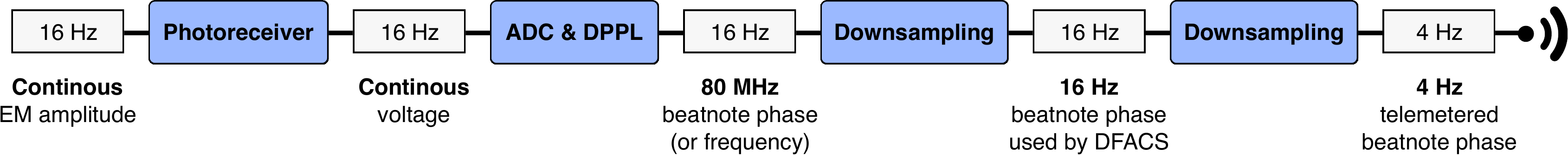}
    \caption{Overview of the real signal sampling rates, from continuous optical and electrical signals, to the cascaded downsampled signals down to the telemetered \SI{4}{\hertz} data. We also indicate the sampling rates used in the simulation (in grey boxes) to represent these signals; continuous and high-frequency signals are represented by discrete \SI{16}{\hertz} simulated quantities, while the telemetry data is simulated at their true \SI{4}{\hertz} rate.}
    \label{fig:sampling-rates}
\end{figure*}

In this section, we describe the processing steps the readout signals undergo on board the spacecraft, and in particular the filtering and downsampling steps. The sampling rates used in our simulation are shown schematically in~\cref{fig:sampling-rates}. We then give the expression of the main measurement signals, which are the main outputs of the simulation.

\subsection{Filtering and downsampling}
\label{sec:filtering-and-downsampling}

\subsubsection{Physics sampling rate}

As described in \cref{sec:phase-readout} and following the current mission design, the onboard phasemeters track the phase (or, equivalently, the instantaneous frequency) of sampled and digitized versions of the \si{\mega\hertz} beatnotes using \glspl{dpll} running at \SI{80}{\mega\hertz}.

For performance reasons, we cannot simulate continuous analog signals nor \gls{dpll} signals at their real sampling rate. Instead, we use a discretized representation and rely on high-level models to capture the most significant effects. In our simulations, continuous quantities, as well as photoreceiver signals and beatnote measurements, are simulated at the physics rate
\begin{equation}
    f_s^\text{phy} = \SI{16}{\hertz}
    \qs
\end{equation}
Note that this physics rate matches the penultimate downsampling step of the real onboard decimation chain (described below), which is used by the \gls{dfacs}.

\subsubsection{Antialiasing filters}
\label{sec:antialiasing-filters}

The current mission design suggests that the raw \SI{80}{\mega\hertz} phasemeter beatnote signals are then filtered and downsampled to various lower sampling rates, and ultimately to the final measurement rate of \SI{4}{\hertz}. This last measurement sampling rate is in line with the mission instrument design, and compatible with the limited telemetry budget and the required bandwidth for on-ground processing. The \SI{4}{\hertz} data are then telemetered down to Earth.

High-order digital low-pass \gls{fir} filters, as well as cascading filters, are expected to be used to prevent noise aliasing in the frequency band relevant for \gls{lisa} data analysis, between \SI{e-4}{\hertz} and \SI{1}{\hertz}~\cite{GerberdingPhD}. These filters must strongly attenuate the signals above the Nyquist frequency, while maintaining a gain close to unity and low phase distortion below \SI{1}{\hertz}. Their precise implementation is still under development.

In the simulation, we use a single digital symmetrical \gls{fir} filter to go from $f_s^\text{phy}$ to the final measurement sampling rate of
\begin{equation}
    f_s^\text{meas} = \SI{4}{\hertz}
    \qs
\end{equation}

The default implementation of the anti-aliasing filter is described in \cref{sec:default-aa-filter}.

\subsubsection{Decimation}

Once the beatnote frequency measurements are filtered, we use a four-fold decimation (we select 1 sample out of 4) to produce the final \SI{4}{\hertz} telemetry data. They are the main output of the simulation.

Analytically, we model the filtering and downsampling step with the continuous, linear filter operator $\filter$, which is applied to the beatnote frequency measurements.

\subsection{Telemetered beatnote measurements}
\label{sec:telemetered-beatnote-measurements}

We summarize here the downsampled, filtered beatnote measurements output by the phasemeter, i.e., the interspacecraft, test-mass, reference carrier and sideband beatnote frequencies. They are ultimately telemetered down to Earth\footnote{As mentioned before, there are other data streams, such as the angular readouts provided by \gls{dws}, which we do not model here. The \gls{mpr} measurements are described in \cref{sec:pseudoranging}.}.

\subsubsection{Beatnote measurement notation}

For these beatnote measurements, we introduce a clear notation that uses the name of the associated interferometer and its index, complemented by the type of beam (carrier or sideband). The real phasemeter will only produce the total frequency or the total phase of the signal. For our studies, however, it is often useful to also have access to the underlying offsets and fluctuations in two separate variables, which is why we give here the signals in this form. The simulation will provide an additional output for the total frequency, given as the sum of the two components.

For readability's sake, we drop all time arguments. We use delay operators to account for time shifts that appear when propagating signals. We denote $\delay{12}$ the delay operator associated with the \gls{ppr} $d^o_{12}(\tau)$ defined in \cref{sec:distant-beams}, such that for any signal $s(\tau)$,
\begin{equation}
\delay{12} s(\tau) = s(\tau - d^o_{12}(\tau))
\qs
\end{equation}
Furthermore, we introduce the Doppler-delay operator, which is defined as
\begin{equation}
\ddelay{12} s(\tau) = (1 - \dot d^o_{12}(\tau)) s(\tau - d^o_{12}(\tau))
\qs
\end{equation}
We also make use of the timestamping operators~$\dtimestamp{i}$ introduced in \cref{sec:signal-sampling}, and the downsampling and filtering operator~$\filter$.

We will also use a shorthand notation for the beatnote frequency offsets in the \gls{tps}, which we define by
\begin{subequations}
\begin{align}
	a^\text{c}_{12} &\equiv \nu_{\text{isi}_{12},\text{c}}^o = \ddelay{12} O_{21} - \nu_0 \dot d^o_{12} - O_{12} \qc
	\\
	a_{12}^\text{sb} &\equiv \nu_{\text{isi}_{12},\text{sb}}^o = a^\text{c}_{12} + \ddelay{12} \qty[\nu_{21}^m \qty(1 + \dot q_2^o)] - \nu_{12}^m \qty(1 + \dot q_1^o) \qc
	\\
	b^\text{c}_{12} &\equiv \nu_{\text{rfi}_{12},\text{c}}^o = O_{13} - O_{12}\qc
	\\
	b_{12}^\text{sb} &\equiv \nu_{\text{rfi}_{12},\text{sb}}^o = b^\text{c}_{12} + \qty(\nu_{13}^m - \nu_{12}^m) \qty(1 + \dot q_1^o)
	\qs
\end{align}
\end{subequations}

In addition, most of the laser-related terms $p_{12}$, $O_{12}$ will be determined by the laser locking scheme, as described in \cref{sec:locking}.

\subsubsection{Interspacecraft interferometer beatnote frequencies}

The carrier-carrier beatnote frequency measurement in the \glspl{isi} contains the delayed distant and local laser frequency fluctuations $\dot{p}_{21}$ and $\dot{p}_{12}$, as well as the delayed distant and local optical-bench path length noises appearing as Doppler shifts  $\dot{N}^\text{ob}_{\text{isi}_{12} \leftarrow 21}$ and $\dot{N}^\text{ob}_{\text{isi}_{12} \leftarrow 12}$. The effect of the gravitational-wave signal $\dot{H}_{12}$ also appears as an extra Doppler shifts on the distant beam. Lastly, the readout $\dot{N}^\text{ro}_{\text{isi}_{12},\text{c}}$ and clock noise $ a^\text{c}_{12} \dot q_1^\epsilon/(1 + q_1^o)$ terms are added.
\begin{subequations}
	\begin{align}
	\text{isi}_{12,\text{c}}^o ={}& \filter \dtimestamp{1}a^\text{c}_{12}
	\qc
	\label{eq:isi-c-offsets-final}
	\\
	\begin{split}
    \text{isi}_{12,\text{c}}^\epsilon ={}& \filter \dtimestamp{1} \biggl\{ \ddelay{12} \dot{p}_{21} - \qty(\nu_0 + \delay{12} O_{21}) \dot{H}_{12}
    \\
    &+ \frac{\nu_0}{c} \dot{N}^\text{ob}_{\text{isi}_{12} \leftarrow 21} - \qty(\dot{p}_{12} + \frac{\nu_0}{c} \dot{N}^\text{ob}_{\text{isi}_{12} \leftarrow 12})
    \\
    &+ \dot{N}^\text{ro}_{\text{isi}_{12},\text{c}} - \frac{a^\text{c}_{12} \dot{q}_1^\epsilon}{1 + \dot{q}_1^o} \biggr\}
    \qc
	\end{split}
    \label{eq:isi-c-fluctuations-final}
    \\
    \text{isi}_{12,\text{c}} ={}& \text{isi}_{12,c}^o + \text{isi}_{12,\text{c}}^\epsilon
    \qs
    \label{eq:isi-c-final}
	\end{align}
\end{subequations}
The sideband-sideband beatnote frequency measurement is similar with two main differences. First, the distant and local laser frequency fluctuations are affected by the coupling of the modulation frequency with clock jitter and modulation noise $\nu_{21}^m \qty(\dot{q}_2^\epsilon + \dot{M}_{21})$ and $\nu_{12}^m \qty(\dot{q}_1^\epsilon + \dot{M}_{12})$. Secondly, the distant sideband beatnote frequency offsets are shifted by the modulation frequency affected by out-of-band clock errors~$\nu^m_{21} \qty(1 + \dot{q}^o_2)$. Overall, we get
\begin{subequations}
	\begin{align}
	\text{isi}_{12,\text{sb}}^o ={}& \filter \dtimestamp{1} a_{12}^\text{sb}
	\qc
	\label{eq:isi-usb-offsets-final}
	\\
	\begin{split}
	\text{isi}_{12,\text{sb}}^\epsilon ={}& \filter \dtimestamp{1} \Biggl\{ \ddelay{12} \qty(\dot{p}_{21} + \nu_{21}^m \qty(\dot{q}_2^\epsilon + \dot{M}_{21}))
    \\
    &- \qty(\nu_0 + \delay{12} \qty[O_{21} + \nu^m_{21} \qty(1 + \dot{q}^o_2)]) \dot{H}_{12} + \frac{\nu_0}{c} \dot{N}^\text{ob}_{\text{isi}_{12} \leftarrow 21}
	\\
	&- \qty(\dot{p}_{12} + \nu_{12}^m \qty(\dot{q}_1^\epsilon + \dot{M}_{12}) + \frac{\nu_0}{c} \dot{N}^\text{ob}_{\text{isi}_{12} \leftarrow 21})
	\\
	&+ \dot{N}^\text{ro}_{\text{isi}_{12},\text{sb}} - \frac{a_{12}^\text{sb} \dot{q}_1^\epsilon}{1 + \dot{q}_1^o} \Biggr\}
	\qc
	\end{split}
	\label{eq:isi-usb-fluctuations-final}
	\\
	\text{isi}_{12,\text{sb}} ={}& \text{isi}_{12,\text{sb}}^o + \text{isi}_{12,\text{sb}}^\epsilon
	\qs
	\label{eq:isi-usb-final}
	\end{align}
\end{subequations}

\subsubsection{Reference interferometer beatnote frequencies}

The carrier-carrier beatnote frequency measurement in the \glspl{rfi} contains the frequency fluctuations of the adjacent and local laser beams $\dot{p}_{13}$ and $\dot{p}_{12}$, as well as the associated optical-bench path length noises $\dot{N}^\text{ob}_{\text{rfi}_{12} \leftarrow 13}$ and $\dot{N}^\text{ob}_{\text{rfi}_{12} \leftarrow 12}$. The adjacent beam that travels through the optical fiber picks up the backlink noise $\dot{N}^\text{bl}_{12}$. The readout $\dot{N}^\text{ro}_{\text{rfi}_{12},\text{c}}$ and clock noise $b^\text{c}_{12} \dot{q}_1^\epsilon/(1 + q_1^o)$ terms are then added.
\begin{subequations}
	\begin{align}
	\text{rfi}_{12,\text{c}}^o ={}& \filter \dtimestamp{1} b^\text{c}_{12}
	\qc
	\label{eq:rfi-c-offsets-final}
	\\
	\begin{split}
	\text{rfi}_{12,\text{c}}^\epsilon ={}& \filter \dtimestamp{1} \Biggl\{ \dot{p}_{13} + \frac{\nu_0}{c} \qty(\dot{N}^\text{ob}_{\text{rfi}_{12} \leftarrow 13} + \dot{N}^\text{bl}_{12})
	\\
	&- \qty(\dot{p}_{12} + \frac{\nu_0}{c}
	\dot{N}^\text{ob}_{\text{rfi}_{12} \leftarrow 12})
    \\
    &+ \dot{N}^\text{ro}_{\text{rfi}_{12},\text{c}} - \frac{b^\text{c}_{12} \dot{q}_1^\epsilon}{1 + \dot{q}_1^o} \Biggr\}
	\qc
	\end{split}
	\label{eq:rfi-c-fluctuations-final}
	\\
	\text{rfi}_{12,\text{c}} ={}& \text{rfi}_{12,\text{c}}^o + \text{rfi}_{12,\text{c}}^\epsilon
	\qs
	\label{eq:rfi-c-final}
	\end{align}
\label{eq:rfi-c-final-global}
\end{subequations}
The expression for the sideband-sideband beatnote frequency measurement follows the same logic, with the adjacent and local laser frequency fluctuations affected by the in-band clock and modulation noises $\nu_{13}^m \qty(\dot{q}_1 + \dot{M}_{13})$ and $\nu_{12}^m \qty(\dot{q}_1 + \dot{M}_{12})$,
\begin{subequations}
	\begin{align}
	\text{rfi}_{12,\text{sb}}^o ={}& \filter \dtimestamp{1} b^\text{sb}_{12}
	\qc
	\label{eq:rfi-usb-offsets-final}
	\\
	\begin{split}
	\text{rfi}_{12,\text{sb}}^\epsilon ={}& \filter \dtimestamp{1} \Biggl\{ \dot{p}_{13} + \nu_{13}^m \qty(\dot{q}_1 + \dot{M}_{13}) + \frac{\nu_0}{c} \qty(\dot{N}^\text{ob}_{\text{rfi}_{12} \leftarrow 13} + \dot{N}^\text{bl}_{12})
	\\
	&- \qty(\dot{p}_{12} + \nu_{12}^m \qty(\dot{q}_1 + \dot{M}_{12}) + \frac{\nu_0}{c} \dot{N}^\text{ob}_{\text{rfi}_{12}\leftarrow 12})
	\\
	&+ \dot{N}^\text{ro}_{\text{rfi}_{12},\text{sb}} - \frac{b^\text{sb}_{12} \dot{q}_1^\epsilon}{1 + \dot{q}_1^o} \Biggr\}
	\qc
	\end{split}
	\label{eq:rfi-usb-fluctuations-final}
	\\
	\text{rfi}_{12,\text{sb}}^o ={}& \text{rfi}_{12,\text{sb}}^o + \text{rfi}_{12,\text{sb}}^\epsilon
	\qs
	\label{eq:rfi-usb-final}
	\end{align}
\end{subequations}

\subsubsection{Test-mass interferometer beatnote frequencies}

The carrier-carrier beatnote frequency measurements in the \gls{tmi} have the same form as for the \gls{rfi}, with the exception of the additional local test-mass noise term $\dot{N}^\delta_{12}$,
\begin{subequations}
	\begin{align}
	\text{tmi}_{12,\text{c}}^o ={}& \filter \dtimestamp{1} b^\text{c}_{12}
	\qc
	\label{eq:tmi-c-offsets-final}
	\\
	\begin{split}
\text{tmi}_{12,\text{c}}^\epsilon ={}& \filter \dtimestamp{1} \Biggl\{ \dot{p}_{13} + \frac{\nu_0}{c} \qty(\dot{N}^\text{ob}_{\text{tmi}_{12} \leftarrow 13} + \dot{N}^\text{bl}_{12})
	\\
	&- \qty(\dot{p}_{12} + \frac{\nu_0}{c} \qty(\dot{N}^\text{ob}_{\text{tmi}_{12}\leftarrow 12} + 2 \dot{N}^\delta_{12}))
	\\
	&+ \dot{N}^\text{ro}_{\text{tmi}_{12},\text{c}} - \frac{b^\text{c}_{12} \dot{q}_1^\epsilon}{1 + \dot{q}_1^o} \Biggr\}
	\qc
	\end{split}
	\label{eq:tmi-c-fluctuations-final}
	\\
\text{tmi}_{12,\text{c}} ={}& \text{tmi}_{12,\text{c}}^o + \text{tmi}_{12,\text{c}}^\epsilon
	\qs
	\label{eq:tmi-c-final}
	\end{align}
\end{subequations}
As mentioned previously, we do not model sideband-sideband beatnote measurements in the \glspl{tmi}.

\section{Laser locking and frequency planning}
\label{sec:locking}

As mentioned in \cref{sec:local-beams}, each laser source is either frequency-locked to a resonant cavity or phase-locked to another laser source using a specific interferometric beatnote. In this section, we describe how we simulate these laser locking control loops. We then list the various locking configurations available for \gls{lisa} in its baseline configuration.

\subsection{Frequency planning}

The beatnote frequencies that can be measured by the \gls{lisa} phasemeters are limited to between \SI{5}{\mega\hertz} and \SI{25}{\mega\hertz}\footnote{The exact frequency range remains to be defined. In addition, some margins are required for both the upper and lower bounds to account for the sideband beatnotes, which are offset by \SI{1}{\mega\hertz} from the carrier beatnotes.}. As a consequence, all beatnote frequencies need to be controlled to fall in this range, which is achieved by introducing pre-determined offset frequencies in the laser locking control loops. A set of these frequency offsets for all lasers over the whole mission duration is called a \textit{frequency plan}.

The problem of finding such frequency plans has recently been studied (G.~Heinzel, \gls{lisa} Consortium internal technical note, November 2018), and exact solutions have been found. We will use these solutions as an input to the simulation.

\subsection{Locking condition}

Laser locking is achieved by controlling the frequency of a locked laser, such that a given beatnote frequency $\nu_\text{PD}(\tau)$ remains equal to a pre-programmed reference value $\nu_\text{plan}(\tau)$ provided in the frequency plan.

We do not simulate the actual control loop, but instead directly compute the correct frequency offsets and fluctuations of the locked laser for this locking condition to be satisfied. In reality, the locking control loops will have finite gain and bandwidth, such that the locking beatnotes can still contain out-of-band glitches and noise residuals. Here, we consider the frequency lock to be perfect. This means that the locking beatnote offset is exactly equal to the desired value.

Locking control loops run according to their local clocks, such that the locking condition is fulfilled in the local clock time frame $\hat\tau_1$. In addition, the frequency-plan locking frequencies are interpreted as functions of the same local clock time frame. In terms of total phase, the result of this control is that the measured beatnote phase is controlled to be exactly equal to the frequency-plan phase,
\begin{equation}
    \Phi^{\hat\tau_1}_\text{PD}(\tau) = \Phi_\text{plan}(\tau)
    \qs
    \label{eq:locking-condition-the}
\end{equation}
Note that the control loop operates on data delivered by the phasemeter at a high frequency of \SI{80}{\mega\hertz} (K.~Yamamoto, personal communication, May 2021). As such, we simulate the locking before applying any filtering or downsampling.

The previous locking condition is expressed in the local time frame, but we really want to solve for it in the \gls{tps}. We can use \cref{eq:phase-time-coordinate-conversion,eq:spacecraft-clock-time,eq:clock-two-variable-definition} to relate the measured beatnote phase to its equivalent in the \gls{tps},
\begin{equation}
    \Phi^{\tau_1}_\text{PD}(\tau)
    = \Phi^{\hat\tau_1}_\text{PD}(\tau + q_1^o(\tau) + q_1^\epsilon(\tau) + \delta\hat\tau_{1,0})
    \qs
\end{equation}
Using this result, we can write the locking condition from \cref{eq:locking-condition-the} as
\begin{equation} \label{eq:locking-condition-phase}
    \Phi^{\tau_1}_\text{PD}(\tau)
    = \Phi_\text{plan}(\tau + q_1^o(\tau) + q_1^\epsilon(\tau) + \delta\hat\tau_{1,0})
    \qs
\end{equation}
We expand the previous equation to first order in $q_1^\epsilon(\tau)$,
\begin{equation}
    \begin{split}
        \Phi^{\tau_1}_\text{PD}(\tau)
        ={}& \Phi_\text{plan}(\tau + q_1^o(\tau) + \delta\hat\tau_{1,0})
        \\
        &+ \nu_\text{plan}(\tau + q_1^o(\tau) + \delta\hat\tau_{1,0}) q_1^\epsilon(\tau)
        \qs
    \end{split}
    \label{eq:locking-condition-phase-expanded}
\end{equation}
The second-order term is proportional to the product $q_1^\epsilon(\tau)^2 \dot\nu_\text{plan}(\tau + q_1^o(\tau) + \delta\hat\tau_{1,0})$ of the square of the clock fluctuations and the time derivative of the frequency-plan locking frequency. To evaluate the order of magnitude of this term, we compute the average clock time deviation~\cite{Riley:2008} after a time corresponding to its saturation frequency (described in \cref{sec:noise-models}); we find a value of the order of $\SI{E-9}{\second}$. In addition, all currently available frequency plans verify $\dot\nu_\text{plan}(\tau) < \SI{3}{\hertz\per\second}$. Therefore, we neglect terms of the order of \SI{3E-18}{cycles}, far below the $\mu$-cycle level of gravitational-wave signals.

From \cref{eq:locking-condition-phase-expanded}, one directly obtains the usual decomposition in phase drifts and fluctuations,
\begin{subequations}
    \begin{align}
        \phi^{\tau_1, o}_\text{PD}(\tau) &= \Phi_\text{plan}(\tau + q_1^o(\tau) + \delta\hat\tau_{1,0})
        \qc
        \\
        \phi^{\tau_1, \epsilon}_\text{PD}(\tau) &= \nu_\text{plan}(\tau + q_1^o(\tau) + \delta\hat\tau_{1,0}) q_1^\epsilon(\tau)
        \qs
    \end{align}
\end{subequations}
Indeed, the current baseline is to use piecewise linear functions with daily inflexions as frequency-plan locking frequencies. As a consequence, the latter are slowly varying, i.e., only consist in large out-of-band frequency offsets, such that $\nu_\text{plan}(\tau) = \nu^o_\text{plan}(\tau)$. This also applies to the pre-programmed reference phase $\Phi_\text{plan}(\tau) = \phi^o_\text{plan}(\tau)$.

We denote the (local) locked laser phase drifts and fluctuations as $\phi_l^{\tau_1,o}(\tau)$ and $\phi_l^{\tau_1,\epsilon}(\tau)$, and the (distant or adjacent) reference laser phase drifts and fluctuations as $\phi_r^{\tau_1,o}(\tau)$ and $\phi_r^{\tau_1,\epsilon}(\tau)$. Using \cref{eq:two-variable-photodiode-phase}, we have
\begin{subequations}
    \begin{align}
        \phi^{\tau_1, o}_\text{PD}(\tau) &= \phi_r^{\tau_1,o}(\tau) - \phi_l^{\tau_1,o}(\tau)
        \qc
        \\
        \phi^{\tau_1, \epsilon}_\text{PD}(\tau) &= \phi_r^{\tau_1,\epsilon}(\tau) - \phi_l^{\tau_1,\epsilon}(\tau) + N_\text{PD}^\text{ro}(\tau)
        \qs
    \end{align}
    \label{eq:two-variable-beatnote-pd}
\end{subequations}
It is now straightforward to write the resulting locked laser phase drifts and fluctuations,
\begin{subequations}
    \begin{align}
        \phi_l^{\tau_1,o}(\tau) ={}& \phi_r^{\tau_1,o}(\tau) - \Phi_\text{plan}(\tau + q_1^o(\tau) + \delta\hat\tau_{1,0})
        \qc
        \\
        \begin{split}
        \phi_l^{\tau_1,\epsilon}(\tau) ={}& \phi_r^{\tau_1,\epsilon}(\tau) - \Phi_\text{plan}(\tau + q_1^o(\tau) + \delta\hat\tau_{1,0}) q_1^\epsilon(\tau)
        \\
        &+ N_\text{PD}^\text{ro}(\tau)
        \qs
        \end{split}
    \end{align}
\end{subequations}
For frequency, we start by taking the derivative of \cref{eq:locking-condition-phase} and expand it once again in $q_1^\epsilon$,
\begin{equation}
    \begin{split}
        \nu^{\tau_1}_\text{PD}(\tau)
        ={}& (1 + \dot{q}_1^o(\tau)) \nu_\text{plan}(\tau + q_1^o(\tau) + \delta\hat\tau_{1,0})
        \\
        &+ \nu_\text{plan}(\tau + q_1^o(\tau) + \delta\hat\tau_{1,0}) \dot{q}_1^\epsilon(\tau)
        \\
        &+ (1 + \dot{q}_1^o(\tau) + \dot{q}_1^\epsilon(\tau)) \dot\nu_\text{plan}(\tau + q_1^o(\tau) + \delta\hat\tau_{1,0}) q_1^\epsilon(\tau)
        \qs
    \end{split}
\end{equation}
The last term of the previous equation has a similar form as the small correction neglected in \cref{eq:propagated-frequency-fluctuations}. We study such a term in \cref{sec:dotnu-discussion} and find that it is several orders of magnitude below the main noise term. Therefore, we will neglect it in the rest of this derivation.

Using the same two-variable decomposition along with the frequency equivalent of \cref{eq:two-variable-beatnote-pd},
\begin{subequations}
    \begin{align}
        \nu^{\tau_1, o}_\text{PD}(\tau) &= \nu_r^{\tau_1,o}(\tau) - \nu_l^{\tau_1,o}(\tau)
        \qc
        \\
        \nu^{\tau_1, \epsilon}_\text{PD}(\tau) &= \nu_r^{\tau_1,\epsilon}(\tau) - \nu_l^{\tau_1,\epsilon}(\tau) + \dot{N}_\text{PD}^\text{ro}(\tau)
        \qc
    \end{align}
\end{subequations}
we finally obtain the resulting locked laser frequency offset and fluctuations,
\begin{subequations}
    \begin{align}
        \begin{split}
            \nu_l^{\tau_1,o}(\tau) ={}& \nu_r^{\tau_1,o}(\tau) - (1 + \dot{q}_1^o(\tau))
            \\
            &\times \nu_\text{plan}(\tau + q_1^o(\tau) + \delta\hat\tau_{1,0})
            \qc
        \end{split}
        \label{eq:locking-condition-frequency-offsets}
        \\
        \begin{split}
            \nu_l^{\tau_1,\epsilon}(\tau) ={}& \nu_r^{\tau_1,\epsilon}(\tau) - \nu_\text{plan}(\tau + q_1^o(\tau) + \delta\hat\tau_{1,0})
            \\
            &\times \dot{q}_1^\epsilon(\tau) + \dot{N}_\text{PD}^\text{ro}(\tau)
        \qs
        \end{split}
    \end{align}
\end{subequations}

Note that these equations describe the locked laser \textit{at the photodiode}. To properly simulate this effect, we need the locked lasers frequency \textit{at the laser source}, which we denote here as $\bar\nu_l(\tau)$. In \cref{sec:local-beams}, we add to the local beam frequency fluctuations an optical path length noise term $\dot N^\text{ob}_\text{PD}(\tau)$ during its propagation from the laser source to the photodiode. As a consequence, we have
\begin{equation}
\begin{split}
    \bar{\nu}_l^\epsilon(\tau) ={}& \nu_r^\epsilon(\tau) - \nu_\text{plan}(\tau + q_1^o(\tau) + \delta\hat\tau_{1,0}) \dot{q}_1^\epsilon(\tau)
    \\
    &+ \dot N^\text{ro}_\text{PD}(\tau) - \frac{\nu_0}{c} \dot N^\text{ob}_\text{PD}(\tau)
    \label{eq:locking-condition-frequency-fluctuations}
\end{split}
\end{equation}
for the local locked lasers fluctuations.

\subsection{Locking configurations}
\label{sec:locking-configurations}

In total, 5 of the 6 lasers in the constellation are locked (directly or indirectly) to one \textit{primary laser}. Each of the locked lasers is locked to either the adjacent laser, using the \gls{rfi}, so that \cref{eq:locking-condition-frequency-offsets,eq:locking-condition-frequency-fluctuations} read
\begin{subequations}
\begin{align}
    \begin{split}
        O_{12}(\tau) ={}& \nu_{\text{rfi}_{12} \leftarrow 13}^o(\tau) - (1 + \dot q_1^o(\tau))
        \\
        &\times \nu_{\text{fplan},\text{rfi}_{12}}(\tau + q_1^o(\tau) + \delta\hat\tau_{1,0})
	   \qc
    \end{split}
	\\
	\begin{split}
    	\dot{p}_{12}(\tau) ={}& \nu_{\text{rfi}_{12} \leftarrow 13}^\epsilon(\tau) - \nu_{\text{fplan},\text{rfi}_{12}}(\tau + q_1^o(\tau) + \delta\hat\tau_{1,0}) \dot q_1^\epsilon(\tau)
    	\\
    	&+ \dot N^\text{ro}_{\text{rfi}_{12}}(\tau) - \frac{\nu_0}{c}
    	\dot N^\text{ob}_{\text{rfi}_{12}\leftarrow 12}(\tau)
    	\qc
	\end{split}
\end{align}
\end{subequations}
or to the distant laser, using the \gls{isi}, such that we get
\begin{subequations}
\begin{align}
    \begin{split}
    	O_{12}(\tau) ={}& \nu_{\text{isi}_{12} \leftarrow 21}^o(\tau) - (1 + \dot q_1^o(\tau))
        \\
        &\times \nu_{\text{fplan},\text{isi}_{12}}(\tau + q_1^o(\tau) + \delta\hat\tau_{1,0})
    	\qc
    \end{split}
	\\
	\begin{split}
    	\dot{p}_{12}(\tau) ={}& \nu_{\text{isi}_{12} \leftarrow 21}^\epsilon(\tau) - \nu_{\text{fplan},\text{isi}_{12}}(\tau + q_1^o(\tau) + \delta\hat\tau_{1,0}) \dot q_1^\epsilon(\tau)
        \\
        &+ \dot N^\text{ro}_{\text{isi}_{12}}(\tau) - \frac{\nu_0}{c}
    	\dot N^\text{ob}_{\text{isi}_{12}\leftarrow 12}(\tau)
    	\qs
	\end{split}
    \label{eq:locking-condition-distant}
\end{align}
\end{subequations}
These expressions can be substituted into the equations of \cref{sec:telemetered-beatnote-measurements} to derive the telemetered beatnote measurements with locked lasers.

\begin{figure*}
	\centering
	\includegraphics[width=\textwidth]{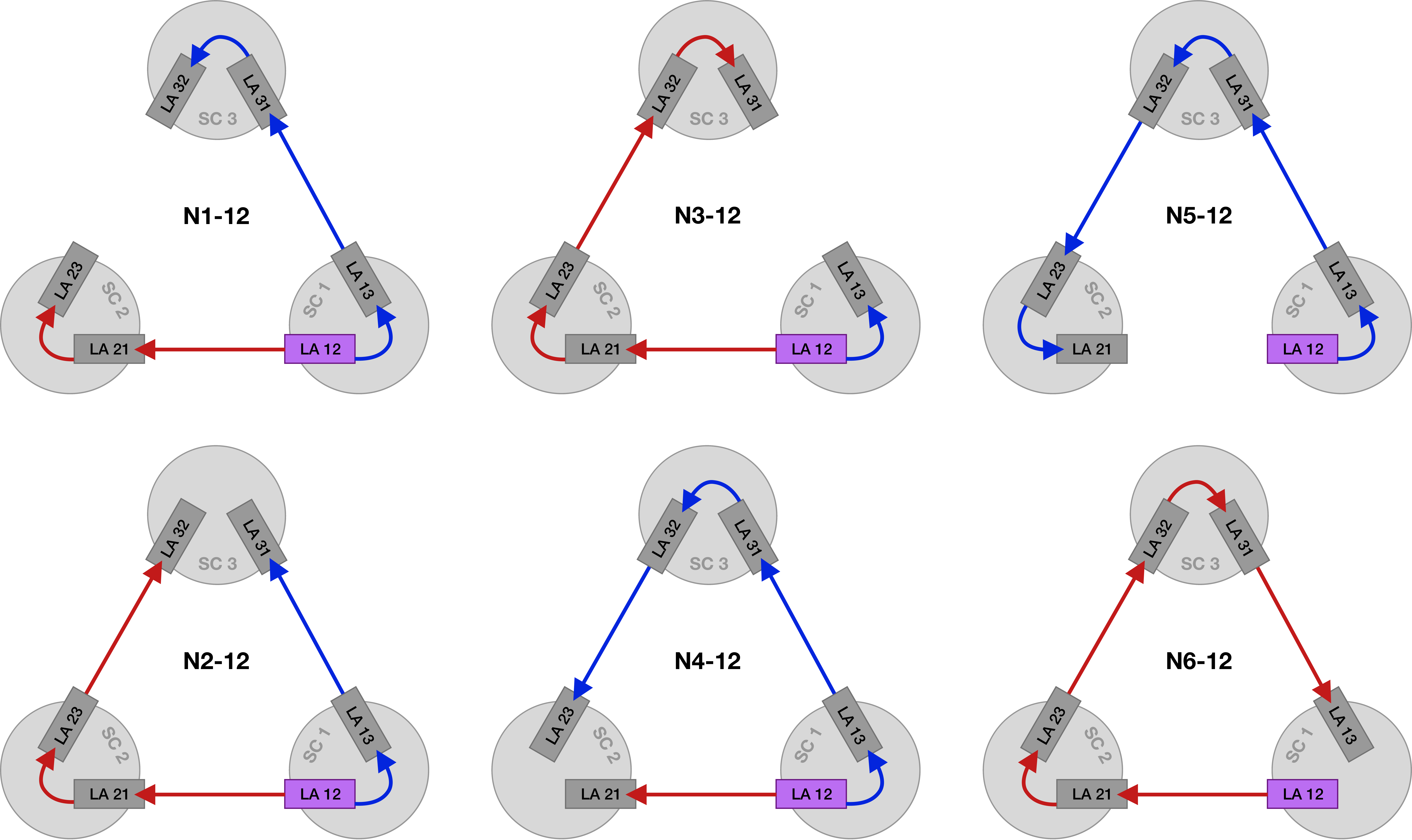}
	\caption{Laser locking configurations for laser 12 as primary laser.}
	\label{fig:locking-configurations}
\end{figure*}

The \gls{lisa} model presented here permits 6 distinct \textit{locking topologies}. For each of them, we have the freedom to choose the primary laser, such that, in total, we have 36 possible \textit{locking configurations}. We plot the 6 configurations with laser~$12$ as the primary laser in \cref{fig:locking-configurations}. The other 30 combinations can be deduced by applying permutations of the indices.

% \begin{table}[t]
% 	\centering
% 	\begin{tabular}{ccccccc}
% 		\toprule
% 		Configuration & LA 23 & LA 31 & LA 13 & LA 32 & LA 21 \\
% 		\midrule
% 		N1-12 & \gls{rfi} & \gls{isi} & \gls{rfi} & \gls{rfi} & \gls{isi} \\
% 		N2-12 & \gls{rfi} & \gls{isi} & \gls{rfi} & \gls{isi} & \gls{isi} \\
% 		N3-12 & \gls{rfi} & \gls{rfi} & \gls{rfi} & \gls{isi} & \gls{isi} \\
% 		N4-12 & \gls{isi} & \gls{isi} & \gls{rfi} & \gls{rfi} & \gls{isi} \\
% 		N5-12 & \gls{isi} & \gls{isi} & \gls{rfi} & \gls{rfi} & \gls{rfi} \\
% 		N6-12 & \gls{rfi} & \gls{rfi} & \gls{isi} & \gls{isi} & \gls{isi} \\
% 		\bottomrule
% 	\end{tabular}
% 	\caption{Definition of 6 fundamental non-swapping locking configurations, with laser 12 as primary laser. For each locking configuration, we indicate the interferometer used to lock each locked laser.}
% 	\label{tab:locking-configurations}
% \end{table}

% \section{Auxiliary measurements}

\section{Pseudoranging}
\label{sec:pseudoranging}

In addition to the \si{\giga\hertz} sideband modulation, each laser beam will also carry an additional modulation with a pre-determined \gls{prn} code used for absolute ranging and timing synchronization. The basic measurement principle is to correlate the received \gls{prn} code in each \gls{isi} with a local copy generated on the receiving spacecraft. The result of this measurement is the \glsfirst{mpr}, which contains information on both the light travel time between the spacecraft and the clock desynchronization.

\subsection{Pseudoranging modulation}

The \gls{prn} modulation is performed at a relatively high frequency of around \SI{2}{\mega\hertz}, far outside our simulation bandwidth. We therefore do not model the actual phase modulation. This modulation also causes a small additional noise in our measurement band, at a level below \SI{1}{\pico\meter\hertz\tothe{-0.5}} in units of displacement~\cite{EstebanPhD}, which we do not model. In addition, we only model the \gls{prn} measurement in the \gls{isi}, and completely ignore the presence of the \gls{prn} codes in the other interferometers.

Instead, we model this measurement by directly propagating the time deviations of each spacecraft timer with respect to their \glspl{tps}, alongside the laser beams. The \gls{mpr} is then computed as the difference between the received and local timer.

Similarly to the main interferometric measurements and as described in \cref{sec:filtering-and-downsampling}, pseudoranging simulation is performed at $f_s^\text{phy}$, while the \glspl{mpr} are ultimately filtered and downsampled to a lower rate $f_s^\text{meas}$.

\subsection{Pseudoranging as clock time difference}
\label{sec:pseudoranging-as-clock-time-difference}

We consider in the following paragraphs a beam received by optical bench~$12$ at the receiver \gls{tps} $\tau$, which was emitted from optical bench~$21$ at emitter \gls{tps} $\tau - d_{12}(\tau)$. Here, the \gls{ppr} $d_{12}(\tau)$ contains the light time of flight, as well as the conversion between the two proper times.

Conceptually, the \gls{mpr} measures the pseudorange, given as the difference between the time $\hat\tau_1^{\tau_1}(\tau)$ shown by the local clock of the receiving spacecraft at the event of reception of the beam, and the time $\hat\tau_2^{\tau_2}(\tau - d_{12}(\tau))$ shown by the local clock of the sending spacecraft at the event of emission of the beam. In reality, the \gls{mpr} only measures the pseudorange up to the repetition period of the \gls{prn} code, which is around \SI{1}{\milli\second}. The full pseudorange is then recovered by combining the \gls{mpr} measurements with ground-based observations.

At the moment, we do not simulate this effect and assume that the \gls{mpr} directly gives the pseudorange without ambiguity. In addition, we assume that the vacuum between the satellites is sufficiently good that we can neglect (or compensate for) any dispersion effects, such that the \gls{prn} code suffers exactly the same delay as the carrier and sidebands.

Thus, we can model the \gls{mpr} as the difference
\begin{equation}
    R_{12}(\tau) = \hat\tau_1(\tau) - \hat\tau_2(\tau - d_{12}(\tau)) + N^R_{12}(\tau)
    \qc
\label{eq:pseudoranging}
\end{equation}
where $N^R_{12}(\tau)$ is a ranging noise term modeling imperfections in the overall correlation scheme.

\subsection{Pseudoranging in terms of timer deviations}
\label{sec:pseudoranging-in-terms-of-timer-deviations}

As explained in \cref{sec:timer-model}, we do not simulate the total clock time $\hat\tau_1(\tau)$ for each spacecraft, but only deviations $\delta\hat\tau_1(\tau)$ from the associated \gls{tps},
\begin{equation}
    \hat\tau_1(\tau) = \tau + \delta\hat\tau_1(\tau)
    \qand
    \hat\tau_2(\tau) = \tau + \delta\hat\tau_2(\tau)
    \qs
\end{equation}
Inserting these definitions into \cref{eq:pseudoranging} yields
\begin{equation}
    R_{12}(\tau) = \delta\hat\tau_1(\tau) - \qty[\delta\hat\tau_2 \qty(\tau - d_{12}(\tau)) - d_{12}(\tau)] + N^R_{12}(\tau)
    \qs
\end{equation}

Let us define the clock time of the sending spacecraft propagated to the photodiode of the distant interspacecraft interferometer as
\begin{equation}
    \delta\hat\tau_{\text{isi}_{12} \leftarrow 2}(\tau) \approx \delta\hat\tau_3(\tau - d_{12}(\tau)) - d_{12}(\tau)
    \qs
\label{eq:time-deviation-propagation}
\end{equation}

We can then express the \gls{mpr} as the simple difference
\begin{equation}
    R_{12}(\tau) \approx \delta\hat\tau_1(\tau) - \delta\hat\tau_{\text{isi}_{12} \leftarrow 2}(\tau)  + N^R_{12}(\tau)
    \qs
\label{eq:implemented-pseudoranging}
\end{equation}
In our simulation, we make the additional assumption that $d_{12} \approx d_{12}^o$ for this measurement. This is valid since the terms contained in $d_{12}^\epsilon$ (in our simulation model, only $H_{12}$) create timing jitters much less than a nanosecond.

Notice that in \cref{eq:implemented-pseudoranging}, we compute the \gls{mpr} as a function of the receiving \glspl{tps}, so that formally $R_{12} = R_{12}^{\tau_1}$. In reality, the \gls{mpr} is measured according to the clock time of the receiving spacecraft, $R_{12}^{\hat\tau_1}$. Similarly to all other measurements, we simulate this by first generating $R_{12}^{\tau_1}$ and then resampling the resulting time series to obtain $R_{12}^{\hat\tau_1}$, as described in \cref{sec:signal-sampling}.

\section{Implementation}
\label{sec:implementation}

The model presented in the previous sections has been implemented independently in two \gls{lisa} Consortium simulators, namely \lisainstrument and \lisanode.

In this section, we briefly describe the structure of both simulators and highlight the key differences between them. Results obtained from these simulators are presented in \cref{sec:results}.

\subsection{LISA Instrument}

\lisainstrument~\cite{lisainstrument} is a Python-based implementation of the simulation model described in this paper. It is designed to facilitate fast exploratory studies, run quick or partial simulations (instrumental effects and noises can easily be toggled on and off), and prototype new features.

\lisainstrument ships as a standalone Python package. As a consequence, it is easy to install, use, and integrate in traditional workflows, such as Jupyter Notebooks. \lisainstrument does not require a custom installation and can be used out-of-the-box on most computing clusters.

\lisainstrument relies strongly on traditional numerical libraries, such as \textsc{Numpy} and \textsc{Scipy}~\cite{Harris:2020xlr,Virtanen:2019joe}, and therefore benefits from fast optimized vectorized operations as it handles large arrays of data. Its runtime performance is studied and compared to that of \lisanode in \cref{fig:performance}.

\begin{figure}
    \centering
    \includegraphics[width=0.9\columnwidth]{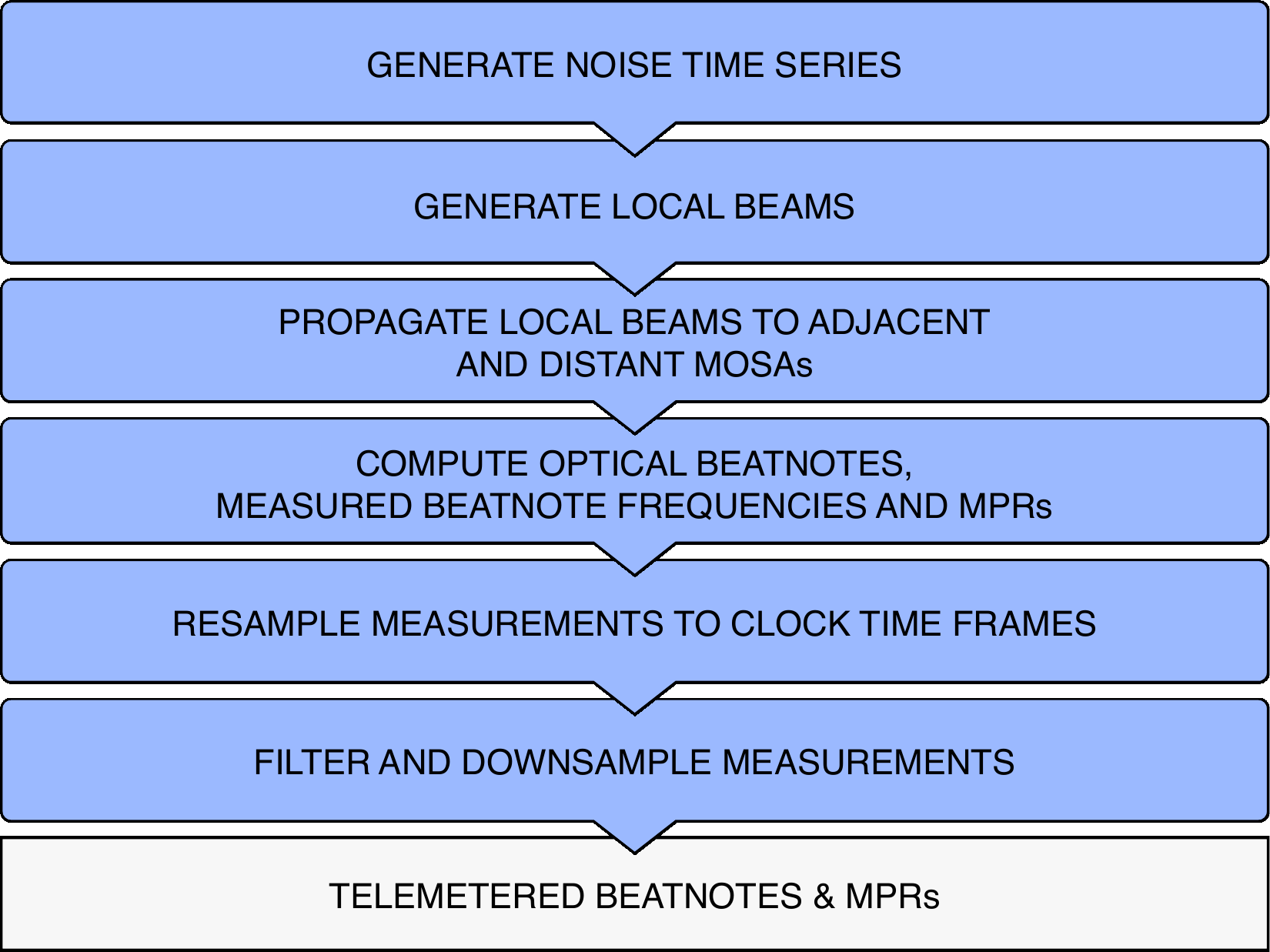}
    \caption{\lisainstrument simulation workflow. At each stage, entire time series are generated. In the end, the simulation products comprise the telemetered beatnotes and the \glspl{mpr}.}
    \label{fig:lisainstrument-flow}
\end{figure}

\lisainstrument runs stage-by-stage simulations, where time series are generated for the entire simulation duration at each stage. The main stages of a simulation are represented in \cref{fig:lisainstrument-flow}. First, time series are generated for all noises enabled in the simulation, following the prescription of \cref{sec:noise-models}. \lisainstrument uses \gls{fir} and cascaded RC filters~\cite{Plaszczynski:2005yp} to generate the noise time series. Then, local beam frequencies are computed (see \cref{sec:local-beams}). Local beams from locked laser sources are obtained by substituting the results of locking condition equations found in \cref{sec:locking-configurations}. These local beams are then propagated to obtain the adjacent and distant beam frequency time series (see \cref{sec:adjacent-beams,sec:distant-beams}). Optical beatnotes and measured beatnote frequencies are obtained from the equations derived in \cref{sec:interferometers,sec:phase-readout}. \Glspl{mpr} are also computed according to the model described in \cref{sec:pseudoranging}. At this point, both beatnote frequencies and \glspl{mpr} are expressed as functions of their respective \glspl{tps}; they are resampled at the next stage to their associated clock time frames, following the methodology given in \cref{sec:signal-sampling}. Finally, all measurements are filtered and downsampled (c.f., \cref{sec:filtering-and-downsampling}) to obtain the telemetered beatnote frequencies and \gls{mpr} measurements described in \cref{sec:telemetered-beatnote-measurements}.

A downside of this simple implementation is that memory usage increases drastically with the simulation length. Memory pressure can become limiting for long simulations (typically more than a few months, see \cref{sec:performance}) if many noises and instrumental effects are enabled. The alternative implementation of the same simulation model, described in the next section, is overall less flexible but is optimized for long simulations.

\subsection{LISANode}

\lisanode~\cite{Bayle:2019dfu,lisanode} is a simulation framework that allows the user to build modular simulation graphs out of atomic computational units, called nodes. This is realized using a mix of Python and \cpp, where the Python code is responsible for defining the graph structure and interconnecting the different nodes. The nodes themselves are implemented in \cpp, such that the final executable is a \cpp command line program.

Using \cpp offers the advantage that compiler optimizations produce a fast executable, and allows us to reuse legacy code from previous \cpp \gls{lisa} simulators, such as \textsc{LISACode}~\cite{Petiteau:2008zz}. Naturally, the cost is reduced readability, usability, and slower development times. Another consequence is that \lisanode needs to be compiled on each machine it runs. To work around this last difficulty, we offer containerization solutions, in the form of Docker and Singularity images that contain optimized compiled versions of \lisanode along with the software environment necessary to run them. These images can be downloaded and used on local machines and on most computing infrastructures, and do not hinder the runtime performance of the simulations.

In \lisainstrument, data is generated for the whole simulation length. On the contrary, \lisanode creates data one step at a time: a sample at time $t_n$ is computed for all quantities before repeating the same instructions for the next samples $t_{n+1}$. New samples are therefore simulated on the fly, keeping only in memory the data that is required for the current and future samples. This way, memory usage remains roughly constant regardless of the simulation length (see \cref{sec:performance}). This allows long simulations to run on memory-constrained machines.

\subsection{Runtime and memory performance}
\label{sec:performance}

\begin{figure*}
    \centering
    \includegraphics[width=\textwidth]{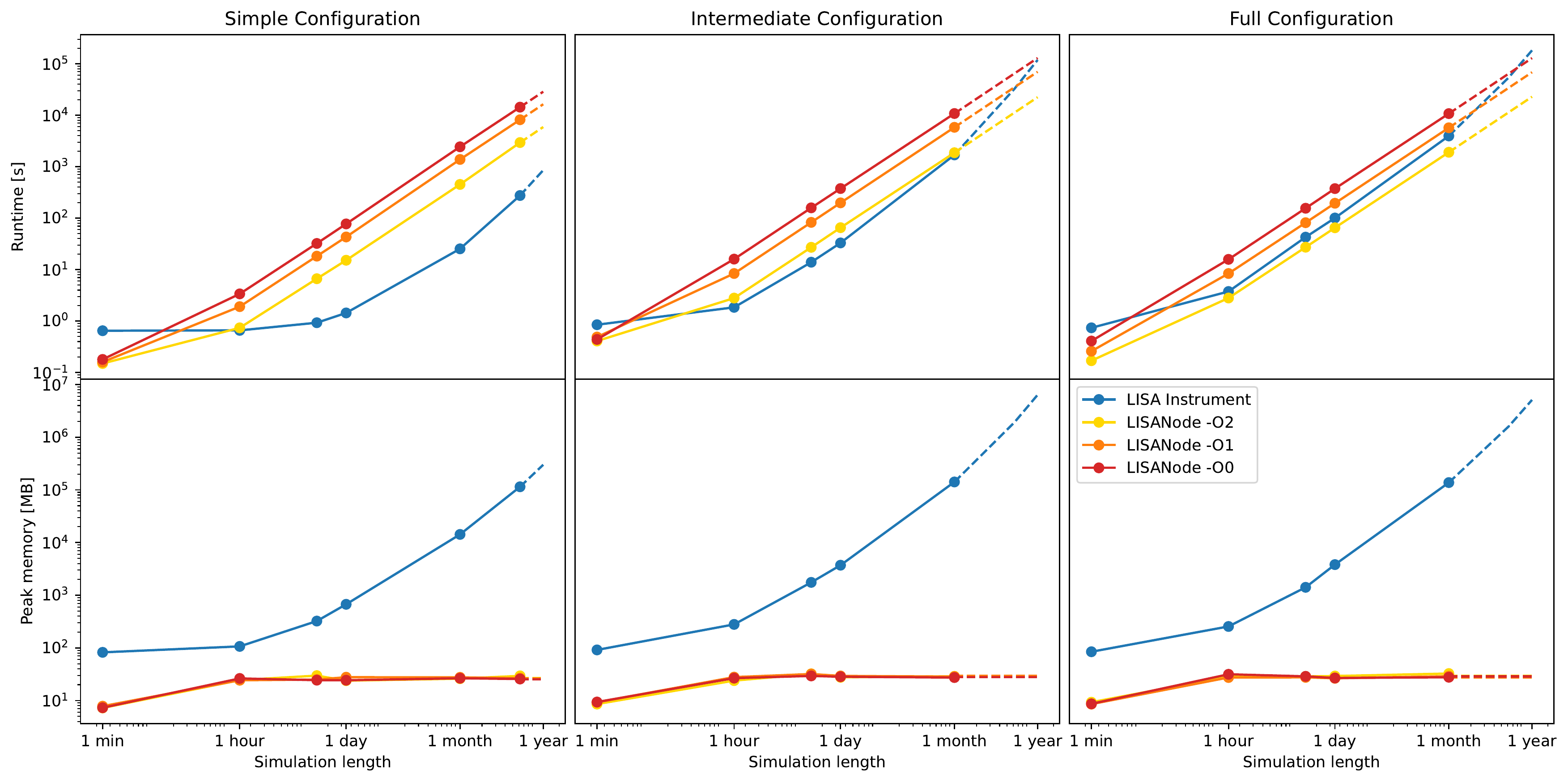}
    \caption{Runtime and memory performance of \lisainstrument (in blue) and \lisanode (in yellow, orange, and red for the various compiler optimization levels), for three instrumental configurations of increasing complexity (from left to right) and simulations durations. Dashed lines are used for extrapolated data.}
    \label{fig:performance}
\end{figure*}

We evaluate the runtime and memory performance of \lisainstrument and \lisanode for three instrumental configurations of increasing complexity. In the \textit{simple configuration}, all noises but laser noise are neglected. Most instrument effects are disabled, as we use a static constellation with constant arm lengths, do not lock the lasers, assume perfect clocks, and set $f_s^\text{meas} = f_s^\text{phy} = \SI{4}{\hertz}$ (no filtering or downsampling). In the \textit{intermediate configuration}, all noises and effects are activated except for clock errors (no resampling of the measurement to clock time frames). We use realistic orbits and frequency plan for the locking configuration N1-12. We filter all measurements and use the nominal sampling frequencies described in \cref{sec:filtering-and-downsampling}. Lastly, the \textit{full configuration} includes the effects of imperfect clocks.

We run simulations of increasing durations, ranging from \SI{1}{\hour} (\num{1.4E4}~telemetered samples at \SI{4}{\hertz} for each channel) to \SI{6}{months} (\num{6.2E7}~samples). Missing points indicates that the simulation did not complete on our test machine (MacBook Pro M1, 2021, 64 GB of RAM) because of excessive memory pressure or runtime. We extrapolate the results to \SI{1}{year} (dashed lines) using a linear (\lisanode) or quadratic (\lisainstrument) fit to the existing data points.

We used the latest version of \lisainstrument and \lisanode, and compiled three different \lisanode executables to study the impact of optimizations: one with no compiler optimization (compiler flag \texttt{-O0}), one with some optimizations (\texttt{-O1}), and one with most optimizations (\texttt{-O2}) enabled. In each case, we measure the runtime and peak memory usage. Results are reported in \cref{fig:performance}. Note that we do not include compilation time for \lisanode in these figures, which strongly depends on the chosen optimization level (around \SI{5}{\second} for \texttt{-O0}, \SI{1}{min} for \texttt{-O1}, and \SI{10}{min} for \texttt{-O2}).

In terms of runtime, as expected, \lisainstrument is significantly (up to several orders of magnitude) faster for simple simulations. For intermediary configurations, \lisainstrument remains faster than \lisanode up to simulation length of a month. Considering full instrumental configurations, highly optimized versions of \lisanode are faster than \lisainstrument irrespective of simulation duration. This is especially true for simulations longer than a month, where \lisanode runs roughly twice as fast. In addition, memory usage can become limiting for simulations of a day or longer on typical machines with a few \si{GB} of memory when using \lisainstrument. \lisanode caps memory peak usage to low values of about \SI{100}{MB} irrespective of the simulation length.

\subsection{Simulation parameters and simulation products}

For reference, we give in \cref{tab:simulation-parameters} the list of options accepted by \lisainstrument to configure the simulations. They parametrize the instrumental configuration (type of orbits, choice of a laser locking configuration, design of the onboard filters, etc.), the various noise models (noise amplitudes and spectral shapes), and the length of the simulation. Note that similar options can be used with \lisanode, with some slight variations in their names.

We also list in \cref{tab:simulation-products} the quantities output by both simulators, alongside their units, and reference equations.

\addtolength{\tabcolsep}{4pt}
\begin{table*}
	\centering
	\begin{tabular}{cccc}
		\toprule
		Parameter & Description & Unit & Reference \\
		\midrule % Simulation run parameters
		\texttt{size} & Number of samples to simulate & - & - \\
        \texttt{dt} & Measurement sampling period $f_s^\text{meas}$ & \si{\second} & \Cref{sec:filtering-and-downsampling} \\
        \texttt{t0} & Initial simulation time & \si{\second} & - \\
        \texttt{physics\_upsampling} & Ratio $f_s^\text{phy} / f_s^\text{meas}$ of physics and measurement sampling frequencies & - & \Cref{sec:filtering-and-downsampling} \\
        \texttt{clockinv\_tolerance} & Convergence criterion for clock noise inversion & \si{\second} & \Cref{sec:signal-sampling} \\
        \texttt{clockinv\_maxiter} & Maximum number of iterations for clock noise inversion & - & \Cref{sec:signal-sampling} \\
        \midrule % Instrument setup
        \texttt{aafilter} & Antialiasing filter design specifications & - & \Cref{sec:filtering-and-downsampling} \\
        \texttt{orbits} & Path to orbit file, or \glspl{ppr} $d^o_{ij}(\tau)$ & \si{\second} & \Cref{sec:distant-beams} \\
        \texttt{gws} & Path to gravitational-wave file, or link responses $H_{ij}(\tau)$ & - & \Cref{sec:distant-beams} \\
        \texttt{lock} & Laser locking configuration (e.g., N1-12) & - & \Cref{sec:locking} \\
        \texttt{fplan} & Path to frequency-plan file, or locking beatnote frequencies $\nu_{\text{PD},r}(\tau)$ & \si{\hertz} & \Cref{sec:locking} \\
        \texttt{central\_freq} & Laser central frequency $\nu_0$ & \si{\hertz} & \Cref{sec:laser-beam-model} \\
        \midrule % Noise models
        \texttt{laser\_asds} & Laser noise \gls{asd} & \si{\hertz\hertz\tothe{-0.5}} & \Cref{sec:noise-models} \\
        \texttt{modulation\_asds} & Modulation noise \gls{asd} & \si{\second\hertz\tothe{-0.5}} & \Cref{sec:noise-models} \\
        \texttt{modulation\_freqs} & Modulation frequencies $\nu^m_{ij}$ & \si{\hertz} & \Cref{sec:local-beams} \\
        \texttt{clock\_asds} & Clock noise \gls{asd} & \si{\hertz\tothe{-0.5}} & \Cref{sec:noise-models} \\
        \texttt{clock\_offsets} & Clock offsets $\delta \hat\tau_{i,0}$ from \gls{tps} $\tau_i$ & \si{\second} & \Cref{sec:timer-model} \\
        \texttt{clock\_freqoffsets} & Clock frequency offsets $y_{i,0}$ & \si{\per\second} & \Cref{sec:noise-models} \\
        \texttt{clock\_freqlindrifts} & Clock frequency linear drifts $y_{i,1}$ & \si{\second\tothe{-2}} & \Cref{sec:noise-models} \\
        \texttt{clock\_freqquaddrifts} & Clock frequency quadratic drifts $y_{i,2}$ & \si{\second\tothe{-3}} & \Cref{sec:noise-models} \\
        \texttt{backlink\_asds} & Backlink noise \gls{asd} & \si{\meter\hertz\tothe{-0.5}} & \Cref{sec:noise-models} \\
        \texttt{backlink\_fknees} & Backlink noise knee frequency & \si{\hertz} & \Cref{sec:noise-models} \\
        \texttt{testmass\_asds} & Test-mass noise \gls{asd} & \si{\meter\second\tothe{-2}\hertz\tothe{-0.5}} & \Cref{sec:noise-models} \\
        \texttt{testmass\_fknees} & Test-mass noise knee frequency & \si{\hertz} & \Cref{sec:noise-models} \\
        \texttt{oms\_asds} & Readout noise \gls{asd} & \si{\meter\hertz\tothe{-0.5}} & \Cref{sec:noise-models} \\
        \texttt{oms\_fknees} & Readout noise knee frequency & \si{\hertz} & \Cref{sec:noise-models} \\
        \texttt{ranging\_biases} & Ranging systematic bias $N_i^{R,o}$ & \si{\second} & \Cref{sec:noise-models} \\
        \texttt{ranging\_asds} & Ranging noise \gls{asd} & \si{\second\hertz\tothe{-0.5}} & \Cref{sec:noise-models} \\
		\bottomrule
	\end{tabular}
	\caption{Simulation parameters available to configure \lisainstrument, alongside with their units and reference sections.}
	\label{tab:simulation-parameters}
\end{table*}
\addtolength{\tabcolsep}{-4pt}

\addtolength{\tabcolsep}{4pt}
\begin{table*}
	\centering
	\begin{tabular}{cccc}
		\toprule
		Dataset & Description & Unit &  Reference \\
		\midrule % ISI Carrier Beatnotes
		\texttt{isi\_carrier\_offsets} & \Gls{isi} Carrier Beatnote Frequency Offsets & \si{\hertz} & \Cref{eq:isi-c-offsets-final} \\
		\texttt{isi\_carrier\_fluctuations} & \Gls{isi} Carrier Beatnote Frequency Fluctuations & \si{\hertz} & \Cref{eq:isi-c-fluctuations-final} \\
		\texttt{isi\_carriers} & \Gls{isi} Carrier Beatnote Total Frequency & \si{\hertz} & \Cref{eq:isi-c-final} \\
		\midrule % ISI Sideband Beatnotes
		\texttt{isi\_usb\_offsets} & \Gls{isi} Upper-Sideband Beatnote Frequency Offsets & \si{\hertz} & \Cref{eq:isi-usb-offsets-final} \\
		\texttt{isi\_usb\_fluctuations} & \Gls{isi} Upper-Sideband Beatnote Frequency Fluctuations & \si{\hertz} & \Cref{eq:isi-usb-fluctuations-final} \\
		\texttt{isi\_usbs} & \Gls{isi} Upper-Sideband Beatnote Total Frequency & \si{\hertz} & \Cref{eq:isi-usb-final} \\
		\midrule % RFI Carrier Beatnotes
		\texttt{rfi\_carrier\_offsets} & \Gls{rfi} Carrier Beatnote Frequency Offsets & \si{\hertz} & \Cref{eq:rfi-c-offsets-final} \\
		\texttt{rfi\_carrier\_fluctuations} & \Gls{rfi} Carrier Beatnote Frequency Fluctuations & \si{\hertz} & \Cref{eq:rfi-c-fluctuations-final} \\
		\texttt{rfi\_carriers} & \Gls{rfi} Carrier Beatnote Total Frequency & \si{\hertz} & \Cref{eq:rfi-c-final} \\
		\midrule % RFI Sideband Beatnotes
		\texttt{rfi\_usb\_offsets} & \Gls{rfi} Upper-Sideband Beatnote Frequency Offsets & \si{\hertz} & \Cref{eq:rfi-usb-offsets-final} \\
		\texttt{rfi\_usb\_fluctuations} & \Gls{rfi} Upper-Sideband Beatnote Frequency Fluctuations & \si{\hertz} & \Cref{eq:rfi-usb-fluctuations-final} \\
		\texttt{rfi\_usbs} & \Gls{rfi} Upper-Sideband Beatnote Total Frequency & \si{\hertz} & \Cref{eq:rfi-usb-final} \\
		\midrule % TMI Carrier Beatnotes
		\texttt{tmi\_carrier\_offsets} & \Gls{tmi} Carrier Beatnote Frequency Offsets & \si{\hertz} & \Cref{eq:rfi-c-offsets-final} \\
		\texttt{tmi\_carrier\_fluctuations} & \Gls{tmi} Carrier Beatnote Frequency Fluctuations & \si{\hertz} & \Cref{eq:tmi-c-fluctuations-final} \\
		\texttt{tmi\_carriers} & \Gls{tmi} Carrier Beatnote Total Frequency & \si{\hertz} & \Cref{eq:tmi-c-final} \\
		\midrule % MPRs
		\texttt{mprs} & \Glspl{mpr} & \si{\second} & \Cref{eq:implemented-pseudoranging} \\
		\bottomrule
	\end{tabular}
	\caption{Simulation products, alongside their units, and reference equations. All quantities are output at the measurement sampling rate $f_s^\text{meas}$.}
	\label{tab:simulation-products}
\end{table*}
\addtolength{\tabcolsep}{-4pt}

\section{Results and discussion}
\label{sec:results}

An example code snippet for simple simulation and on-ground processing is given in \cref{sec:simulation-snippet} for the current versions of \lisainstrument and \pytdi. In the following sections, we describe the results of more complete simulations.

\subsection{Telemetry measurements}
\label{sec:telemetry-measurements}

We present here the results of numerical simulations performed with \lisainstrument. The simulations include all noises described in the previous sections, in addition to a gravitational-wave signal from the loudest verification binary listed in \cite{Kupfer:2018jee}. Its orbital period is about \SI{569.4}{\second} and its 4-year \gls{snr} is estimated at \num{113}. We simulated 3 days (about \SI{E6}{\second}) of measurements at the final rate of $f_s^\text{meas} = \SI{4}{\hertz}$. We have scaled the amplitude of the gravitational-wave signal such that its 3-day \gls{snr} matches the expected 4-year \gls{snr}. The light travel times between the spacecraft are computed from orbits files provided by \gls{esa}~\cite{lisaorbits}. They are treated as time varying but remain roughly constant over the simulation duration, with values for each link between \SI{8.17}{\second} and \SI{8.32}{\second}. Lasers are locked in the N1-12 configuration (c.f.,~\cref{sec:locking}) with a frequency plan computed accordingly (G.~Heinzel, 2020, private communication).

\begin{figure}
    \centering
    \includegraphics[width=\columnwidth]{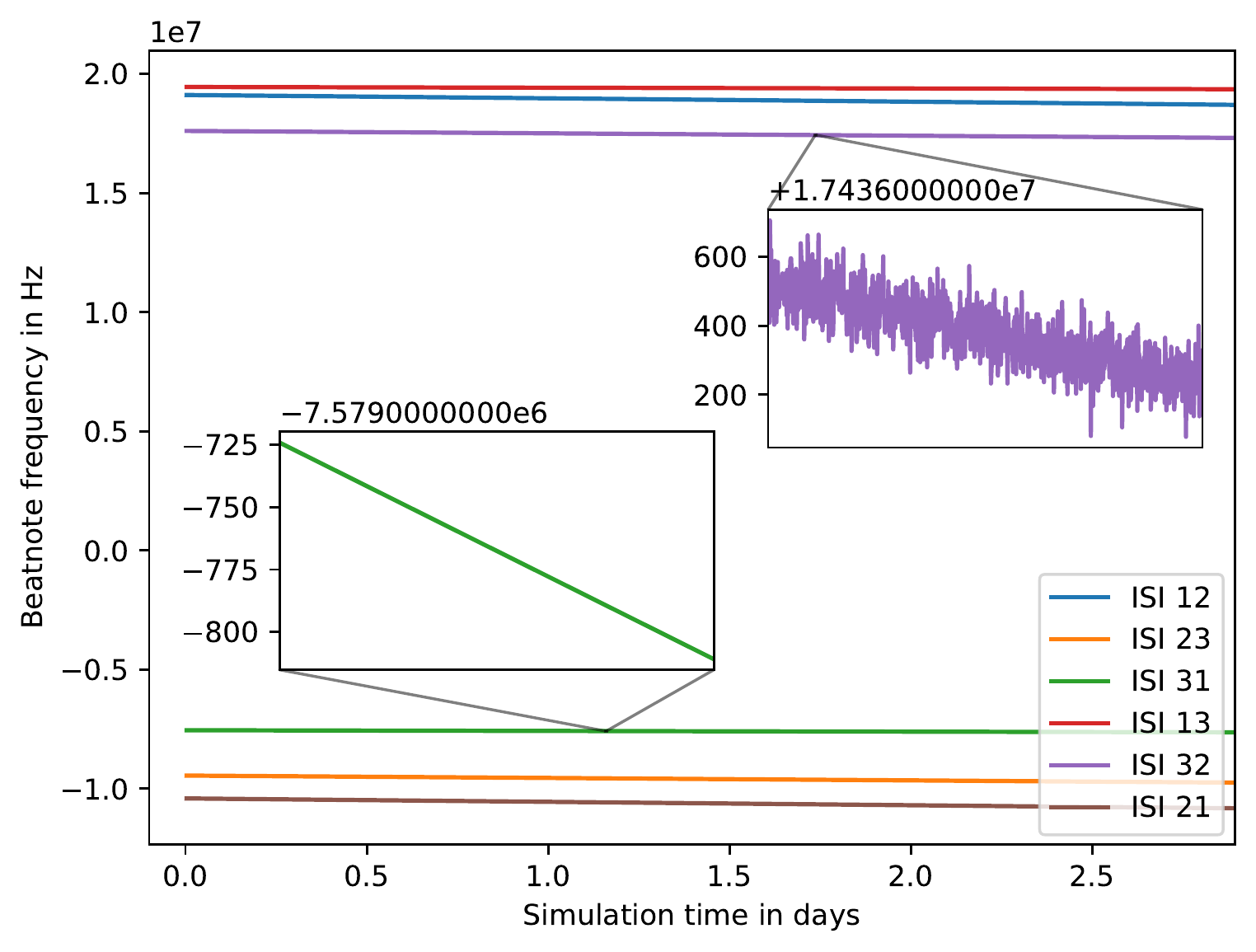}
    \caption{Time series of the \gls{isi} carrier beatnote frequencies. Locking \gls{isi} carrier beatnote frequencies (green 31 and brown 21) are piecewise linear functions driven by the frequency plan, free of any small fluctuations (c.f.~left focus on 31). Non-locking \gls{isi} carrier beatnotes (blue 12, orange 23, red 13, and purple 32) have large trends driven by the frequency plan and Doppler effect, and small in-band fluctuations (c.f.~right focus on 32).}
    \label{fig:isi-timeseries}
\end{figure}

\Cref{fig:isi-timeseries} shows the time evolution of the 6 \gls{isi} beatnotes in terms of total frequency. In the chosen locking configuration, \gls{isi} 31 and 21 beatnotes (in green and brown, respectively) are locking beatnotes, and therefore are piecewise linear functions entirely determined by the frequency plan (c.f.,~\cref{sec:locking}). They do not contain any noise since we assume a perfect laser phase-lock loop at the frequencies we study. Conversely, the remaining \gls{isi} beatnotes (in blue, orange, red, and purple) are non locking. Therefore, they have large \si{\mega\hertz} trends driven by the frequency plan and the relative motion of the spacecraft, in addition to a number of noises, dominated by the $\sim \SI{30}{\hertz\hertz\tothe{-0.5}}$ laser noise.

As expected, the frequency plan ensures that the beatnotes remain in the valid range of the phasemeter, i.e., between \num{5} and \SI{25}{\mega\hertz} in absolute value.

\begin{figure*}
    \centering
    \includegraphics[width=\textwidth]{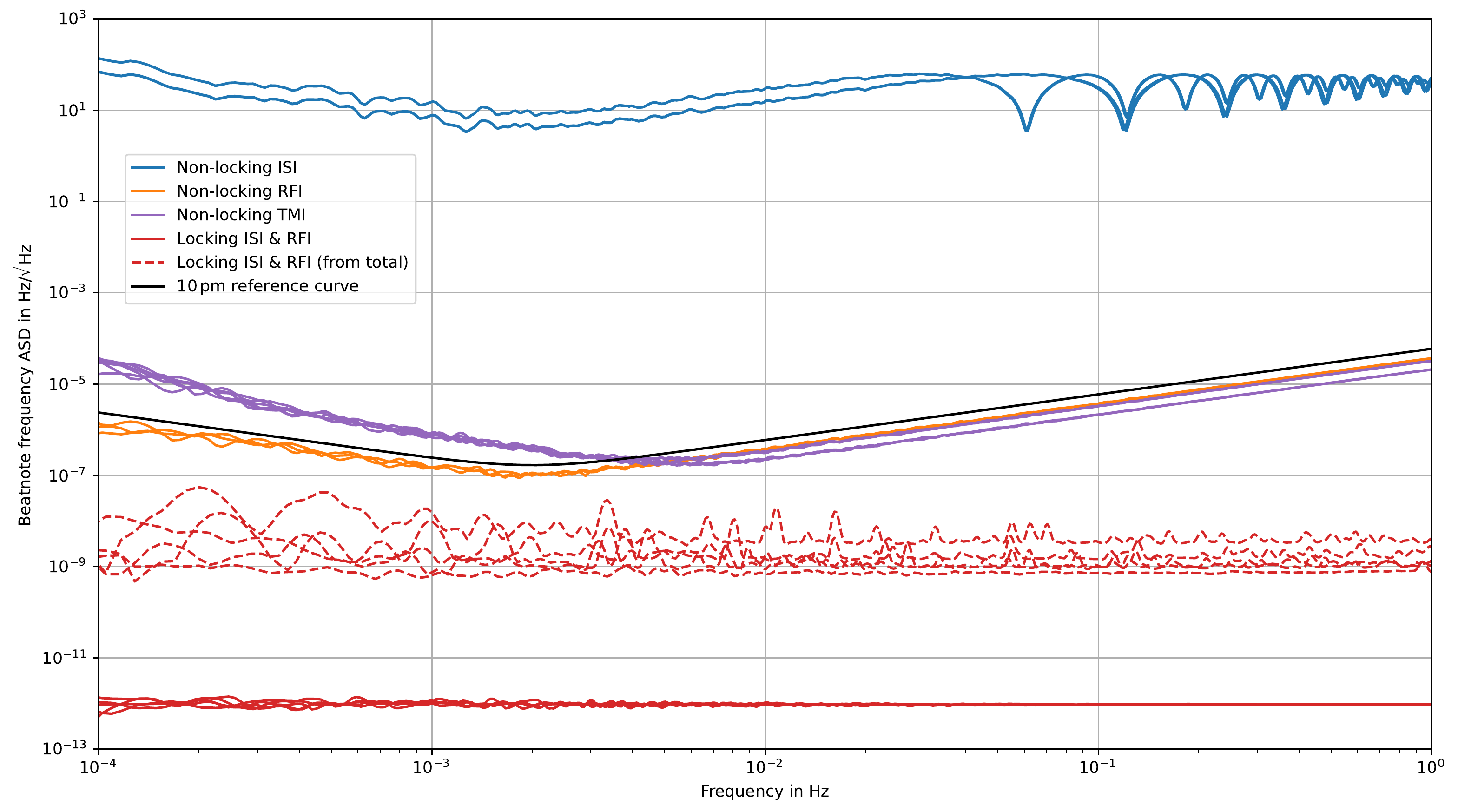}
    \caption{Amplitude spectral densities of all carrier beatnote frequency fluctuations. Non-locking \gls{isi} beatnotes (blue lines) are dominated by laser noise, while non-locking \gls{rfi} and \gls{tmi} beatnotes (orange and purple lines) contain mostly readout and test-mass acceleration noises. The \SI{10}{\pico\meter} reference curve is represented as a bold black line. Locking beatnotes (plain red lines) should be vanishing, but represent here the numerical noise floor at about \SI{1E-12}{\hertz\hertz\tothe{-0.5}}. Non-locking beatnote fluctuations computed from the total beatnote frequencies (dashed red lines) have a 1000x larger numerical noise floor and contain other numerical artifacts.}
    \label{fig:beatnote-spectrum}
\end{figure*}

The amplitude spectral densities of all carrier beatnote frequency fluctuations are presented in \cref{fig:beatnote-spectrum}. We used a Python implementation the \gls{lpsd} method~\cite{Trobs:2006ou} developed by C.~Vorndamme with Kaiser windows. We overlay the \SI{10}{\pico\meter} noise reference curve (in black), which is a typical target noise level for metrology noise in a single \gls{lisa} link~\cite{LISA:2017pwj}. Its \gls{psd} in units of frequency reads
\begin{equation}
    \qty(\frac{2 \pi f}{\SI{1064}{\nano\meter}} \frac{\SI{10}{\pico\meter}}{\si{\sqrt{\hertz}}})^2 \qty[1 + \qty(\frac{\SI{2}{\milli\hertz}}{f})^4].
    \label{eq:ten-picometer-allocation}
\end{equation}
We expect test-mass acceleration noise to remain above this reference curve at low frequencies.

The non-locking \gls{isi} beatnotes (blue lines) are dominated by laser noise (at about \SI{30}{\hertz\hertz\tothe{-0.5}}), only modulated by the one or two-way transfer function. Further processing is required to reduce this laser noise to below the noise requirements, which will reveal the presence on the injected gravitational-wave signal, c.f., \cref{sec:processed-measurements}.

Non-locking \gls{rfi} beatnotes (orange lines) contain mostly readout noises, and therefore remain below the \SI{10}{\pico\meter} noise reference curve (refer to \cref{sec:noise-models} for the noise models used in the simulation). Non-locking \gls{tmi} beatnotes (purple lines) contain, in addition to the same readout noises, test-mass acceleration noises, which become dominant below $\sim \SI{5}{\milli\hertz}$. At these frequencies, the test-mass acceleration noise is clearly above the noise reference curve, as expected. At higher frequencies, we see that the different \glspl{tmi} have different noise levels. On optical benches where the \gls{rfi} is used for locking most common noises in the beams cancel, and the \gls{tmi} is dominated by its own readout noise only. On the adjacent optical benches, on the other hand, we see an increased noise level due to the fibre backlink noise added to the locked laser during propagation between the benches.

\Gls{isi} and \gls{rfi} locking beatnotes are represented as plain red lines. Since we assume perfect laser phase-lock loops, these beatnotes should be vanishing, and we measure here the numerical noise floor of our simulations at about \SI{E-12}{\hertz\hertz\tothe{-0.5}}, well below the expected gravitational-wave signals at about $\nu_0 \times \num{E-21} \approx \SI{E-7}{\hertz}$. Such a low numerical noise floor can be achieved despite the large dynamic range of the quantities in play thanks to the two-variable decomposition described in \cref{sec:two-variable-decomposition} (the precision of beatnote frequency fluctuations are only limited by the magnitude of the laser noise).

We compare these results to what can be obtained with a single-variable model. The same non-locking beatnote fluctuations have been computed by linearly detrending the total beatnote frequencies (to remove large out-of-band trends), and are plotted as dashed red lines. We see a numerical noise floor between \num{E-8} and \SI{E-9}{\hertz\hertz\tothe{-0.5}}, leaving little margin with respect to the expected magnitude of the gravitational signals and secondary noises that we wish to simulate and study.

% p_12 = p
% p_21 = D p + N^isi_21
% p_23 = D p + N^isi_21 + N^rfi_23 + B_23
% p_13 = p + N^rfi_23 + B_13
% p_31 = D p + D N^rfi_23 + N^isi_31
% p_32 = D p + D N^rfi_23 + N^isi_31 + N^rfi_32 + B_32

% isi_21 = D p_12 - p_21 + N^isi_21 = D p - p_21 + N^isi_21 = 0
% rfi_23 = p_21 - p_23 + N^rfi_23 + B_23 = 0
% rfi_13 = p_12 - p_13 + N^rfi_23 + B_13 = 0
% isi_31 = D p_13 - p_31 + N^isi_31 = 0
% rfi_32 = p_31 - p_32 + N^rfi_32 + B_32 = 0

% tmi_12 = p_13 - p_12 + N^tmi_12 + B_12 = N^rfi_23 + N^tmi_12 + B_12 + B_13
% tmi_13 = p_12 - p_13 + N^tmi_13 + B_13 = - N^rfi_23 + N^tmi_13
% tmi_21 = p_23 - p_21 + N^tmi_21 + B_21 = N^rfi_23 + N^tmi_21 + B_21 + B_23
% tmi_23 = p_21 - p_23 + N^tmi_23 + B_23 = -N^rfi_23 + N^tmi_23
% tmi_31 = p_32 - p_31 + N^tmi_31 + B_31 = N^rfi_32 + N^tmi_31 + B_31 + B_32
% tmi_32 = p_31 - p_32 + N^tmi_32 + B_32 = -N^rfi_32 + N^tmi_32

\begin{figure}
    \centering
    \includegraphics[width=\columnwidth]{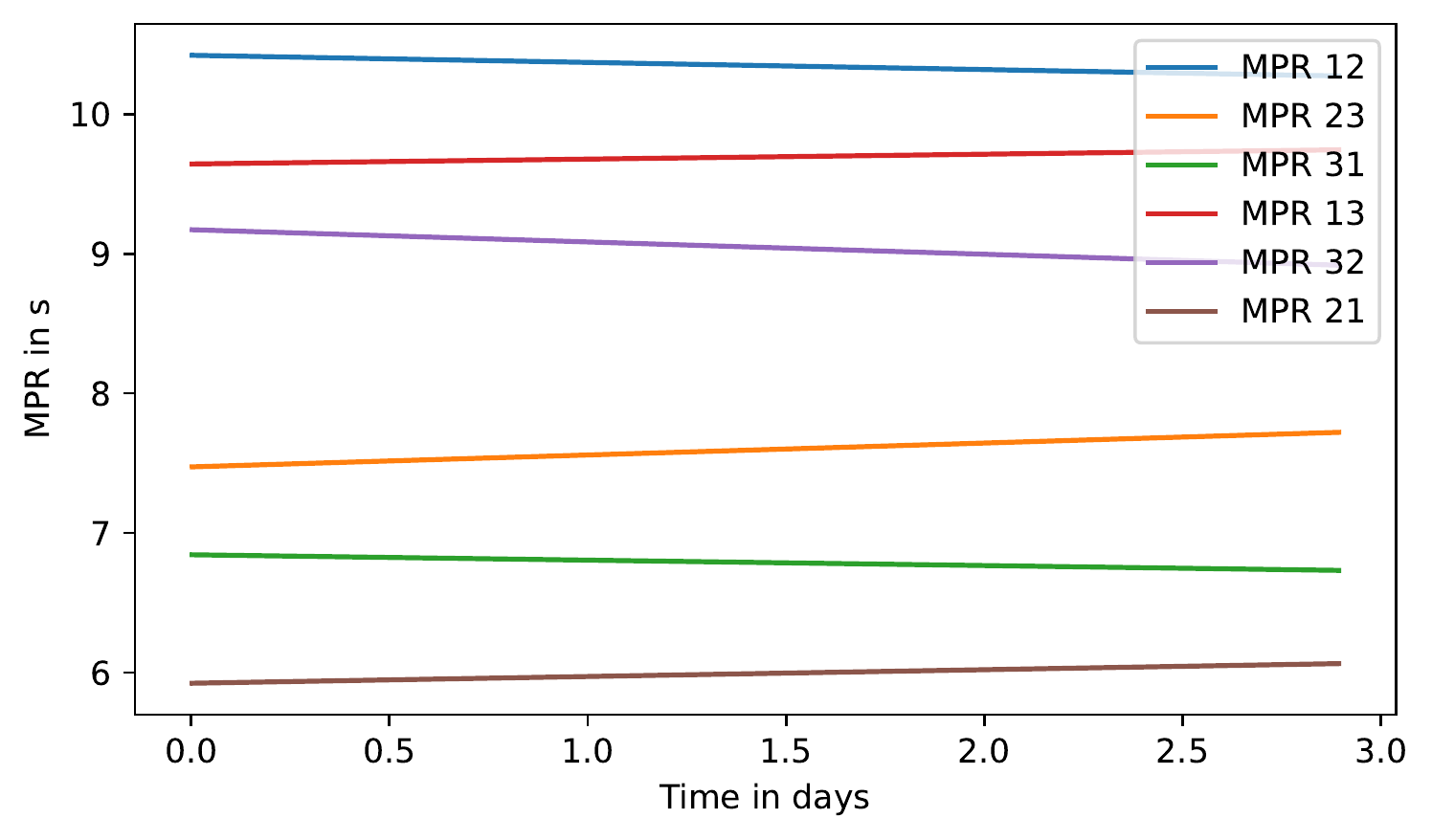}
    \caption{Time series of the \glspl{mpr}. We can observe large deterministic errors w.r.t. the expected light travel time of $\approx \SI{8}{\second}$ due to initial timer offsets and clock drifts.}
    \label{fig:mpr-timeseries}
\end{figure}

Next, \cref{fig:mpr-timeseries} shows time series of the 6 \glspl{mpr}, as described in \cref{sec:pseudoranging}. In addition to the expected light travel times of about \SI{8}{\second}, we can observe that they also include the differential initial timer offsets (of a few seconds) and clock drifts (a few tens of milliseconds per day).

\begin{figure}
    \centering
    \includegraphics[width=\columnwidth]{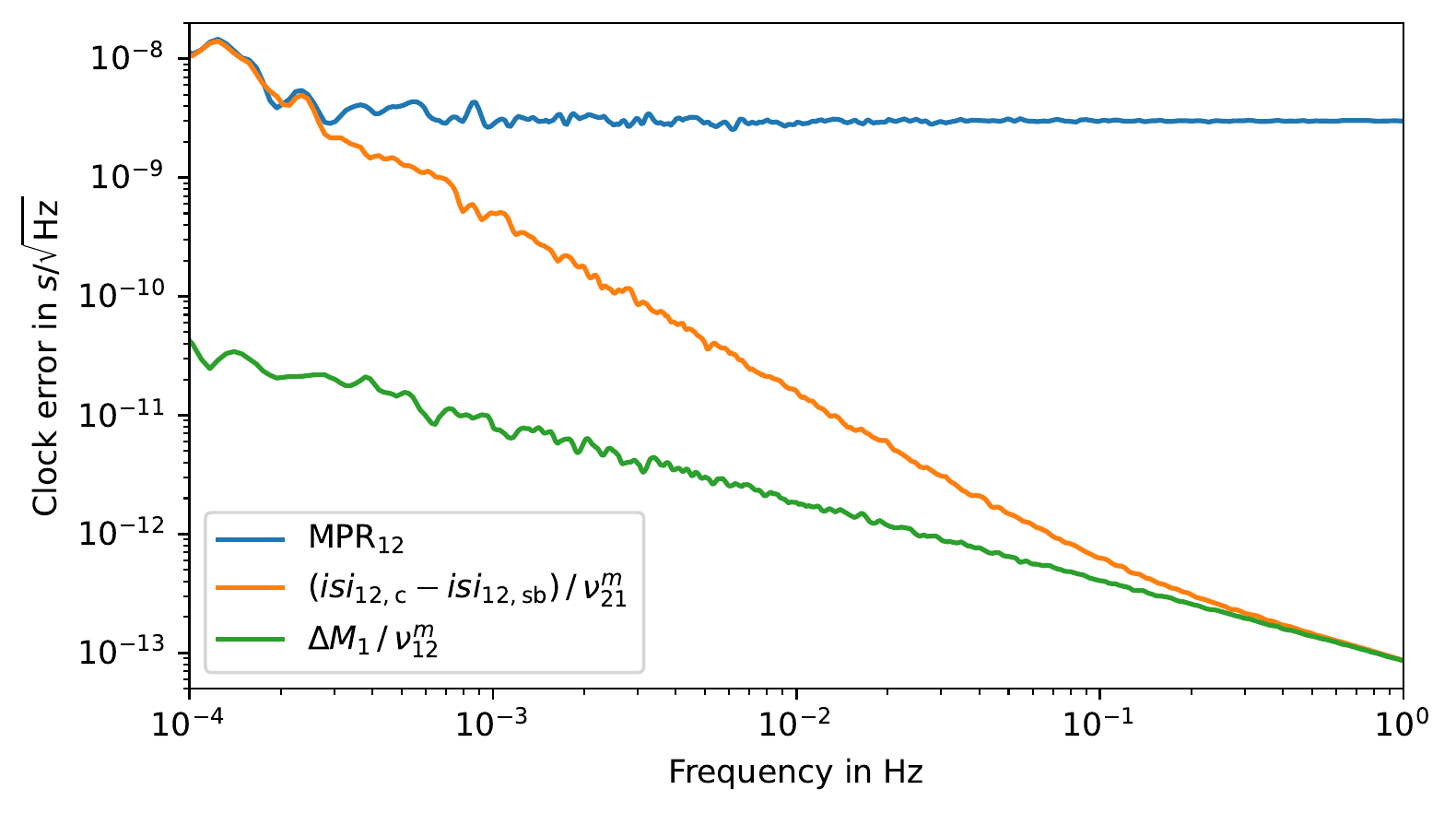}
    \caption{PSDs of clock-noise-related measurements. We plot the \gls{mpr} (blue line) alongside measurements derived from the \gls{isi} sidebands (orange line) and the \gls{rfi} sidebands (green line). See main text for details.}
    \label{fig:clock-signals-spectra}
\end{figure}

Finally, we show in \cref{fig:clock-signals-spectra} the \glspl{psd} of the different clock-noise-related measurements we simulate, all converted to units of \si{\second\hertz\tothe{-0.5}}. The \gls{mpr} (blue) is dominated by a white noise down to the lowest frequencies. Around \SIrange{0.1}{0.3}{\milli\hertz}, the noise level coincides with that of the clock noise measured by the \gls{isi} sideband measurements (orange). Here, we plot a signal combination rejecting common mode noise between carrier and sideband, following~\cite[eq.~B1]{Hartwig:2022yqw}, such that the plotted curve is dominated by the actual clock noise at most of the frequency band. Similarly, using~\cite[eq.~B9]{Hartwig:2022yqw} of the same reference, we can combine the \gls{rfi} sideband beatnotes to give a measurement of the larger of the two modulation noise terms, labeled $\Delta M_1$ (green). We see that the modulation noise is orders of magnitude smaller than the clock noise in most of the band.% The sideband measurements are rescaled by a factor $2\pi f$ to convert them to an equivalent timing jitter.

\subsection{Processed measurements}
\label{sec:processed-measurements}

We have seen that the raw telemetered beatnotes can be grouped in three categories. Locking beatnotes do not contain any gravitational-wave signal or noise (assuming perfect laser locking) and are dominated by numerical noises in our simulations. Non-locking \gls{rfi} beatnotes are also signal-free and are dominated by secondary (readout and test-mass) noises. Only non-locking \gls{isi} beatnotes carry useful gravitational-wave information, but contain laser noise at many orders of magnitude above the expected signals, alongside other noise sources.

In order to detect and analyze the gravitational-wave signals, we must therefore reduce these sources of noise to reasonable levels. This is achieved by a processing technique called \gls{tdi}, in which multiple interferometric readouts are time-shifted and combined to cancel the main noise sources.

\begin{figure}
    \centering
    \includegraphics[width=\columnwidth]{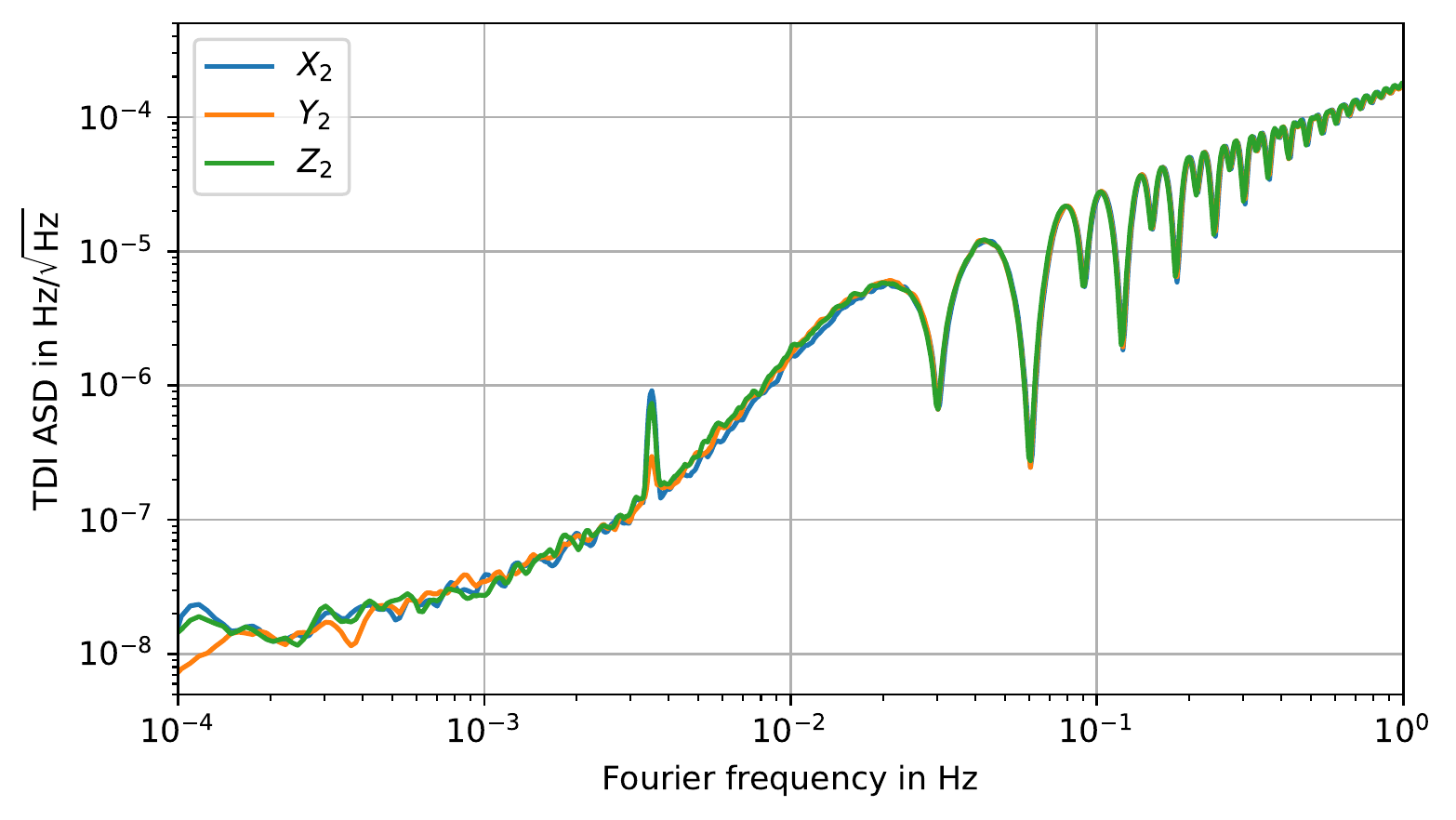}
    \caption{Amplitude spectral densities of second-generation \gls{tdi} combinations $X_2, Y_2$ and $Z_2$. Note that laser noise, overwhelming in the telemetered beatnote frequencies, is suppressed by many orders of magnitude, such that the injected verification binary (\gls{snr} of \num{113}) is now clearly visible around \SI{3.5}{\milli\hertz}.}
    \label{fig:tdi-spectrum}
\end{figure}

To demonstrate that this kind of processing is possible using our simulated data, we apply the algorithm described in~\cite{Hartwig:2022yqw}, using \pytdi~\cite{pytdi}, to reduce the limiting noise sources included in our simulation (laser and clock noise) to below the required level. \Cref{fig:tdi-spectrum} shows the spectra of the second-generation \gls{tdi} combinations $X_2, Y_2$ and $Z_2$, in which the laser and clock noises have been suppressed. The gravitational-wave signal is clearly visible at the expected frequency of \SI{3.5}{\milli\hertz}, with an \gls{snr} of about 100.

Following the conclusions of \cite{Hartwig:2022yqw}, the biggest contributors to the residual noise are the \gls{isi} readout noise at high frequencies and the test-mass acceleration noise at low frequencies; other nonsuppressed noise sources have a smaller but non-negligible contribution.

\section{Conclusion}
\label{sec:conclusion}

In this paper, we proposed a model of the \gls{lisa} measurement chain, which includes the propagation of optical signals (modulated laser beams, each containing a carrier and an upper sideband) across the constellation and on the optical benches; the phase readout of the different interferometers; as well as the on-board processing of the beatnote signals. We also included a high-level model for the \gls{mpr} auxiliary measurements, which are used to estimate the interspacecraft distances, necessary for the on-ground processing (such as \gls{tdi}). This model accounts for laser locking control loops, and properly treats different time frames and clock errors.

We presented two implementations of the model, along with a comparison of their runtime and memory performance that highlights their respective advantages and drawbacks. \lisainstrument is a Python implementation that is easy to use, and very efficient for short simulations (a few months or less), while we recommend \lisanode for longer and more complex simulations.

Some results obtained with \lisainstrument are presented to demonstrate the correctness of the implementation. In particular, we check that the beatnote measurements exhibit the expected behavior, in terms of total frequency or frequency fluctuations. For the latter, we show that our model keeps numerical noise to acceptable levels for the study of instrumental noises and gravitational-wave signals. We also check that laser and clock noise reduction by \gls{tdi} performs as expected by computing the second-generation Michelson combinations. We confirm that the residual noises in these channels matches their expected levels, and that a typical gravitational signal hidden in the raw telemetry data becomes clearly apparent at the expected frequency and magnitude.

The injection of gravitational-wave signals is possible through the multiple interfaces of \lisainstrument and \lisanode with other simulation tools. These tools include in particular \lisaorbits~\cite{lisaorbits} for realistic spacecraft orbits, \lisaglitch~\cite{lisaglitch} for injection of instrumental artifacts, \lisagwresponse~\cite{lisagwresponse} for injection of gravitational signals, and \pytdi~\cite{pytdi} for further on-ground processing.

At the time this paper is written, some important instrumental effects are still under development. While some of them are already implemented in the simulation, we do not include them in this description of the model but refer to future dedicated publications. In particular, we do not include any dynamical effects or dynamical control loops to actuate on the spacecraft and test masses~\cite{Inchauspe:2022hlu}. We also do not include tilt-to-length effects, which occur as an apparent path length change due to any misalignment of optical elements and that will be partially mitigated on ground using \gls{dws} measurements~\cite{Paczkowski:2022nrt}. We do not simulate any of the ground-based observations that will be used to determine the spacecraft positions and velocities, as well as the offsets of the onboard clocks with respect to a global timescale. These are required inputs to the further processing and data analysis steps, and therefore will be included in a future version of the simulation.

Finally, we currently produce phasemeter measurements expressed as total frequencies in \si{\hertz}. There are ongoing discussions to choose the best representation for telemetry data; in order to capture any effects related to this choice, we plan to update the simulators to use the official data format once it is agreed upon. In particular, if phase data must be produced, the model should be updated to include an initial phase for each optical beam and clock signal, which do not affect the current frequency data.

\appendix

\section{Symbol glossary}
\label{sec:symbols}

This paper defines a large number of quantities and uses or introduces many symbols to represent them. To facilitate the reading, we have listed the main conventions in \cref{tab:conventions} and the main quantities in \cref{tab:quantities}.

\addtolength{\tabcolsep}{4pt}
\begin{table*}
	\centering
	\begin{tabular}{cl}
		\toprule
		Symbol & Description \\
		\midrule
		$\dottedsquare^t(\tau)$ & Quantity expressed in the \glsfirst{tcb} (global time frame) \\
		$\dottedsquare^{\tau_i}(\tau)$ & Quantity expressed in the \glsfirst{tps} $i$ (related to $t$ by relativistic corrections) \\
		$\dottedsquare^{\hat\tau_i}(\tau)$ & Quantity expressed in the clock time on board spacecraft $i$ (related to $\tau_i$ by instrumental imperfections) \\
		\midrule
		$\dot\dottedsquare$ or $\dv{\tau}$ & Time derivative of quantity (in the specified time frame) \\
		\midrule
		$\dottedsquare_i$ & Quantity related to spacecraft $i$ \\
		$\dottedsquare_{ij}$ & Quantity related to \gls{mosa} $ij$ \\
		$\dottedsquare_{A \leftarrow B}$ & Quantity measured on $A$ propagated from $B$ \\
		$\dottedsquare_\text{PD}$ & Quantity related to a generic photodetector (no indices)\\
		\midrule
		$\dottedsquare_\text{c}$ & Quantity related to the main carrier \\
		$\dottedsquare_m$ or $\dottedsquare^m$ & Quantity related to the phase-modulation signal \\
		$\dottedsquare_\text{sb}$ or $\dottedsquare_{\text{sb}^+}$ & Quantity related to the upper sideband \\
		$\dottedsquare_{\text{sb}^-}$ & Quantity related to the lower sideband \\
		\midrule
		$\dottedsquare^o$ & Large out-of-band component of a quantity (offsets or drifts) \\
		$\dottedsquare^\epsilon$ & Small in-band component of a quantity (fluctuations) \\
		\bottomrule
	\end{tabular}
	\caption{Summary of the conventional notations used in this paper.}
	\label{tab:conventions}
\end{table*}
\addtolength{\tabcolsep}{-4pt}

\addtolength{\tabcolsep}{4pt}
\begin{table*}
	\centering
	\begin{tabular}{ccl}
		\toprule
		Symbol & Unit & Description \\
		\midrule
		$c$ & \si{\meter\per\second} & Speed of light in a vacuum \\
		$\nu_0$ & \si{\hertz} & Optical frequency of the lasers (\SI{281.6}{\tera\hertz}) \\
		\midrule
		$\Phi_{ij}$ & Cycles & Total phase of local beam at laser source $ij$ \\
		$\nu_{ij}$ & \si{\hertz} & Total frequency of local beam at laser source $ij$ \\
		$O_{ij}$ & \si{\hertz} & Carrier frequency offset of local beam at laser source $ij$ \\
		$p_{ij}$ & Cycles & Carrier phase fluctuations of local beam at laser source $ij$ \\
		\midrule
		$\Phi_{ij,m}$ & Cycles & Total phase of modulating signal on \gls{mosa} $ij$ \\
		$\nu_{ij,m}$ & Cycles & Total frequency of modulating signal on \gls{mosa} $ij$ \\
		$\nu^m_{ij}$ & \si{\hertz} & Nominal frequency of the modulating signal on \gls{mosa} $ij$ \\
		\midrule
		$\Phi_{ij \leftarrow ji}$ & Cycles & Total phase of beam $ij$ propagated to \gls{mosa} $ji$ \\
		$\nu_{ij \leftarrow ji}$ & \si{\hertz} & Total frequency of beam $ij$ propagated to \gls{mosa} $ji$ \\
		$\Phi_{\text{ifo}_{ij} \leftarrow {kl}}$ & Cycles & Total phase of the beam $kl$ propagated to photodetector ifo (isi, rfi, or tmi) \\
		$\nu_{\text{ifo}_{ij} \leftarrow {kl}}$ & \si{\hertz} & Total frequency of the beam $kl$ propagated to photodetector ifo (isi, rfi, or tmi) \\
		\midrule
		$\Phi_{\text{ifo}_{ij}}$ & Cycles & Optical beatnote total phase of ifo (isi, rfi, or tmi) \\
		$\nu_{\text{ifo}_{ij}}$ & \si{\hertz} & Optical beatnote frequency of ifo (isi, rfi, or tmi) \\
		\midrule
		$\hat\tau^{\tau_i}_i$ & \si{\second} & Instrumental clock time on board spacecraft $i$ as a function of the \gls{tps} \\
		$\delta\hat\tau_i$ & \si{\second} & Deviations of instrumental clock time on board spacecraft $i$ from \gls{tps} $i$\\
		$\tau^{\hat\tau_i}_i$ & \si{\second} & \gls{tps} of spacecraft $i$ as a function of instrumental clock time on board spacecraft $i$ \\
		\midrule
		$\filter$ & None & Filter operator, modeling the filtering and decimation stages from \SI{16}{\hertz} to \SI{4}{\hertz} \\
		$\timestamp{i}$ & None & Phase timestamping operator, transforming a phase quantity from \gls{tps} $i$ to instrumental clock time $i$ \\
		$\dtimestamp{i}$ & None & Frequency timestamping operator, transforming a frequency quantity from \gls{tps} $i$ to instrumental clock time $i$ \\
		$\delay{ij}$ & None & Delay operator, time-shifting a phase quantity by $d_{ij}$ \\
		$\ddelay{ij}$ & None & Doppler-delay operator, time-shifting a frequency quantity by $d_{ij}$ (including Doppler corrections) \\
		\midrule
		$a_{ij}$ & \si{\hertz} & Short-hand notation for the optical beatnote frequency offsets in the \gls{isi} $ij$ \\
		$b_{ij}$ & \si{\hertz} & Short-hand notation for the optical beatnote frequency offsets in the \gls{rfi} and \gls{tmi} $ij$ \\
		\midrule
		$\text{ifo}_{ij}$ & \si{\hertz} & Frequency readout of the ifo (isi, rfi or tmi) \\
		\midrule
		$q_i$ & \si{\second} & Noise of instrumental clock on board spacecraft $i$ with respect to \gls{tps} \\
		$M_{ij}$ & \si{\second} & Modulation noise on \gls{mosa} $ij$ \\
		$N^\text{ob}_{\text{ifo}_{ij} \leftarrow {kl}}$ & \si{\second} & Noise of beam $kl$ propagated to ifo (isi, rfi, or tmi) due to optical path length variations on the optical bench \\
		$N^\text{ro}_{\text{ifo}_{ij}}$ & \si{\second} & Readout noise in ifo (isi, rfi, or tmi) \\
		$N^\text{bl}_{\text{ifo}_{ij}}$ & \si{\second} & Backlink noise in ifo (rfi or tmi) \\
		$N^\delta_{ij}$ & \si{\second} & Test-mass displacement noise \\
		$N^R_{ij}$ & \si{\second} & Ranging noise \\
		\midrule
		$d_{ij}$ & \si{\second} & \Glsfirst{ppr} as the difference of \glspl{tps} $i$ (at reception) and $j$ (at emission) \\
		$R_{ij}$ & \si{\second} & \Glsfirst{mpr} $ij$ as the difference of clock times $i$ (at reception) and $j$ (at emission) \\
		$H_{ij}$ & \si{\second} & Integrated fluctuations of the \gls{ppr} $ij$ due to gravitational waves \\
		\bottomrule
	\end{tabular}
	\caption{Symbols for quantities used in this model alongside the units used in the simulation.}
	\label{tab:quantities}
\end{table*}
\addtolength{\tabcolsep}{-4pt}

\section{Noise models}
\label{sec:noise-models}

We describe here the different noise sources that we include in our simulations. For each noise, we give a short description and its mathematical expression. That includes spectral shapes in the form of their \gls{psd} for stochastic terms, as well as any deterministic effects.

Noise models are derived from allocations or \glspl{cbe} given in the performance model (\gls{lisa} Consortium internal technical note, 2019), where applicable\footnote{Some noise sources do not match the \gls{cbe} given in latest performance model, and we plan to update them as soon as possible to reflect the currently expected LISA performance.}.

We give here a continuous description of these noise models; however, they are actually implemented as discrete noise sources at $f_s^\text{phy} = \SI{16}{\hertz}$.

\paragraph{Laser noise.}

Laser noise describes the optical phase fluctuations in the electromagnetic field of a free-running laser stabilized to a cavity (fluctuations in the field amplitude are not included here). It is given in the performance model by the allocation for the \textit{laser frequency stability} in units of frequency,
\begin{equation}
	\psd{\dot N^p}(f) = \qty(\SI{30}{\hertz\hertz\tothe{-0.5}})^2 \qty[1 + \qty(\frac{\SI{2E-3}{\hertz}}{f})^4]
	\qs
\end{equation}

\paragraph{Modulation noise.}

Modulation noise describes any mismatch in the phase of the modulation sidebands (transmitted to the distance optical bench) and the pilot tone (used as a local timing reference)~\cite{Barke:2015srr}.

Both the \SI{2.4}{\giga\hertz} and the \SI{2.401}{\giga\hertz} sideband signals used for modulation are generated from the local \gls{uso}. The pilot tone is derived from the electrical \SI{2.4}{\giga\hertz} signal using a series of low noise frequency dividers. The optical modulation is performed using an \gls{eom} followed by a fibre amplifier.

The fiber amplifier is the dominating part for the \SI{2.4}{\giga\hertz} signal~\cite{Barke:2015srr}. We fit a rough model to the blue curve in~\cite[fig.~5.13]{Barke:2015srr} to obtain a timing jitter power spectral density of
\begin{equation}
	\psd{M}(f) = \qty(\SI{1E-14}{\second\hertz\tothe{-0.5}})^2 \qty[ 1 + \qty(\frac{\SI{1.5e-2}{\hertz}}{f})^2 ]
	\qs
\end{equation}
Note that the blue curve used for the fit corresponds to a \SI{1}{\watt} fibre amplifier, which allows for lower noise levels than the more recent measurements cited in~\cite{Hartwig:2022yqw} for a \SI{2}{\watt} amplifier. This noise is therefore underestimated in the current version of the simulation.

For the \SI{2.401}{\giga\hertz} signal, we expect a higher noise level due to the electrical conversion chain, which can no longer be realized by simple frequency dividers. Following~\cite{Hartwig:2022yqw}, we model this by increasing the modulation noise in the right-hand side optical benches by a factor 10.

\paragraph{Test-mass acceleration noise.}

Test-mass acceleration noise describes the optical path length variations due to the test-mass motion with respect to its nominal position inside its housing.

It is given in the performance model by the allocation value for the \textit{single test-mass acceleration noise} in acceleration units. We include an extra factor 2 in order to account for the beam reflection onto the test mass, and neglect the high-frequency component because it is smaller than the \gls{oms} displacement noise (see below). Moreover, we whiten the noise at below \SI{E-4}{\hertz} to prevent numerical overflow. We get
\begin{equation}
\begin{split}
	\psd{N^\delta}(f) ={}& \qty(2 \times \SI{2.4E-15}{\meter\second\tothe{-2}\hertz\tothe{-0.5}})^2
    \\
    &\times \qty[1 + \qty(\frac{\SI{0.4E-3}{\hertz}}{f})^2]
	\qs
\end{split}
\end{equation}

Note that this is an this is an out-of-loop value, ignoring the coupling of test mass to spacecraft motion introduced by \gls{dfacs}.

\paragraph{Backlink noise.}

Beams are transmitted between adjacent optical benches using optical fibres. During this transmission, the beams can pick up an additional phase noise term. We model only the nonreciprocal noise terms, i.e., the difference between the phase shift of a beam propagating from optical bench $ij$ to $ik$ vs. that of the beam propagating from $ik$ to $ij$.

Backlink noise is given in the performance model by the allocation for the \textit{reference backlink} in displacement,
\begin{equation}
\begin{split}
	\psd{N^{\text{bl}}}(f) ={}& \qty(\SI{3E-12}{\meter\hertz\tothe{-0.5}})^2
    \\
     &\times \qty[ 1 + \qty(\frac{\SI{2E-3}{\hertz}}{f})^4 ]
	\qs
\end{split}
\end{equation}
We use the same value for the \glspl{tmi} and \glspl{rfi}.

\paragraph{Readout noise.}

We summarize as readout noise the equivalent positional readout error due to technical noise sources, such as shot noise.

The \gls{oms} displacement noise is given in the performance model by the allocation value for the overall displacement long-arm, test-mass, and reference noise entries; in terms of displacement\footnote{The overall displacement noise in the performance model summarizes multiple noise sources, some of which are already accounted for independently in this model. The values for the \gls{oms} displacement noise are therefore overestimates.},
\begin{equation}
	\psd{N^{\text{ro}}}(f) = A^2 \qty[ 1 + \qty(\frac{\SI{2E-3}{\hertz}}{f})^4 ]
	\qc
\end{equation}
where $A = \SI{6.35E-12}{\meter\hertz\tothe{-0.5}}$ for the \glspl{isi}, \SI{1.42E-12}{\meter\hertz\tothe{-0.5}} for the \glspl{tmi}, and \SI{3.32E-12}{\meter\hertz\tothe{-0.5}} for the \glspl{rfi}.

The performance model does not give values for the sideband beatnotes. We approximate them using $\epsilon = \num{0.15}$ instead of \num{0.85} in the shot noise formula to account for the lower power level. This yields $A = \SI{1.25e-11}{\meter\hertz\tothe{-0.5}}$ for the \glspl{isi}, and \SI{7.90E-12}{\meter\hertz\tothe{-0.5}} for the \glspl{rfi}.

\paragraph{Optical bench path length noise.}

Optical bench path length noise summarizes different optical path length noises due to, for example, jitters of optical components in the path of the different beams.

Optical bench path length noise in terms of displacement is given by the performance model as
\begin{equation}
    \psd{N^{\text{ob}}}(f) = A^2
    \qc
\end{equation}
where $A = \SI{4.24E-12}{\meter\hertz\tothe{-0.5}}$ for local beams in \glspl{tmi} and \SI{2e-12}{\meter\hertz\tothe{-0.5}} for local beams in \glspl{rfi}.

\paragraph{Ranging noise.}

Pseudoranging is performed by correlating local and distant \gls{prn} signals, c.f.~\cref{sec:pseudoranging}. Ranging noise describes the imperfection of the overall ranging measurement scheme in a single link due to technical noise sources. Note that this does not include any noise appearing in the pseudorange itself, such as clock noise or changes in the optical path length between the spacecraft.

Pseudoranging is given by an \textit{ad hoc} model, combining a systematic bias $N_i^{R,o}$ and a zero-mean stochastic Gaussian white noise $N_i^{R,\epsilon}(\tau_i)$,
\begin{equation}
	N_i^R(\tau_i) = N_i^{R,o} + N_i^{R,\epsilon}(\tau_i)
	\qc
\end{equation}
with default values of $\psd{N^{R,\epsilon}} = \SI{0.9}{\meter\hertz\tothe{-0.5}}$ and $N_i^{R,o} = \SI{0}{\second}$.

\paragraph{Clock noise.}

\Glspl{uso} on each spacecraft act as central time references for all onboard systems. As described in~\cref{sec:phase-readout}, we actually use the pilot tone as the timing reference for all phasemeter measurements. \textit{Clock noise} here models any deviations of these pilot tones from the corresponding \glsfirst{tps}.

Clock noise is given by the model described in \cite{Hartwig:2022yqw}, expressed in terms of fractional frequency deviations as the sum of a random jitter, and constant deterministic frequency offset, linear drift, and quadratic drift,
\begin{equation}
	\dot q_i(\tau) = \dot N^q_i(\tau) + y_{0,i} + y_{1,i} \tau + y_{2,i} \tau^2
	\qc
\end{equation}
$\dot N^q_i(\tau)$ is a random jitter, generated as a flicker noise with a \gls{psd} between \SI{E-5}{\hertz} and $f_s^\text{phy} / 2 = \SI{8}{\hertz}$ given by
\begin{equation}
    \psd{\dot N^q_i}(f) = \qty(\num{6.32e-14})^2 f^{-1}
    \qc
\end{equation}
The deterministic coefficients are
\begin{align}
    y_0 &\approx \SI{5e-7}{\second\per\second}
    \qc
    \\
    y_1 &\approx \SI{1.6e-14}{\second\per\second\squared}
    \qc
    \\
    y_2 &\approx \SI{9e-23}{\second\per\second\cubed}
    \qs
    \label{eq:clock-offsets-orders-of-magnitude}
\end{align}
These values should be seen as orders of magnitude, and will be different for all 3 \glspl{uso}.

\section{Magnitude of \texorpdfstring{$\dot\nu$}{nu-dot}}
\label{sec:dotnu-discussion}

\begin{figure}
    \includegraphics[width=\columnwidth]{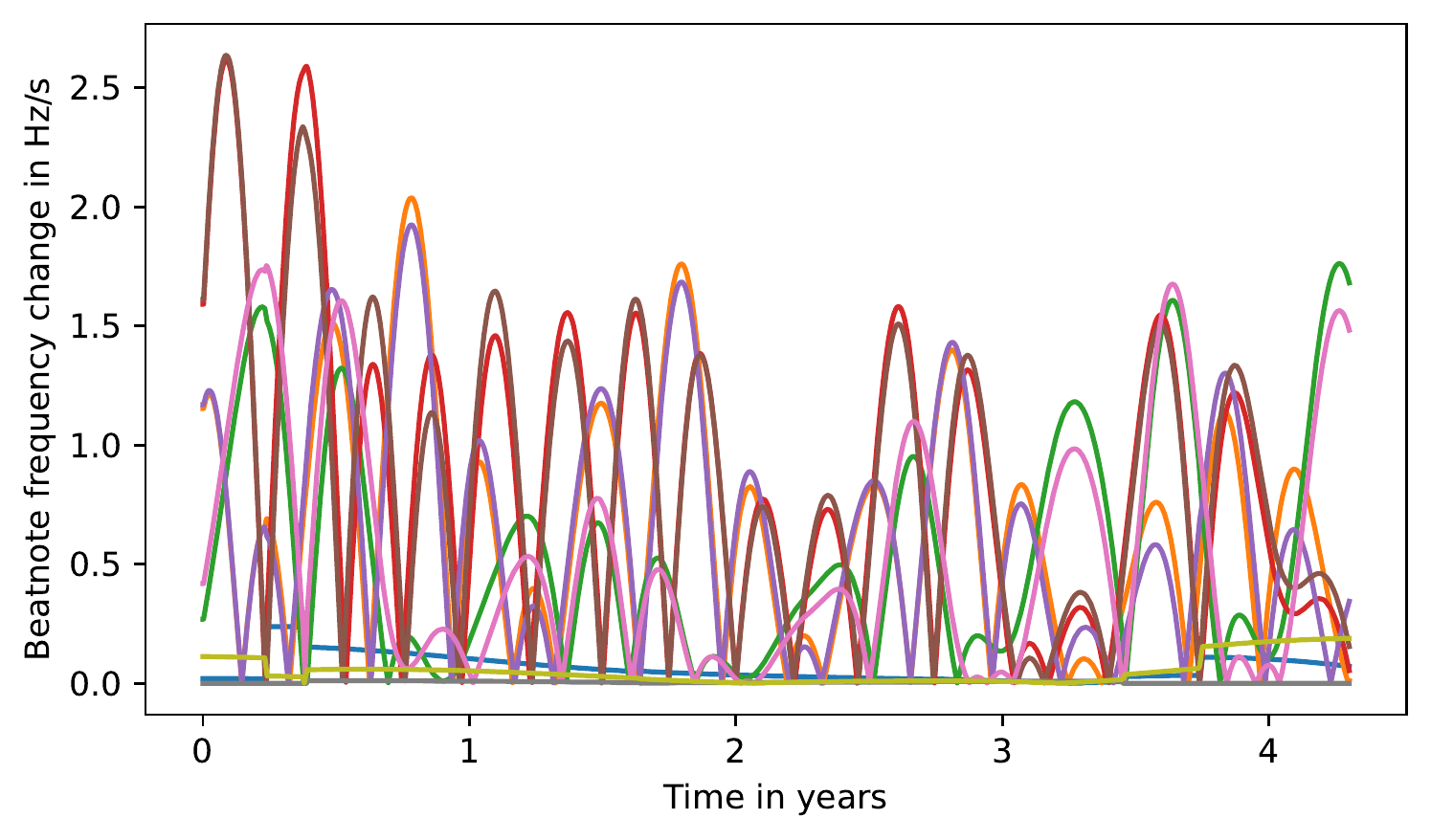}
	\caption{Magnitude of beatnote frequency derivatives for all 9 beatnotes in the example frequency plan. Data provided by G.~Heinzel.}
    \label{fig:frequency-plan-derivs}
\end{figure}

To estimate the order of magnitude of the term $\dot \nu_A^o  H_{12}(\tau)$ we neglected compared to the term $\nu_A^o \dot H_{12}(\tau)$ we included, we can observe the rate of change in the example frequency plan presented in \cref{fig:frequency-plan-derivs}. This is plotted in \cref{fig:frequency-plan-derivs}. As we can see, we have $\dot \nu_A^o < \SI{3}{\hertz\per\second}$ for the whole 4 year duration. On the other hand, $\nu_A^o$ is of the order of \SI{10}{\mega\hertz}. We consider both $\nu_A^o$ and $\dot \nu_A^o$ as constant scaling factors for this estimate.

Note that $H_{12}(\tau)$ and $\dot H_{12}(\tau)$ are noise terms that we can evaluate in the frequency domain. We have
\begin{equation}
    \ft{\dot H_{12}}(f) = 2 \pi f \times \ft{H_{12}}(f)
    \qs
\end{equation}
The usual \gls{lisa} measurement band extends from \SI{E-4}{\hertz} to \SI{1}{\hertz}, such that even at the lower limit of \SI{E-4}{\hertz}, we have
\begin{equation}
\begin{split}
    \nu_A^o \ft{\dot H_{12}}(f) &\approx \SI{E4}{\hertz\per\second} \times \ft{H_{12}}(f)
    \\
    &\gg \dot \nu_A^o \ft{H_{12}}(f) \approx \SI{4}{\hertz\per\second} \times \ft{H_{12}}(t)
    \qs
\end{split}
\end{equation}

Note that the term $\nu_A^o\dot H_{12}(\tau)$ is already a very small correction to the dominant term $\nu_0\dot H_{12}(\tau)$, such that we can safely neglect these additional terms.

\section{Default implementation of the anti-aliasing filter}
\label{sec:default-aa-filter}

\begin{figure}
	\includegraphics[width=\columnwidth]{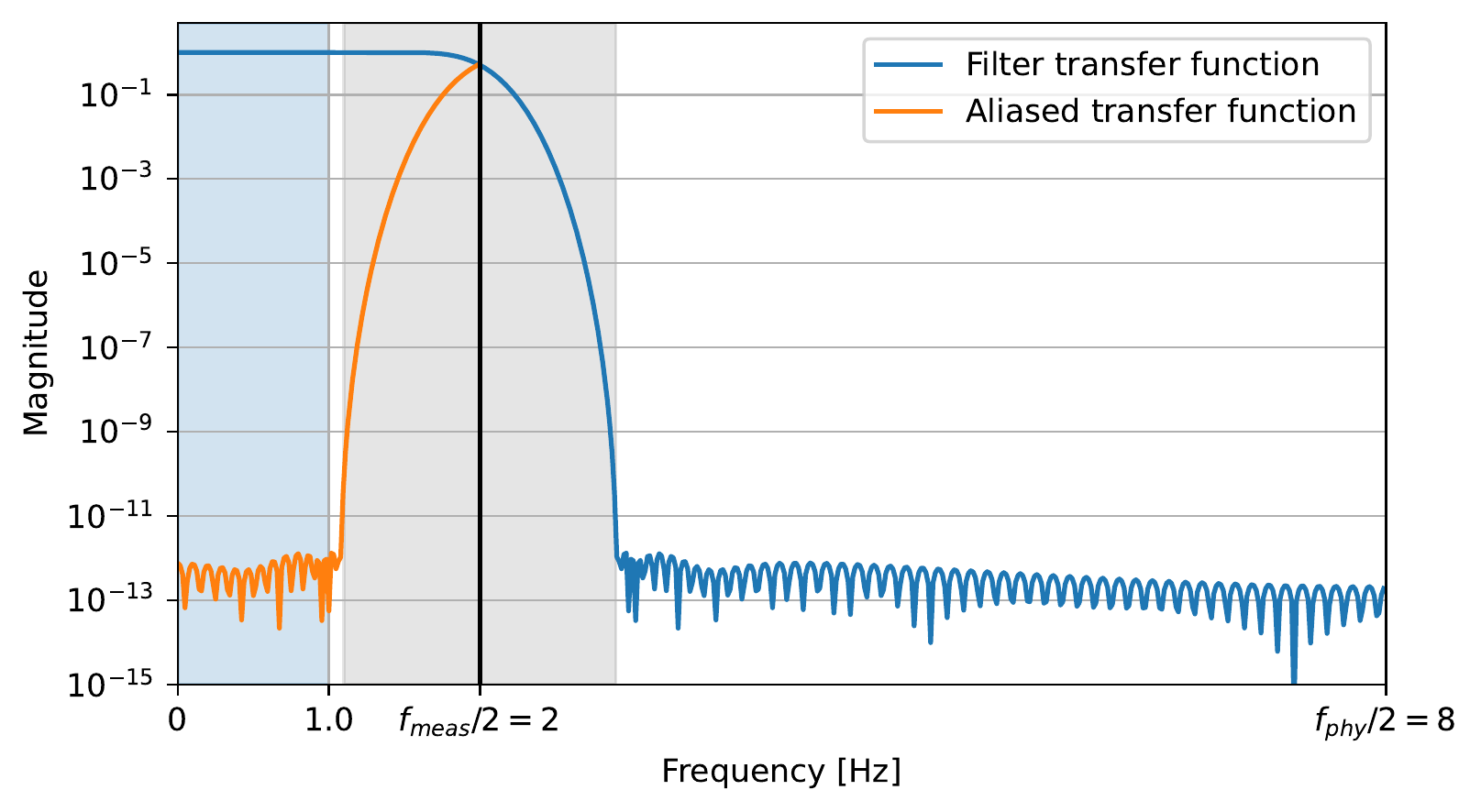}
	\caption{Antialiasing filter transfer function magnitude. The transition band (grey) is chosen to avoid aliasing into the measurement band (blue). Unlike what is often presented, the frequency axis uses a linear scale here.}
    \label{fig:filter-transfer-function}
\end{figure}

By default, the anti-aliasing filter is designed from a Kaiser windowing function, with a transition band that extends from $\SI{1.1}{\hertz}$ to $\SI{2.9}{\hertz}$,
and minimum attenuation above \SI{2.9}{\hertz} of \SI{240}{\decibel}. Note that the filter transition band extends above the Nyquist frequency, such that there will be a significant amount of aliasing during downsampling, as depicted in \cref{fig:filter-transfer-function}. However, since aliasing happens by reflection across the Nyquist frequency, any noise in the band $[f_s^\text{meas}/2, f_s^\text{meas} - \SI{1}{\hertz}]$ will be aliased into the band $[\SI{1}{\hertz}, f_s^\text{meas}/2]$, such that it stays outside our measurement band of $[\SI{E-4}{\hertz}, \SI{1}{\hertz}]$.

Analytically, we model this digital filter with the continuous, linear filter operator $\filter$, which is applied to the beatnote frequency measurements. In the frequency domain, this is equivalent to multiplying our signals by the filter transfer function $\ft{\filter}$, pictured in \cref{fig:filter-transfer-function} and given by
\begin{equation}
    \ft{\filter}(\omega) = \sum_{k=0}^N{c_k e^{-i \omega k / f_s^\text{phy}}}
    \qc
\label{eq:filter-tf-definition}
\end{equation}
where $c_k$ are the filter coefficients.

\section{Example simulation code snippet}
\label{sec:simulation-snippet}

We give here an example simulation and processing code snippet for a simple instrumental setup and minimal processing. The simulation duration is fixed to 10 days. We inject the gravitational-wave signal  of one of the strongest verification binaries expected in the \gls{lisa}~\cite{Kupfer:2018jee}, as described in \cref{sec:telemetry-measurements}. Note that the signal amplitude is rescaled such that the cumulated \gls{snr} over the simulation duration matches the full expected 4-year \gls{snr}. The relative frequency shifts induced by the gravitational-wave strain on the 6 \gls{lisa} links are computed using \lisagwresponse~2.1.2~\cite{lisagwresponse}.

The instrument and the measurements are computed using \lisainstrument~1.1.1~\cite{lisainstrument}. The spacecraft follow a set of Keplerian orbits with an average constellation arm length of \SI{2.5E9}{\meter}. The locking configuration N1-12 is used. Only a white laser frequency noise and the injected signal are present in the data, as all other effects are disabled. We compute the second-generation Michelson \gls{tdi} channels $X_2, Y_2$, and $Z_2$ using \pytdi~1.2.1~\cite{pytdi}, and plot them.

\begin{minted}[
    frame=lines,
    framesep=2mm,
    fontsize=\footnotesize
]{python}
#!/usr/bin/env python3
# -*- coding: utf-8 -*-

import numpy as np
import matplotlib.pyplot as plt
import h5py
import scipy

from lisagwresponse import VerificationBinary
from lisainstrument import Instrument
from pytdi import Data
from pytdi.michelson import X2, Y2, Z2

# Use a standard set of Keplerian orbits
# These are provided by LISA Orbits
orbit_file = 'keplerian-orbits.h5'
with h5py.File(orbit_file) as f:
    orbits_t0 = f.attrs['t0']

# Simulation runs for 3 days at 4 Hz,
# and starts 10 s after the orbit file
dt = 0.25 # s
fs = 1 / dt # Hz
duration = 60 * 60 * 24 * 3 # s
size = duration * fs # samples
t0 = orbits_t0 + 10 # s

# Compute link responses to signal
source = VerificationBinary(
    period=569.4,
    distance=2089,
    masses=(0.8, 0.117),
    glong=57.7281,
    glat=6.4006,
    iota=60 * (np.pi / 180),
    orbits=orbit_file,
    size=size,
    dt=dt,
    t0=t0,
)

# Plot and write the link responses to disk
source.plot(source.t[:8000])
source.write('verification-binary.h5')

# Define the instrumental setup, simulate
# and write the measurements to disk
instru = Instrument(
    orbits=orbit_file,
    gws='verification-binary.h5',
    laser_shape='white',
    size=size,
    dt=dt,
    t0=t0,
)
instru.disable_all_noises(but='laser')
instru.write('measurements.h5')

# Compute TDI Michelson channels
data = Data.from_instrument('measurements.h5')
X2 = X2.build(**data.args)(data.measurements)
Y2 = Y2.build(**data.args)(data.measurements)
Z2 = Z2.build(**data.args)(data.measurements)

# Plot the TDI Michelson channels
psd = lambda tseries: scipy.signal.welch(
    tseries,
    fs=fs,
    nperseg=2**16,
    window=('kaiser', 30),
    detrend=None
)
freq, X2_psd = psd(X2)
freq, Y2_psd = psd(Y2)
freq, Z2_psd = psd(Z2)

plt.loglog(freq, np.sqrt(X2_psd), label='X2')
plt.loglog(freq, np.sqrt(Y2_psd), label='Y2')
plt.loglog(freq, np.sqrt(Z2_psd), label='Z2')

plt.xlabel('Frequency [Hz]')
plt.ylabel('ASD [Hz/, Hz$^{-1/2}$]')
plt.legend()
\end{minted}

\begin{acknowledgments}
The authors thank the LISA Simulation Working Group and the LISA Simulation Expert Group for the lively discussions on all simulation-related activities. They would like to personally thank A.~Petiteau, G.~Heinzel, V.~Müller, A.~Hees, M.~Staab, and J.N.~Reinhardt for their insightful feedbacks.

J.B.B. gratefully acknowledges support from UK Space Agency (grant ST/X002136/1), the Centre National d'\'Etudes Spatiales (CNES), the Central National pour la Recherche Scientifique (CNRS) and the Université Paris-Diderot. J.B.B. has been supported by an appointment to the NASA Postdoctoral Program at the Jet Propulsion Laboratory, California Institute of Technology, administered by Universities Space Research Association under contract with NASA. Part of this research was carried out at the Jet Propulsion Laboratory, California Institute of Technology, under a contract with the National Aeronautics and Space Administration (80NM0018D0004).

O.H. gratefully acknowledges support by Centre National d'\'Etudes Spatiales (CNES) and by the Deutsches Zentrum für Luft- und Raumfahrt (DLR, German Space Agency) with funding from the Federal Ministry for Economic Affairs and Energy based on a resolution of the German Bundestag (Project Ref.~No.~50OQ1601 and 50OQ1801). This work was supported by the Programme National GRAM of CNRS/INSU with INP and IN2P3 co-funded by CNES.
\end{acknowledgments}

%% References
% \bibliographystyle{apsrev4-2}
\bibliography{references}

\end{document}